\documentclass[twoside,twocolumn,english,aps,showpacs,prx,reprint,amsmath,amssymb,notitlepage,superscriptaddress,longbibliography]{revtex4-1}

\usepackage[T1]{fontenc}
\usepackage[utf8]{inputenc}
\usepackage[english]{babel}
\usepackage{amsmath,amssymb}
\usepackage{bm}
\usepackage{bbm}
\usepackage{mathrsfs}
\usepackage{bigints}
\usepackage{braket}
\usepackage{extarrows}
\usepackage{esint}

\usepackage{graphicx}
\usepackage{subfigure} 
\usepackage{booktabs}
\usepackage{multirow}
\usepackage{makecell}
\usepackage{longtable}
\usepackage{dcolumn}
\usepackage{afterpage}

\usepackage{xcolor}
\usepackage{times}

\usepackage{pdfpages}

\usepackage{etoolbox}
\makeatletter
\AtBeginDocument{\let\LS@rot\@undefined}
\@ifundefined{@outputpage@head}{}{%
  \patchcmd{\@outputpage@head}
    {\@ifx{\LS@rot\@undefined}{}{\LS@rot}}
    {}{}{}
}
\patchcmd{\@outputpage}
  {\@ifx{\LS@rot\@undefined}{}{\LS@rot}}
  {}{}{}
\makeatother

\includepdfset{turn=false,angle=0,fitpaper=false}

\usepackage[
  plainpages=false,
  pdfpagelabels,
  bookmarks,
  bookmarksopen=true,
  bookmarksnumbered=true,
  breaklinks=true,
  linktocpage,
  colorlinks=true,
  linkcolor=blue,
  urlcolor=blue,
  citecolor=blue,
  anchorcolor=green,
  hyperindex=true,
  hyperfigures
]{hyperref}

\makeatletter

\newcommand*{\rom}[1]{\expandafter\@slowromancap\romannumeral #1@}
\makeatother

\renewcommand{\vec}{\mathbf}

\DeclareMathOperator{\sgn}{sgn}

\DeclareMathOperator{\tr}{tr}

\DeclareMathOperator{\sech}{sech}

\DeclareMathAlphabet{\mathbbold}{U}{bbold}{m}{n}

\newcommand{\sket}[1]{\left.\left| {#1} \right\rangle\right\rangle}
\newcommand{\sbra}[1]{\left\langle\left\langle {#1} \right|\right.}
\newcommand{\sbraket}[1]{\left\langle\left\langle {#1}\right\rangle\right\rangle}

\begin{document}

\title{Non-Commutative weak measurements: Entanglement, Symmetry Breaking, and the Role of Readout}

\author{Yuanchen Zhao}
\affiliation{State Key Laboratory of Low Dimensional Quantum Physics, Department of Physics, Tsinghua University, Beijing, 100084, China}
\affiliation{Frontier Science Center for Quantum Information, Beijing 100184, China}

\author{Li Rao}
\affiliation{State Key Laboratory of Low Dimensional Quantum Physics, Department of Physics, Tsinghua University, Beijing, 100084, China}
\affiliation{Frontier Science Center for Quantum Information, Beijing 100184, China}

\author{Dong E. Liu}
\email{Corresponding to: dongeliu@mail.tsinghua.edu.cn}
\affiliation{State Key Laboratory of Low Dimensional Quantum Physics, Department of Physics, Tsinghua University, Beijing, 100084, China}
\affiliation{Frontier Science Center for Quantum Information, Beijing 100184, China}
\affiliation{Beijing Academy of Quantum Information Sciences, Beijing 100193, China}
\affiliation{Hefei National Laboratory, Hefei 230088, China}

\begin{abstract}
The preparation of long-range entangled (LRE) states via quantum measurements is a promising strategy, yet its stability against realistic, non-commuting measurement noise remains a critical open question. Here, we systematically investigate the rich phase structure emerging from a minimal model of competing, non-commuting weak measurements: nearest-neighbor Ising ($Z_iZ_j$) and single-qubit transverse ($X_i$) operators. We analyze three experimentally relevant scenarios based on which measurement outcomes are read out: complete readout, no readout, and partial readout. Using a replica mean-field theory for higher dimensions, complemented by numerical simulations in one dimension, we derive the complete finite-time and stationary phase diagrams. Our analysis reveals a striking dependence on the readout protocol. Complete readout  yields a direct transition between a short-range entangled (SRE) phase and a pure LRE phase.
No readout (pure decoherence) precludes entanglement but exhibits a strong-to-weak spontaneous symmetry breaking (SWSSB) transition into a classically ordered mixed state. Most intriguingly, partial readout interpolates between these limits, featuring a mixed-state phase transition where the system can become trapped in the SWSSB phase or, for weaker non-commutativity, undergo successive symmetry breaking to reach a mixed LRE phase. A novel technical contribution is the use of a channel-fidelity-based partition function that allows us to simultaneously characterize both entanglement and SWSSB order, revealing a deep interplay between them in the replica limit. These results provide a cohesive picture for understanding measurement phase transitions, SWSSB, and mixed-state phase transitions, offering crucial insights for designing robust state preparation protocols on noisy quantum devices.
\end{abstract}


\date{\today}

\maketitle
\tableofcontents
\section{Introduction}
Long-range entanglement (LRE) in many-body systems is not only known as an important resource for various tasks in the field of quantum information, such as quantum metrology \cite{metrology} and quantum error correction \cite{ref:dennis,color_code,Haah,hastings_dynamically_2021}, but also a characteristic feature of certain quantum phases such as topological order \cite{chen_LRE}. An efficient method for experimentally preparing certain LRE states on quantum devices is to introduce measurements into the quantum circuit, which requires only constant depth  \cite{verresenEfficientlyPreparingSchrodingers2022}, or in other words constant time, compared with purely unitary circuits that take super constant depth due to the limitation of an effective light-cone \cite{hastingsLocalityQuantumSystems2010,fisherRandomQuantumCircuits2023,yiComplexityOrderApproximate2023}. The measurement preparation method is widely utilizable for stabilizer codes \cite{gottesman1997stabilizercodesquantumerror,ref:dennis,iqbalTopologicalOrderMeasurements2024,tantivasadakarnLongRangeEntanglementMeasuring2024}, Floquet codes \cite{hastingsDynamicallyGeneratedLogical2021,zhangXCubeFloquetCode2022,davydovaFloquetCodesParent2023,davydovaQuantumComputationDynamic2023,duaEngineeringFloquetCodes2023}, and non-Abelian anyons \cite{tantivasadakarnHierarchyTopologicalOrder2023,tantivasadakarnShortestRouteNonAbelian2023,iqbalNonAbelianTopologicalOrder2024,renEfficientPreparationSolvable2024,lyonsProtocolsCreatingAnyons2025}. 


However, the practical implementation of quantum measurements unavoidably suffers from noise. Realizing quantum measurements requires coupling the target quantum system to an external quantum probe. Subsequently, the unitary evolution of this combined system must be precisely controlled. Crucially, the coupling and control is typically non-ideal, leading to the phenomenon of weak measurement~\cite{wisemanQuantumMeasurementControl2009,clerkIntroductionQuantumNoise2010,jacobsQuantumMeasurementTheory2014,ref:guoyi,ref:fisher}, or equivalently positive operator-valued measurement (POVM) \cite{nielsenQuantumComputationQuantum2010}. Under these circumstances, a key question is whether the preparation protocol faithfully generates the desired many-body state. In Ref. \cite{ref:guoyi,ref:fisher}, they developed a framework to study weak measurement-prepared topological stabilizer codes and showed the stability of 2-dimensional Greenberger-Horne-Zeilinger (GHZ) state (or repetition code) and 3-dimensional toric code as phases of matter. Their result has been tested experimentally on a superconducting quantum computer \cite{chenNishimoriTransitionError2025}. Nonetheless, their method only works for commutative weak measurement operators, which may become non-commutative under more realistic noise models \cite{zhao2023latticegaugetheorytopological} or for more complicated 
tasks as in Floquet codes \cite{zhu_qubit_2023}. The current understanding of non-commutativity in weak measurement preparation remains severely limited, making the development of rigorous new analytical methods an urgent necessity.

The question we address also naturally relates to a recently fast-developing topic named measurement-induced phase transition (MIPT), where the competition between unitary evolution and non-unitary measurement is studied under various settings such as monitored quantum circuits \cite{nahumOperatorSpreadingRandom2018a,li2019measurement,Chan_2019,baoTheoryPhaseTransition2020,Turkeshi_2020,vijay2020measurementdrivenphasetransitionvolumelaw,jianMeasurementinducedCriticalityRandom2020,Nahum_2021,PhysRevX.10.041020,PhysRevLett.125.070606,PhysRevB.101.060301,PhysRevLett.128.050602,Han_2022,PhysRevB.104.104305,Lavasani_2021,PhysRevLett.127.235701,Weinstein_2022,Block_2022,liEntanglementDomainWalls2023,fisherRandomQuantumCircuits2023}, continuous dynamics in fermion \cite{
buchholdEffectiveTheoryMeasurementInduced2021,Chen_2020,Alberton_2021,Buchhold_2021,Jian_2021,altlandDynamicsMeasuredManybody2022,buchhold2022revealingmeasurementinducedphasetransitions,Altland_2022,M_ller_2022,Alberton_2021,yangKeldyshNonlinearSigma2023,favaNonlinearSigmaModels2023,poboikoTheoryFreeFermions2023,poboikoMeasurementInducedPhaseTransition2024,müller2025monitoredinteractingdiracfermions} or spin \cite{Rossini_Ising,Turkeshi_Ising,Biella_Ising,Tirrito_Ising,Murciano_Ising} systems and dissipative systems \cite{Sierant_dissipative,Diehl_dissipative}. 
Recently experimental demonstrations of MIPT have been performed on noisy intermediate-scale quantum (NISQ) devices quantum computers \cite{noelMeasurementinducedQuantumPhases2022,kohMeasurementinducedEntanglementPhase2023,hokeMeasurementinducedEntanglementTeleportation2023,kamakariExperimentalDemonstrationScalable2025}.

Here, we examine non-commutative weak measurements, with particular attention to the interplay among mutually non-commuting measurement processes and their competition with environmental decoherence. 
Non-commuting measurement problems are central to quantum information processing because they arise in imperfect stabilizer readout of topological codes and in adaptive feed-forward protocols; understanding their mutual interference and associated errors is therefore important for achieving truly fault-tolerant quantum computation. Within the MIPT framework, introducing non-commuting monitoring can help to enrich the phase diagram as well.
Earlier studies of non--commuting protocols~\cite{langEntanglementTransitionProjective2020,ippolitiEntanglementPhaseTransitions2021,botzungRobustnessMeasurementinducedPercolation2023,tikhanovskayaUniversalityCrossentropySymmetric2024,Zhu_2024_structured,
Sriram_Kitaev,
Lavasani_Kitaev} centred on stochastic projective measurements, which deviate from our motivation of the imperfectness of measurements.
Those models exhibit universality classes related to classical percolation, whereas the weak--measurement imperfections we considered here can map naturally onto the Nishimori line of a disordered Ising theory.
Moreover, most prior work focuses on the infinite--time steady state, yet finite--time behavior is crucial for realistic state--preparation protocols.  Apart from the numerical study of Floquet codes in Ref.~\cite{zhu_qubit_2023}, this regime remains largely unexplored.  We provide an analytical treatment that captures both the finite--time dynamics and the eventual stationary phases of a minimal model governed by non--commuting weak measurements.



\begin{figure*}[t]
    \includegraphics[width=2\columnwidth]{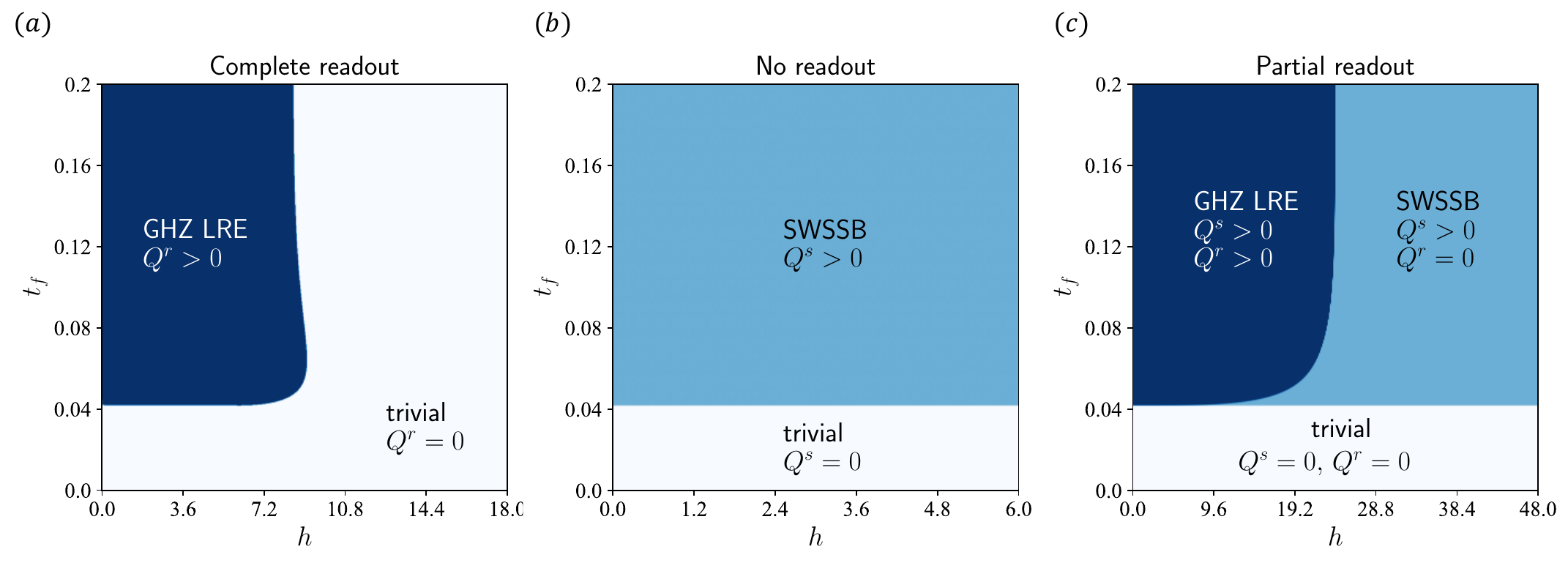}
    \caption{\textbf{Phase diagrams numerically obtained using the mean-field solutions as functions of measurement strength $h$ (which controls the degree of non-commutativity) and the evolution time $t_f$, with fixed parameters $J=1$ and spatial dimension $d=6$.} In all three diagrams, the navy blue marks the GHZ-type lRE state, the light blue marks the SWSSB state, and the white marks the trivial disordered state. From the perspective of SSB, the trivial phase is fully symmetric $\mathbb{Z}_2 \times \mathbb{Z}_2$, the SWSSB phase has weak $\mathbb{Z}_2$ symmetry remained, and the LRE phase has no symmetry. (a) The complete readout phase diagram when both $ZZ$ and $X$ measurement outcomes are collected where the SRE-to-LRE transition manifests. Here the entanglement transition order parameter $Q^r>0$ indicates the LRE phase. (b) The no readout phase diagram when both $ZZ$ and $X$ measurement outcomes are averaged, and the SWSSB phase appears above critical time. Here the SWSSB order parameter $Q^s>0$ indicates the SWSSB phase. (c) The partial readout phase diagram when $X$ measurement outcomes are averaged and $ZZ$ outcomes are collected. Here all three phases exist in the different parameter regions of a single diagram. The behaviors of both order parameters $Q^s$ and $Q^r$ specifies three different phases.} 
    \label{fig:phase}
\end{figure*}

This work studies a minimal continuous-measurement protocol associated with generating GHZ-type long-range-entangled (LRE) states on a qubit lattice.  The system is subject to two weak, mutually non-commuting monitored observables: nearest-neighbor Ising terms \(Z_i Z_j\) and single-qubit transverse operators \(X_i\).  A key experimental knob is whether the measurement records are retained or discarded, so we analyze three scenarios: (i) \textbf{complete readout}, where both \(Z_i Z_j\) and \(X_i\) outcomes are collected; (ii) \textbf{no readout}, where all outcomes are discared and the measurements act purely as decoherence; and (iii) \textbf{partial readout}, in which only the \(Z_i Z_j\) results are kept. Fixing the Ising-measurement ($Z_i Z_j$) strength and varying the transverse strength ($X_i$), we derive a \(d\)-dimensional mean-field theory that yields the phase diagrams shown in Fig.~\ref{fig:phase}.


In the complete readout case as shown Fig.~\ref{fig:phase} (a), there is a critical $X_i$ measurement strength for the stationary state that separates the LRE phase and the short-range entangled (SRE) trivial phase. The stationary property controls the finite time behavior, which mimics the quantum criticality phenomenon at finite temperature \cite{sachdevQuantumPhaseTransitions2011}. Below the critical measurement strength, the system evolves into the LRE phase across a $\mathcal{O}(1)$ critical time. The critical time increases with measurement strength until above the critical strength where the system eventually stays in the trivial phase. 

In the no readout case as shown Fig.~\ref{fig:phase} (b), the measurements effectively become decoherence which drives the system into a mixed state. It is commonly known that MIPT does not manifest and there will not be a LRE phase under these circumstances. However, another classical order occurs after a critical time which is known as strong-to-weak spontaneous symmetry breaking (SWSSB) \cite{lessaStrongtoWeakSpontaneousSymmetry2024,salaSpontaneousStrongSymmetry2024,huang_hydrodynamics_2024,gu2024spontaneoussymmetrybreakingopen,zhang2025strongtoweakspontaneousbreaking1form}, although it is still SRE. In contrast, the LRE phase should completely break the strong symmetry from this perspective.

In the partial readout case as shown Fig.~\ref{fig:phase} (c), additional quantum-critical measurement strength likewise separates the stationary LRE and SRE phases and governs the accompanying finite-time critical behaviour. Here due to the existence of $X$ decoherence, the system becomes a statistical mixture and the criticality belongs to a ``quantum'' version of the mixed state phase transition \cite{fanDiagnosticsMixedstateTopological2023,dai_steady-state_nodate,sang_mixed-state_2023,lee_quantum_2023,bao_mixed-state_2023,zhao2023extractingerrorthresholdsframework,lee2024exactcalculationscoherentinformation,su_higher-form_2024,huang_mixed_2024,huang2024coherentinformationmixedstatetopological,sang2024stabilitymixedstatequantumphases,niwa2024coherentinformationcsscodes,Lee_2025}.
Intriguingly, when the measurement strength falls below the long-time critical strength, the system’s evolution first enters the SWSSB phase before ultimately reaching the LRE phase; by contrast, for strengths above this long-time critical strength, the dynamics become trapped in the SWSSB phase and never develop long-range entanglement.


One of our key theoretical contributions lies in the formulation of an initial-state-independent partition function based on the channel fidelity. This single framework uniquely generates the order parameters for both the LRE and the SWSSB phases (Eq. \eqref{eq:partition-def}), thereby providing a unified lens to analyze their simultaneous existence and interplay. Within the replica method, this approach reveals a crucial insight: the effective actions for the LRE and SWSSB phases decouple, manifesting at different orders in the replica index (Eq. \eqref{eq:decoupling}). This decoupling uncovers a non-trivial dependency where, in the complete readout protocol, the physical LRE order relies on a non-zero SWSSB order parameter that emerges mathematically from the no readout case in the replica limit, even though this SWSSB order is physically absent when all measurement outcomes are collected.


We also provide theoretical analysis and numerical simulations for the one-dimensional (1D) case which is beyond the mean-field solution. The simulation is done with a discrete setup, which is equivalent to the continuous dynamics for sufficiently small measurement strength. Our work is a preliminary example of a systematic analytical study of both finite-time and stationary continuous non-commutative measurements. it also serves as a bridging example of the cutting-edge topics: MIPT, mixed state phase transition and SWSSB, and reveals their deeper connections and differences.

The paper is organized as follows. In Sec. \ref{sec:model} we introduce our model of non-commutative measurement and its relevant symmetries and phases. We also review the replica method here which is used to deal with the measurement randomness. In Sec. \ref{sec:1d} we simply analyze and provide numerical results for the 1D case. In Sec. \ref{sec:meanfield} we derive the mean-field effective theory for general dimensions and present our main results. In Sec. \ref{sec:discussion} we summarize and discuss further questions.

\section{Model}
\label{sec:model}
\subsection{Model of non-commutative measurements}
We consider a $d$-dimensional lattice with periodic boundary conditions and qubits placed on each vertex. The system is subject to two-qubit nearest neighbor $Z_i Z_j$ measurements and single-qubit $X_i$ measurements, which are non-commutative. This is qualitatively equivalent to repeatedly applying $Z_i Z_j$ and $X_i$ discrete weak measurements when they are sufficiently weak (see Appendix \ref{app:1}). Notice that for strong enough repeated discrete measurements, the behavior could be drastically different. 

Suppose that the outcomes of the detector are continuous variables, denoted by $\xi_{ij}(t) \in \mathbb{R}$ and $\xi_i(t) \in \mathbb{R}$ for the instantaneous $Z_i Z_j$ and $X_i$ outcomes respectively. They are also called quantum trajectories. The noises of the measurements are assumed to be Gaussian-type.
Such measurement dynamics are described by a stochastic master equation \cite{scottQuantumNonlinearDynamics2001,wisemanQuantumMeasurementControl2009,jacobsQuantumMeasurementTheory2014},
\begin{equation}
\begin{aligned}
    \partial_t \hat{\rho} &= - J \sum_{\braket{ij}} \left(2\hat{\rho} - 2 Z_i Z_j \hat{\rho} Z_i Z_j - 4 \xi_{{ij}}(t) \{Z_iZ_j, \hat{\rho}\} \right)\\
    &-  h \sum_{i} \left(2\hat{\rho} - 2 X_i \hat{\rho} X_i - 4 \xi_{i}(t) \{X_i, \hat{\rho}\} \right),
\end{aligned}
\label{eq:dynamics}
\end{equation}
where $J$ and $h$ are measurement rates or measurement strengths. The $\xi_{ij}(t)$'s and $\xi_i(t)$'s formally are treated as independent Gaussian disorders ($\mathbb{E}_G$ denotes Gaussian average)
\begin{equation}
    \begin{aligned}
          &\mathbb{E}_G\{\xi_i(t)\} = \mathbb{E}_G\{\xi_{ij}(t)\}= \mathbb{E}_G\{\xi_i(t) \xi_{ij}(t)\} = 0,\\ 
          & \mathbb{E}_G\{\xi_i(t)\xi_j(t')\} = \frac{1}{8h}\delta_{ij} \delta(t-t'),\\
          & \quad \mathbb{E}_G\{\xi_{ij}(t)\xi_{i'j'}(t')\} = \frac{1}{8J}\delta_{ii'} \delta_{jj'} \delta(t-t').\\
    \end{aligned}
\label{eq:disorder}
\end{equation}
Without leading to confusion, we sometimes abbreviate the trajectory configuration as $\xi$.
Notice that the $\hat{\rho}$ is the unnormalized density matrix since the $\xi$ dependent terms alert its trace. If we denote the solution of Eq. \eqref{eq:dynamics} at time $t_f$ given a particular trajectory configuration (abbreviated as $\xi$) as $\hat{\rho}[\xi,t_f]$, the physical final state should be 
\begin{align}
    &\rho [\xi, t_f] =  \frac{\hat{\rho} [\xi, t_f]}{\tr\left(\hat{\rho} [\xi, t_f]\right)}.
\end{align}
The real probability of obtaining the measurement outcomes $\xi$ on the detector is given by 
\begin{equation}
    P[\xi] = P_G [\xi] \tr(\hat{\rho}[\xi,t_f]),
\end{equation}
where $P_G [\xi]$ denotes the Gaussian probability following from Eq. \eqref{eq:disorder}. Appendix \ref{app:1} provides a detailed review of the continuous measurement formalism. In the following, we will denote the statistical average over $P[\xi]$ as $\mathbb{E}$.

Given trajectory $\xi$, the linear equation Eq. \eqref{eq:dynamics} is solved by
\begin{align}
     \hat{\rho} [\xi, t_f] &= K[\xi] \rho (0) K[\xi]^\dagger, \notag\\
    K[\xi] = \mathcal{T}  \exp &\left[ \int_0^{t_f} dt \left(\sum_{\braket{ij}} \left(4J\xi_{ij}(t)Z_iZ_j-2J\right)\right.\right.\notag\\ 
    &\left.\left.+ \sum_i \left(4h\xi_i(t)X_i-2h\right)\right)\right].
\end{align}
Here the time ordering operator $\mathcal{T}$ appears due to the non-commutativity between $ZZ$ and $X$. Such a evolution obeys a global $\mathbb{Z}_2$ symmetry $[K[\xi],\prod_i X_i]=0$. For a $\mathbb{Z}_2$ symmetric initial pure state $\prod_i X_i \ket{\psi(0)} = \pm \ket{\psi(0)}$, the final state $\ket{\psi[\xi,t_f]} \propto K[\xi]\ket{\psi(0)}$ is still pure and $\mathbb{Z}_2$ symmetric, $\prod_i X_i  \ket{\psi[\xi,t_f]} = \pm  \ket{\psi[\xi,t_f]}$.

Setting the $X$-basis measurements to zero ($h = 0$), we find that in the long-time limit $t_f \to \infty$ the Kraus operator $K[\xi]$ converges to a projective measurement onto the joint eigenspaces of $ZZ$ operators,
\begin{equation}
    K[\xi] \propto \prod_{\braket{ij}}\left( \frac{ I+ Z_i Z_j}{2} \delta(s_{ij}- 1) + \frac{ I- Z_i Z_j}{2} \delta(s_{ij}+ 1)\right),
\end{equation}
where $s_{ij} = (1/t_f) \int_0^{t_f} dt \xi_{ij}(t)$ is the time-averaged measurement outcome and will converge to the eigenvalues of $Z_iZ_j$. Consequently, starting from a product of $X=+1$ eigenstates $\ket{\psi(0)} = \ket{+}^{\otimes n}$ ($n$ is the system size), the final state approaches the form of GHZ state \cite{greenberger_going_2007} which is long-range entangled,
\begin{equation}
    \ket{\psi[\xi,t_f]} \rightarrow \frac{\ket{\{z_i\}} + \ket{\{z_i+1\}}}{\sqrt{2}}.
\end{equation}
Here the $z_i = 0,1$'s label the computational basis ($Z$ basis) and their specific realizations depend on the measurement outcomes. In contrast, activating the $X$ measurements damages the preparation of the GHZ state, because the $X$-basis readout tends to collapse each qubit individually onto a product state, thereby favoring a short-range entangled (SRE) configuration. 

The following sections study the competition between the two non-commutative measurements $Z_i Z_j$ and $X_i$. 
Keeping the $ZZ$-measurement rate fixed, we vary the total evolution time $t_f$ and tune the $X$-measurement rate $h$, the parameter that sets the degree of non-commutativity, thereby revealing a measurement-induced transition from a SRE phase to a LRE phase marked by spontaneous $\mathbb{Z}_2$ symmetry breaking.
Crucially, this transition becomes apparent only in observables that are non-linear functions of the density matrix, such as Rényi entropies~\cite{leeQuantumCriticalityDecoherence2023}.
Notably, the common statistically averaged observables like the Pauli $Z$ correlator
\begin{equation}
    \mathbb{E} \left\{ \braket{Z_i Z_j}\right\} = \int P[\xi] \tr \left(Z_i Z_j\rho[\xi,t_f]\right) = \tr \left(Z_i Z_j \rho_l \right),
\end{equation}
remain trivial and cannot detect the transition due to the absence of entanglement in $\rho_l$. Instead, nonlinear order parameters make it possible to capture the phase transition, e.g. the Edwards-Anderson correlation function (see the review in Ref. \cite{nishimoriStatisticalPhysicsSpin2001}),
\begin{equation}
    \mathbb{E} \left\{ \braket{Z_i Z_j}^2\right\} = \int P[\xi] \tr \left(Z_i Z_j \rho[\xi,t_f]\right)^2.
    \label{eq:Edwards-Anderson}
\end{equation}
In the LRE phase this correlator saturates to a non-zero constant as $|i-j|\rightarrow +\infty$, indicating the long-range order and a non-vanishing Edwards-Anderson order parameter $\mathbb{E} \{ \braket{Z_i}^2\} $ in the presence of an infinitesimal symmetry breaking perturbation. In contrast, such a correlation exponentially decays with distance in the SRE phase. Note that the trajectory average should be performed after evaluating the quantum expectation value in such nonlinear quantities. Physically, this only concerns the properties that manifest after collecting measurement outcomes.


The preparation of a GHZ state demands explicit readout of the detector’s measurement outcomes; executing the measurements without recording their results leaves the system in a mixed state,
\begin{equation}
    \rho_l(t_f)  = \int D\xi P [\xi] \rho[\xi,t_f] = \int D\xi P_G [\xi] \hat\rho[\xi,t_f],
\end{equation}
whose dynamics is governed by the Lindblad equation with the disorder terms in Eq. \eqref{eq:dynamics} averaged out,
\begin{equation}
\begin{aligned}
    \partial_t \rho_l &= - 2J \sum_{\braket{ij}} \left(\rho_l - Z_i Z_j \rho_l Z_i Z_j \right) \\
    &-  2h \sum_{i} \left(\rho_l - X_i\rho_l X_i \right).
\end{aligned}
\label{eq:lindblad}
\end{equation}
Such an evolution has a strong symmetry (or quantum symmetry) \cite{buca_note_2012,albert_symmetries_2014,siebererKeldyshFieldTheory2016,albert_lindbladians_2018,lieu_symmetry_2020,siebererUniversalityDrivenOpen2023}, that is 
\begin{equation}
\begin{aligned}
    & \prod_i X_i \mathcal{L}^{(1)}(\rho_l) =\mathcal{L}^{(1)} \left(\prod_i X_i \rho_l\right),\\
    &  \mathcal{L}^{(1)}(\rho_l) \prod_i X_i =\mathcal{L}^{(1)}\left( \rho_l  \prod_i X_i \right),
\end{aligned}
\end{equation}
where the global symmetry operation can act on the left and right-hand side of the density matrix independently and the symmetry group is $\mathcal{Z}_2 \times \mathcal{Z}_2$. Here $\mathcal{L}^{(1)}$ denotes the Liouvillian operator on the right-hand side of Eq. \eqref{eq:lindblad}.
In the same manner, Eq. \eqref{eq:dynamics} can also be viewed as having strong $\mathbb{Z}_2$ symmetry given a particular trajectory $\xi$. Although the interesting entanglement transition does not manifest in $\rho_l$ since the disorder average smears out the entanglement structure, the dynamics is not completely trivial and a phase transition of SWSSB \cite{lessaStrongtoWeakSpontaneousSymmetry2024,salaSpontaneousStrongSymmetry2024} occurs, where the strong $\mathbb{Z}_2$ breaks down to a weak $\mathbb{Z}_2$ symmetry (or classical symmetry),
\begin{equation}
\begin{aligned}
    \prod_i X_i \mathcal{L}^{(1)}(\rho_l) \prod_i X_i =\mathcal{L}^{(1)} \left(\prod_i X_i \rho_l \prod_i X_i\right)
\end{aligned}
\end{equation}
where the symmetry operation acts on both sides and the symmetry group is $\mathcal{Z}_2$.
Such a phase transition is captured by nonlinear quantities that characterize the correlation between forward and backward evolution branches (left and right-hand side of the density matrix), e.g. the fidelity correlator \cite{lessaStrongtoWeakSpontaneousSymmetry2024}
\begin{equation}
    F(\rho_l, Z_i Z_j \rho_l Z_i Z_j),
    \label{eq:fidelity-correlator}
\end{equation}
where $F(\rho,\sigma) = \tr( \sqrt{\sqrt{\rho} \sigma \sqrt{\rho}})^2$ is the Uhlmann fidelity of quantum states. 
As we will see later, such a transition depends only on time $t_f$ but not $X$ measurement rate $h$. In contrast, the SRE-to-LRE transition governed by Eq. \eqref{eq:dynamics} should be interpreted as a spontaneous breaking of a strong $\mathbb{Z}_2$ symmetry: because the post-measurement conditional states are pure, the dynamics ultimately selects one of the two symmetry sectors, leaving no residual symmetry.


In addition, we also consider the situation of partial readout, where only $ZZ$ measurement outcomes are collected while $X$ measurement outcomes are averaged over. Then the stochastic master equation is written as
\begin{equation}
\begin{aligned}
        \partial_t \hat{\rho}_Z &= - 2J \sum_{\braket{ij}} \left(\hat{\rho}_Z -  Z_i Z_j \hat{\rho}_Z Z_i Z_j - 2 \xi_{{ij}}(t) \{Z_iZ_j, \hat{\rho}_Z\} \right)\\
        & -  2h \sum_{i} \left(\hat{\rho}_Z -  X_i \hat{\rho}_Z X_i  \right),
\end{aligned}
    \label{eq:average-X-master}
\end{equation}
which must be normalised at the final time by dividing through by $\operatorname{tr}\hat{\rho}_{Z}(t_f)$.
The first line describes the measurement-only dynamics that prepares a LRE state, while the second line represents on-site $X$ decoherence.  Tuning $t_f$ and $h$, such decoherence is expected to destroy the LRE order through a mixed-state phase transition. Also, notice that Eq. \eqref{eq:average-X-master} respects the strong $\mathbb{Z}_2$ symmetry, and $\hat{\rho}_Z$ still exhibits the pattern of SWSSB (distinguished from the mixed-state phase transition) as we will see later.

\subsection{Replica method}
To evaluate non-linear observables we must average tensor powers of the conditional state. In that sense, we need to compute the average of $N$ copies of the density matrix 
\begin{equation}
  \mathbb{E}\{ \rho[\xi,t_f]^{\otimes N}\} = \int P_G[\xi] \frac{\hat{\rho}[\xi,t_f]^{\otimes N}}{\tr(\hat{\rho}[\xi,t_f])^{N-1}},
\end{equation}
where $N=2$ for the Edwards-Anderson order parameter.
Because the numerator and the denominator depend on the same stochastic
variables, one cannot factor them apart and average separately. The factor in the denominator can be removed elegantly with the replica trick, which converts the otherwise intractable ratio under the stochastic integral into a tractable product over replicated copies—at the price of performing an analytic continuation in the replica number\cite{favaNonlinearSigmaModels2023,poboikoTheoryFreeFermions2023,poboikoMeasurementInducedPhaseTransition2024}:
\begin{equation}
    \mathbb{E}\{ \rho[\xi,t_f]^{\otimes N}\} = \lim_{R\rightarrow 1} \mathbb{E}_G \left\{ \tr_{a=N+1,\cdots,R}\left( \otimes_{a=1}^R\hat{\rho}[\xi,t_f]\right)\right\}.
\end{equation}
The above formula should be interpreted as first evaluating the Gaussian average for $R>N$ integers. The trace over the $a=N+1, \cdots, R$ copies yields a coefficient $\tr(\hat{\rho}[\xi,t_f])^{R-N}$. Then we perform analytical continuation, take the replica limit $R\rightarrow 1$, and arrive at the right answer.

We define the replica density matrix as 
\begin{equation}
    \rho^{(R)}(t_f) = \mathbb{E}_G \left\{ \hat{\rho}[\xi,t_f]^{\otimes R}\right\}.
\end{equation}
It can be evaluated by applying the super-operator formalism and mapping the density matrix to the doubled Hilbert space,
\begin{equation}
    \rho^{(R)} \mapsto \sket{\rho^{(R)}},
\end{equation}
whose evolution is governed by a super operator.
\begin{equation}
    \sket{\rho^{(R)}(t_f)} = \mathbb{E}_G \left\{ K[\xi,t_f]^{\otimes R} \otimes (K[\xi,t_f]^*)^{\otimes R}\right\}\sket{\rho(0)}^{\otimes R}.
\end{equation}
Now we might directly perform the Gaussian average. Notice that $\mathbb{E}_G$ commutes with the time ordering, since the random variables $\xi(t)$ are independent at each time step so that we can integrate them out accordingly. This results in
\begin{equation}
 \sket{\rho^{(R)}(t_f)} = \exp\{t_f L^{(R)}\} \sket{\rho(0)}^{\otimes R}, 
\label{eq:replica-evolution}
\end{equation}
where for the complete readout case, the Liouvillian becomes
\begin{align}
     &L^{(R)} = -2J R  n d - 2h R n \nonumber\\
     &+ 2J\sum_{\alpha<\beta, \braket{ij}} Z_i^\alpha Z_j^\alpha Z_i^\beta Z_j^\beta + 2h\sum_{\alpha<\beta,i} X_i^\alpha X_i^\beta,
\end{align}
Where $n$ is the system size and $d$ is the spatial dimension.
Here the superscript $\alpha= (\pm,a)$, where $a$ labels $R$ replica copies, and $\pm$ labels the forward evolution branch (left-hand side of density matrix) and the backward evolution branch (right-hand side of density matrix) respectively. $L^{(R)}$ is the replica Liouvillian of the evolution. $\alpha<\beta$ indicated the summation over all different pairs of the $2R$ branches.

If we directly set $R=1$ in $\rho^{(R)}(t_f)$, the Lindblad evolution of the averaged mixed state $\rho_l = \rho^{(1)}$ follows the Liouvillian, 
\begin{equation}
\begin{aligned}
    & L^{(1)} = -2J  n d - 2h n \\&+ 2J\sum_{\braket{ij}} Z_i^+ Z_j^+ Z_i^- Z_j^- + 2h\sum_{i} X_i^+ X_i^-.
\end{aligned}
\label{eq:Liouvillian-l}
\end{equation}
This is to be distinguished from evaluating a nonlinear quantity of density matrix and then taking the $R \rightarrow 1$ limit. This case refers to the fully ensemble-averaged (unconditioned) evolution, or no readout case.

Similarly, we apply the replica method on the partial readout case, 
\begin{equation}
    \rho^{(R)}_Z(t_f) = \mathbb{E}_G \left\{ \hat{\rho}_Z[\xi,t_f]^{\otimes R}\right\},
\end{equation}
whose evolution is given by replacing $L^{(R)}$ in Eq. \eqref{eq:replica-evolution} with
\begin{equation}
\begin{aligned}
    &L^{(R)}_Z = -2J R  nd - 2h Rn  \\&+2J\sum_{\alpha<\beta, \braket{ij}} Z_i^\alpha Z_j^\alpha Z_i^\beta Z_j^\beta + 2h\sum_{a=1}^{R} \sum_i X_i^{(+,a)} X_i^{(-,a)}.
\end{aligned}
\label{eq:average-X}
\end{equation}
Here the $X_i^{(\pm,a)} X_i^{(\pm,b)}$ interaction terms among different replica copies are eliminated compared with Eq. \eqref{eq:replica-evolution}.


\section{One-dimensional case}
\label{sec:1d}
Have introduced the model, we now consider the one-dimensional (1D) case, as an illustrative example. 
Due to noneligible fluctuation in one dimension, the system exhibits significant deviations from the mean-field solution, which will be analyzed in Sec.~\ref{sec:meanfield} for higher dimensional scenarios. 
Meanwhile, the system exhibits distinct finite-time dynamics in 1D. Specifically, 1D systems do not acquire finite time orders (LRE or SWSSB) despite the scenarios of readout. However, the corresponding correlation lengths (of EA or Fidelity correlator) will increase exponentially with time and result in qualitatively similar stationary phase diagram as in Fig. \ref{fig:phase} (see Fig. \ref{fig:6qcase} (e)). Given these unique features, we analyze the 1D case separately, combining concise analytical insights with numerical demonstrations to develop an intuitive understanding of the problem.
Throughout this section, we assume the initial state is the trivial product state in the $X$-basis:
\begin{equation}
    \rho(0) = \ket{+}^{\otimes L} \bra{+}^{\otimes L},
\end{equation}
where $L$ denotes the linear size of the system. 

\subsection{Commuting $ZZ$ measurements}
We begin by analyzing the analytically tractable limit where the non-commuting $X$ measurement is absent ($h=0$), leaving solely the influence of the $ZZ$ measurements. While related phenomena have been investigated in the context of discrete weak measurement protocols \cite{ref:guoyi}, our work provides a crucial extension by examining the continuous measurement regime. This framework allows us to go beyond steady-state analysis and provide a detailed characterization of the system's temporal dynamics. By solving for the correlation length as an explicit function of measurement time, we establish an essential baseline for how measurement-induced order develops before addressing the more complex non-commuting case.

The non-unitary evolution operator in this case can be written as
\begin{equation}
    K[\xi] = \exp{\left[ \int_0^{t_f} dt\sum_i \left(4J\xi_{i,i+1}(t)Z_iZ_{i+1}-2J\right)\right ]}. 
\end{equation}
Notice that the evolution $K[\xi]$ does not depend on the specific choice of quantum trajectory but only on the averaged measurement result $s_{i,i+1} = \frac{1}{t_f}\int_0^{t_f} dt \xi_{i,i+1}(t)$, which has zero mean value and variance $1/(8Jt_f)$ according to the central limit theorem. This phenomenon arises because the measured observables commute with each other. Thus we can re-express the unnormalized state as 
\begin{equation}
    \begin{aligned}
         &K(s) =  \exp{\left[ Jt_f\sum_i \left(4 s_{i,i+1}Z_iZ_{i+1}-2\right)\right ]},\\
    &\hat{\rho} (s, t_f) = K(s) \rho (0) K(s)^\dagger,\\
    &\mathbb{E}_G \{s_{{i,i+1}} s_{{i',i'+1}}\} = \frac{1}{8Jt_f} \delta_{ii'}.
    \end{aligned}
\end{equation}

We first consider reading out the measurement outcomes (complete readout).
The above form of $K(s)$ mimics a disordered 1D classic Ising model with bond disorder, and can be solved by statistical mechanical (SM) mapping \cite{castelnovo_quantum_2008,zhuGaplessCoulombState2019,zhangNonHermitianEffectsIntrinsic2020,ref:guoyi, ref:fisher}. This is done by expanding the initial state in the computational basis. The details of the derivation are summarized in Sec. SI of Supplemental Material (SM) \cite{SupMat}. By applying the SM mapping, we are able to rigorously calculate the expectation values of Pauli observables as physical quantities of 1D Ising model for each measurement outcome $s$. 
Here we consider the
$\mathbb{Z}_2$ spin correlation 
\begin{equation}
    \begin{aligned}
        &\braket{Z_i Z_j}_{s} = \frac{ \tr \left(\hat{\rho} (s, t_f) Z_i Z_j \right)}{\tr \left(\hat{\rho} (s, t_f)\right)},\\
    \end{aligned}
\end{equation}
we can check that the normal correlation function $\mathbb{E} \braket{Z_i Z_j}_{s}$ averaged over measurement results $s$ always yields $0$. The non-trivial order induced by continuous measurement can only be detected by nonlinear expectation values. Thus we consider the Edwards-Anderson correlation function,   
\begin{equation}
    \mathbb{E}\left\{\braket{Z_i Z_j}_{s}^2\right\}  = \int ds P(s)  \braket{Z_i Z_j}_{s}^2\xrightarrow{L\rightarrow \infty}  e^{-|i-j|/\xi}
\end{equation}
where the correlation length is
\begin{equation}
\begin{aligned}
    &\xi = -1/\log\left [ \int ds \sqrt{\frac{4Jt_f}{\pi}}\exp\left(-4Jt_f  s^2-4Jt_f\right)\right.\\ 
    &\times  \cosh (8Jt_f s) \tanh^2 (8Jt_f s)\Bigg].
\end{aligned}
    \label{eq:length-square}
\end{equation}
Here $ P(s) = P_G(s) \tr \left(\hat{\rho} (s, t_f)\right)$ is the physical measurement probability and we kept $|i-j|$ finite and took the thermodynamic limit $L\rightarrow \infty$. $\xi$ denotes the correlation length. In the large time $t_f \rightarrow + \infty$ limit, we can apply saddle point approximation to the Gaussian integral, which leads to
\begin{equation}
    \xi \sim \frac{1}{3} \exp({16 J t_f}).
    \label{eq:length-square2}
\end{equation}
For a sufficiently large time, $\xi$ increases exponentially with $t_f$. This suggests an exponentially fast entanglement generation since Eq. \eqref{eq:length-square} lower bounds the two-point mutual information of the post-measurement state \cite{Cirac_mutual}. 
For the exact $t_f\rightarrow +\infty$ limit, the Edwards-Anderson correlation function acquires long-range order $\xi \rightarrow + \infty$, which can be interpreted as a spontaneous breaking of $\mathbb{Z}_2$ symmetry with the presence of the quenched disorder $s$. It indicates that the post-measurement states $\rho(s)$ with the measurement outcomes $s$ read out are long-range entangled cat states. As long as $t_f$ is finite, $\xi$ is also finite suggesting only short-range entanglement. However, the entanglement generation is exponentially fast. Such an exponential behavior is dominated by the $t_f \rightarrow +\infty$ fix point.

We also examine another kind of correlation called the R\'enyi-2 correlation function \cite{lessaStrongtoWeakSpontaneousSymmetry2024,salaSpontaneousStrongSymmetry2024},
\begin{equation}
    \begin{aligned}
           &R^{(2)}(i,j)= \frac{\int ds P(s)^2  \braket{Z_i Z_j}_{s}^2}{\int ds P(s)^2}
           \rightarrow (\tanh{4Jt_f})^{|i-j|}.
    \end{aligned}
    \label{eq:renyi2}
\end{equation}
Such correlation is not as universal as other more strict order parameters and tends to underestimate the order by providing only a lower bound of the phase transition point. However, it is analytically much more tractable and provides a good approximation especially away from the critical point. 
and the correlation length is also exponentially increasing with time for sufficiently large $t_f$,
\begin{equation}
    \xi = -1/ \log \tanh(4Jt_f) \sim \frac{1}{2} e^{8Jt_f}.
\end{equation}
Compared with Eq. \eqref{eq:length-square2}, the Renyi-2 correlation underestimated the correlation length.

Now we consider discarding the measurement outcomes (no readout) and focus on the density matrix $\rho_l$ averaging over measurement results. Physically it means that we perform measurements but do not read out the outcomes. Its evolution follows
\begin{equation}
    \partial_t \rho_l = 2J \sum_i (Z_iZ_{i+1} \rho_l Z_iZ_{i+1}  - \rho_l).
\end{equation}
The accumulating effect at the final time will be sending the initial state $\rho(0)$ through a stochastic $ZZ$ noise channel.
Such an evolution exhibits strong to weak $\mathbb{Z}_2\times \mathbb{Z}_2 \rightarrow \mathbb{Z}_2$ symmetry breaking, and its behavior of fidelity correlation Eq.\eqref{eq:fidelity-correlator} is captured by the 1D Nishimori universal class \cite{lessaStrongtoWeakSpontaneousSymmetry2024,salaSpontaneousStrongSymmetry2024}. Here for convenience, we evaluate the R\'enyi-2 correlator which is defined as
\begin{equation}
    R^{(2)}_l(i,j) = \frac{\tr(\rho_l Z_iZ_j \rho_l Z_iZ_j)}{\tr \rho_l^2} = \frac{\sbra{\rho_l} Z_i^+ Z_j^+ Z_i^- Z_j^-\sket{\rho_l}}{\sbraket{\rho_l |\rho_l}}.
\end{equation}
A similar SM mapping method can be applied to the super vector (see Sec. SI of SM \cite{SupMat}), and as a result,
\begin{equation}
\begin{aligned}
    R^{(2)}_l(i,j) 
     \xrightarrow{L\rightarrow \infty} (\tanh(4Jt_f))^{|i-j|}.
\end{aligned}
\end{equation}
There is no strong-to-weak symmetry breaking at finite $t$, but the correlation length is also exponentially increasing with time,
\begin{equation}
    \xi = -1/ \log \tanh(4Jt_f) \sim \frac{1}{2} e^{8Jt_f},
\end{equation}
which is dominated by the $t_f \rightarrow +\infty$ SWSSB fixed point. Notice that the R\'enyi-2 correlation of SWSSB coincides with the one of the entanglement transition Eq. \eqref{eq:renyi2} with complete readout. However, these two concepts are not identical. Strong-to-weak symmetry breaking in some sense characterizes a classical order, while the entanglement is a fully quantum phenomenon. Their discrepancy manifests when the non-commuting $X$ measurement is present.

\subsection{Non-commuting $ZZ$ and $X$ measurements}
We now include the non-commutativity from both $ZZ$ and $X$ measurements and first examine the scenario involving a complete readout. Due to the non-commutative nature of the measurements, the analytical method discussed previously becomes invalid. Instead, we directly analyze the effective Liouvillian (up to a constant term):
\begin{equation}
    L^{(R)}= 2J\sum_{\alpha<\beta, i} Z_i^\alpha Z_{i+1}^\alpha Z_i^\beta Z_{i+1}^\beta + 2h\sum_{\alpha<\beta,i} X_i^\alpha X_i^\beta.
    \label{eq:Liouvillian-1d}
\end{equation}
Notably,  certain local terms in the Liouvillian anticommutes, e.g. $Z_i^\alpha Z_{i+1}^\alpha Z_i^\beta Z_{i+1}^\beta$ and $X_{i+1}^\beta X_{i+1}^\gamma$ where $\alpha \neq \gamma$. This non-commutativity generates quantum fluctuations that drive a quantum phase transition. 

Consider the $t_f\rightarrow +\infty$ steady state, which is an eigenstate of $L^{(R)}$ with the largest eigenvalue. By noticing that $L^{(R)}$ has a bond algebra self-duality \cite{cobaneraBondalgebraicApproachDualities2011} under the replacement $Z_i^\alpha Z_{i+1}^\alpha \longleftrightarrow X_{i}^\alpha$, we conclude that the quantum phase transition point locates at $h/J = 1$. 
In particular, the correlation function $\mathbb{E}\{\braket{Z_i Z_j}_{s}^2\}$ distinguishes the phases clearly: for $h/J<1$, it is constant at large distances, indicating LRE; for $h/J>1$, it decays exponentially, representing SRE. Furthermore, the system maps onto a Majorana fermion chain via a Jordan-Wigner transformation, with its steady-state properties previously explored using a nonlinear sigma model~\cite{favaNonlinearSigmaModels2023}.

For finite evolution time $t_f$, the final state never reaches LRE phase. However, the correlation behavior remains significantly influenced by the quantum critical point at $h/J=1$, displaying quantum critical phenomena~\cite{sachdevQuantumPhaseTransitions2011}. Qualitatively, the correlation length $\xi$ of $\mathbb{E}\{\braket{Z_i Z_j}_{s}^2\}$ behaves as:  
\begin{itemize}  
    \item \textbf{$h/J < 1$}: $\xi$ grows exponentially when $t_f$ greatly exceeds the inverse spectral gap of $\mathcal{L}^{(R)}$.  
    \item \textbf{$h/J > 1$}: $\xi$ saturates to a constant as $t_f \rightarrow +\infty$.  
    \item \textbf{$h/J \sim 1$}: $\xi$ grows linearly with $t_f$, reflecting algebraic scaling.  
\end{itemize}  


As for the no readout case, the Liouvillian of the Lindblad evolution of the averaged density matrix $\rho_l$ (up to a constant term) is written as 
\begin{equation}
    L^{(1)}= 2J\sum_{i} Z_i^+ Z_{i+1}^+ Z_i^- Z_{i+1}^- + 2h\sum_{i} X_i^+ X_i^- .
\end{equation}
Unlike Eq.~\eqref{eq:Liouvillian-1d}, here the replica indices are absent, and crucially, all local terms commute with each other. Therefore, the inclusion of $X$ terms does not induce quantum fluctuations sufficient to trigger a quantum phase transition.
Consequently, the behavior of SWSSB is identical to the $h=0$ case discussed before. In conclusion, the SWSSB order is not affected by finite $h$, while the exponentially fast entanglement generation could be prevented by tuning $h/J>1$.


\begin{figure*}[htbp]
  \centering
  \includegraphics[width=1\linewidth]{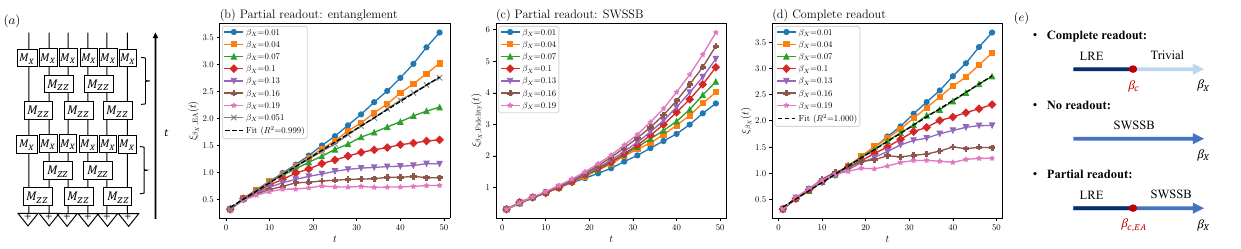}
  \caption{\textbf{Circuit and different behaviors of entanglement and SWSSB setting $L=6$ and $\beta_Z=0.1$}. Panel (b) and (c) correspond to the partial readout scenario, where only the ZZ measurement outcomes are collected while the X measurement outcomes are discarded. Panel (d) represents the complete readout scenario, in which all measurement outcomes are collected. (a) The schematic diagram of the weak measurement process. (b) The entanglement-associated correlation length of the Edwards-Anderson correlation function, $\xi_{\beta_X,\text{EA}}(t)$. There exists a critical value $\beta_{c,\text{EA}}\approx 0.051$, at which $\xi_{\beta_X,\text{EA}}(t) \sim O(t)$. (c) The SWSSB-associated correlation length of the fidelity correlator, $\xi_{\beta_X,\text{Fidelity}}(t)$. For different measurement strengths $\beta_{X}$, the correlation length exhibits exponential growth in all cases, showing no critical behavior. (d) The correlation length $\xi_{\beta_X}(t)=\xi_{\beta_X,\text{EA}}(t)=\xi_{\beta_X,\text{Fidelity}}(t)$ since the Edwards-Anderson correlation function coincides with the fidelity correlator. In this complete readout scenario, entanglement and SWSSB display identical behavior, characterized by a critical value $\beta_c \approx 0.07$, where $\xi_{\beta_X}(t) \sim O(t)$. (e) The stationary $t_f \rightarrow +\infty$ phase diagrams in one dimension.} 
  \label{fig:6qcase}
\end{figure*}

We are also interested in the partial readout case, described by the Liouvillian (up to a constant term)
\begin{equation}
    L^{(R)}= 2J\sum_{\alpha<\beta, i} Z_i^\alpha Z_{i+1}^\alpha Z_i^\beta Z_{i+1}^\beta + 2h\sum_{a,i} X_i^{(+,a)} X_i^{(-,a)}.
    \label{eq:Liouvillian-1d2}
\end{equation}
Similar to the complete readout scenario, no ordered phase emerges at finite evolution time $t_f$. However, in the stationary limit $t_f \rightarrow +\infty$, distinct phases emerge with a critical $X$ measurement strength $h_c$:
\begin{itemize}  
    \item For $h < h_c$, the LRE phase persists.   
    \item For $h > h_c$, the system enters a SWSSB where only the weak $\mathbb{Z}_2$ symmetry is preserved. 
\end{itemize}  
The case above $h_c$ contrasts sharply with the complete readout case, which exhibits a strongly symmetric trivial phase.  
This is to say, above $h_c$, the Edwards-Anderson correlation length saturates while the Fidelity correlation length grows exponentially with $t_f$.
Below $h_c$, both the Edwards-Anderson correlation length and the Fidelity correlation length grow exponentially with $t_f$. The stationary $t\rightarrow +\infty$ phase diagrams for the complete readout, no readout and partial readout cases are summarized in Fig. \ref{fig:6qcase} (e). We demonstrate the behavior of their order parameters numerically below.

\subsection{Numerical simulation}
To elucidate phase transitions in the complete and partial readout regimes, we conduct numerical simulations on a finite-size model. 
For numerical convenience, we fix the system size $L=6$ with open boundary condition. As shown in Fig.~\ref{fig:6qcase} (a), we first initialize the state in the product state $\ket{+}^{\otimes 6}\bra{+}^{\otimes 6}$. Then we perform multiple rounds of measurements. Each round consists of measuring all nearest-neighbour $ZZ$ operators followed by measuring all $X$ operators.
In the complete readout case, all measurement outcomes are recorded. In contrast, the partial readout case retains only the \(ZZ\) measurement outcomes, while the \(X\) outcomes are discarded. 


Consider an initial system state denoted by $\rho_0$. To simulate a single layer of nearest-neighbor $ZZ$ weak measurements with outcome readout, we introduce five ancilla qubits initialized in the $\ket{+}$ state, forming $\rho_{a1}=\ket{+_a}^{\otimes 5}\bra{+_a}^{\otimes 5}$. The combined system then undergoes unitary evolution for some time $t_Z$:
\begin{equation}
   \rho_1 = U_Z \left(\rho_0\otimes \rho_{a1}\right)U_Z^\dagger;\quad  U_Z=e^{-it_Z\sum_{i=1}^{5} Z_iZ_{i+1}\otimes Z_{i_a}}.
\end{equation}
Subsequently, all ancilla qubits are measured along the $Y$ direction, and their outcomes are recorded as a vector $\vec s=(s_1,s_2,\ldots,s_5)$, $s_i\in\{-1,1\}$. After tracing out these ancilla degrees of freedom, the resulting state of the system is denoted as $\rho_2$. This procedure is equivalent to performing one layer of weak nearest-neighbor $ZZ$ measurements on the original state $\rho_0$ while reading out the measurement outcomes (see Appendix~\ref{app:1} for details). To be specific, when the measurement outcome is $\vec s$,
\begin{equation}
   \rho_2 (\vec s) =\frac{M(\vec s) \rho_0 M(\vec s)^\dagger}{\text{tr}(M(\vec s)\rho_0M(\vec s)^\dagger)}.
   \label{eq:ZZWM}
\end{equation}
The Kraus operators $\{M(\vec s)\}$ define a set of positive operator-valued measurements (POVM) and are given by
\begin{equation}
     M(\vec s)=\frac{\text{exp}\left[(\beta_Z/2)\sum_{i=1}^5 s_i Z_iZ_{i+1}\right]}{\sqrt{2\cosh{\beta_Z}}^5},
    \label{eq:ZZWMKraus}
\end{equation}
where $\beta_Z=2\tanh^{-1}(\tan t_Z)$ is the strength of the $ZZ$ weak measurement~\cite{ref:guoyi}. When we set the evolution time $t_Z=\pi/4$, $\beta_Z$ tends to infinity, and the associated Kraus operators approach those of a projection measurement $M(\vec s)=\Pi_i\left(I+s_iZ_iZ_{i+1}\right)$, effectively transforming the weak measurement into a strong, or projective, one. The strength $\beta_Z$ is adjustable through the evolution time $t_Z$; for small values of $t_Z$, we have $\beta_Z \approx 2t_Z$.

Following the $ZZ$ measurements, we then simulate a layer of $X$ weak measurements where the outcomes are discarded. We introduce $6$ auxiliary qubits, all initialized in the $\ket{+}$ state, denoted as $\rho_{a2}=\ket{+_a}^{\otimes 6}\bra{+_a}^{\otimes 6}$. The combined system undergoes the following unitary evolution:
\begin{equation}
    \rho_3 = U_X \left(\rho_2\otimes \rho_{a2}\right)U_X^\dagger;\quad  U_X=e^{it_X\sum_{i=1}^{6} X_i \otimes Z_{i_a}}.
\end{equation}
Discarding the measurement outcomes implies directly tracing out the ancilla qubits without measurement. This results in a mixed state for the 6-qubit system. Similar to the previous case, the strength of the weak $X$ measurement, $\beta_X=2\tanh^{-1}(\tan t_X)$, is controlled by adjusting the evolution time $t_X$, and for small $t_X$, one finds $\beta_X \approx 2t_X$. These strengths $\beta_Z$ and $\beta_X$ are related to the continuous measurement rates $J$ and $h$, respectively.

We denote the number of measurement rounds by $t$. The state of the system after $t$ rounds of measurement run is denoted by $\rho(t)$. By averaging over many such experimental trajectories, we obtain the ensemble-averaged mixed state, $\rho_{l}(t)$. In the limit of small measurement strengths ($\beta_Z, \beta_X \ll 1$), the discrete-time evolution of $\rho_{l}(t)$ approximates the continuous-time dynamics governed by the Liouvillian presented in Eq.~\eqref{eq:Liouvillian-1d2}.

We employ the Edwards-Anderson correlation function Eq. \eqref{eq:Edwards-Anderson} and the fidelity correlator Eq. 
\eqref{eq:fidelity-correlator} to capture the behaviors of entanglement and SWSSB, respectively. Fixing $\beta_{Z} = 0.1$ and for a given $\beta_X$, we conduct multiple experimental runs. At each measurement round in every experimental realization, we calculate both $\braket{Z_i Z_j}_{\rho}^2$ and the fidelity $F(\rho, Z_i Z_j \rho Z_i Z_j)$ for the resulting quantum state. These quantities are then averaged over multiple runs to obtain the desired Edwards-Anderson correlation function $\mathbb{E}\left\{ \braket{Z_i Z_j}^2 \right\}(t)$ and the fidelity correlator $F_{ij}(t)$ with the respective correlation lengths as functions of the measurement round $t$. Finally, by varying $\beta_X$, we obtain the entanglement-associated correlation length $\xi_{\beta_X, \text{EA}}(t)$ (shown in Fig.~\ref{fig:6qcase} (b)) and the SWSSB-associated correlation length $\xi_{\beta_X, \text{Fidelity}}(t)$ (shown in Fig.~\ref{fig:6qcase} (c)). Furthermore, as a comparison, we also consider the case where both the $ZZ$ and $X$ operators are measured and all outcomes are collected (complete readout), as shown in Fig.~\ref{fig:6qcase} (d). In this scenario, since the resulting state $\rho$ becomes pure, we have the identity $\braket{Z_i Z_j}_{\rho}^2=F(\rho,Z_iZ_j\rho Z_i Z_j)$. Consequently, the correlation length characterizing entanglement coincides with that characterizing SWSSB, which we denote as $\xi_{\beta_X}(t)$.

The numerical simulation illustrates the differences in entanglement and SWSSB behaviors between partial readout and complete readout scenarios. In the case of complete readout, entanglement and SWSSB exhibit identical behavior, characterized by the correlation length $\xi_{\beta_X}(t)$. As shown in Fig.~\ref{fig:6qcase} (d), there exists a critical value $\beta_c$ such that: (i) for $\beta_{X}<\beta_c$, the correlation length grows superlinearly with $t$, and as the system scales up, we expect this growth rate to approach exponential; (ii) for $\beta_{X}>\beta_c$, it eventually saturates to a constant; and (iii) near criticality ($\beta_{X}\sim \beta_c$), the correlation length increases linearly with $t$. Fig.~\ref{fig:6qcase} (d) reveals that $\beta_c \approx 0.07$, with the coefficient of determination $R^2=1.000$. However, according to the preceding theoretical analysis, the critical weak measurement strength $\beta_c$ for $X$ should equal the preset weak measurement strength for $ZZ$, i.e., $\beta_c = 0.1$. We attribute the discrepancy between numerical results and theoretical predictions to finite-size effects and non-vanishing measurement strengths. 
As the system size increases and both weak measurement strengths $\beta_Z$ and $\beta_X$ approach zero, 
we expect the critical value $\beta_c$ to converge to preset $\beta_Z$. In contrast, under partial readout, entanglement and SWSSB display fundamentally distinct behaviors. On the one hand, as depicted in Fig.~\ref{fig:6qcase} (b), the entanglement-associated correlation length $\xi_{\beta_X,\mathrm{EA}}(t)$ follows the same trend as in the complete readout case, except with a modified critical value $\beta_{c,\text{EA}} \approx 0.051$ with the coefficient of determination $R^2=0.999$. In particular at the critical $X$ measurement strength $\xi_{\beta_X,\mathrm{EA}}(t)$ also increase linearly with $t$. On the other hand, Fig.~\ref{fig:6qcase} (c) demonstrates that the SWSSB-associated correlation length $\xi_{\beta_X,\text{Fidelity}}(t)$ grows exponentially with $t$ for all $\beta_{X}$ values, showing no critical behavior. The numerical results are thus compatible with the theoretical analysis presented in previous sections.\\


\begin{widetext}
\section{Mean field approach to higher dimensional cases}
\label{sec:meanfield}
In order to study higher-dimensional cases, we employ a mean-field theory, where has been widely utilized in studies of classical and quantum spin glasses (see, e.g. Refs. \cite{nishimoriStatisticalPhysicsSpin2001,suzukiQuantumIsingPhases2013}).
In particular, we introduce the ancillary field, corresponding to the Edwards-Anderson order parameter $Q^{\alpha \beta}_i \sim Z^\alpha_i Z^\beta_i$, to decouple the $ZZZZ$ interaction term in Eq.\eqref{eq:replica-evolution}. This yields the path integral representation:
\begin{equation}
    e^{t_f L^{(R)}}
         \propto \int DQ^{\alpha\beta}_i D\xi_i  \mathcal{T}\exp\left({\int_0^{t_f}dtL_{\text{eff}}^{(R)}(t)}\right)  \exp \int_0^{t_f} dt \left(  -J\sum_{\alpha<\beta,ij} Q_i^{\alpha\beta}(t) K_{ij} Q_j^{\alpha\beta}(t) - \frac{1}{2} \sum_i\xi_i(t)^2  \right),\label{eq:HS}
\end{equation}
with an effective Liouvillian defined as
\begin{equation}
    L_{\text{eff}}^{(R)}(t) =    2J\sum_{\alpha<\beta,ij}  Q_i^{\alpha\beta}(t) K_{ij} Z_j^\alpha  Z_j^\beta + \sqrt{2h}  \sum_{\alpha,j} \xi_j(t)X_j^\alpha, \label{eq:eff}
\end{equation}
where the adjacency matrix is defined as 
\begin{equation}
    K_{ij} = \delta_{\braket{ij}}= \left\{ \begin{aligned}
        &1 & i,j \text{ nearest neighbor}\\
        &0 & \text{otherwise}
    \end{aligned}\right.
\end{equation}
Here, the trajectories associated with the $X-$ measurement have been restored and rescaled as $\xi_i \rightarrow \xi_i/\sqrt{8h}$.
Importantly, due to the intrinsic non-commutativity of the Liouvillian terms, the introduced order parameter field $Q_i^{\alpha\beta}(t)$ explicitly depends on time, indicating the intrinsic quantum nature of this problem.
The decoupled effective qubit model $L_{\text{eff}}$ now only involves correlation between different replica copies and evolution branches. Similarly the effective model $L^{(R)}_{\text{eff},Z}$ for the partial readout case Eq. \eqref{eq:average-X} can be obtained by adding a replica dependence to the $X$ trajectory, $\xi_i(t) \rightarrow \xi_i^a(t)$ and $\mathbb{E}_G \{\xi_i^a(t) \xi_j^b(t')\} = \delta_{ab} \delta_{ij} \delta(t-t')$.

Formally, the above path integral representation Eq. \eqref{eq:HS} resembles imaginary-time evolution with a Hamiltonian given by $-L^{(R)}$ and inverse temperature $t_f$. Thus, finite $t_f$ effectively introduces fluctuations analogous to thermal fluctuations, affecting the macroscopic properties of the system.
However, it is crucial to highlight the distinction between our situation and that of the quenched disordered quantum spin glass at finite temperature \cite{kopecTransverseFreezingQuantum1988}: (1) the order parameter $Q$ field depends only on one time variable $t$ rather than two, since the dynamics of $\hat{\rho}$ is Brownian, i.e. $\xi(t)$s are temporally independent; (2) the replica limit is now $R\rightarrow 1$ rather than $R\rightarrow 0$, due to the presence of a $\tr(\hat{\rho})$ factor in the expression of measurement probability $P[\xi]$.
This scenario resembles the Nishimori condition~\cite{nishimori_internal_1981}  with gauge symmetrization \cite{ref:guoyi}, where the disorder distribution is proportional to the partition function. Under such conditions, replica symmetry breaking is typically absent, even in the presence of vanishing linear order parameters—mirroring the behavior observed in Mattis-type spin glass models~\cite{mattis_solvable_1976,binder_spin_1986}.
\end{widetext}

\subsection{No readout case and SWSSB}
We begin by analyzing the no readout case described by the fully averaged Lindblad evolution described by Eq.~\eqref{eq:Liouvillian-l}. Notice that it corresponds to directly setting $R = 1$ in Eq. \eqref{eq:HS}, to be distinguished with the replica limit $R\rightarrow 1$ where we are more interested in the asymptotic behavior of $R$. Although this case is relatively simple, it establishes a foundation for understanding the more complex, non-commutative scenarios to be discussed soon in this subsection.
To decide the partition function of interest, we look for a quantity that can manifest the SWSSB transition while characterizing the initial-state-independent properties of the measurement dynamics.
Recalling that the formal super operator solution of the evolution $\exp(t_f L^{(1)})$ act as a quantum channel $\exp(t_f \mathcal{L}^{(1)})$ in density matrix representation, we thus consider the Choi channel fidelity (or square root of the average entanglement fidelity) \cite{schumacher_quantum_1996,schumacherSendingEntanglementNoisy1996,barnumInformationTransmissionNoisy1998,horodeckiGeneralTeleportationChannel1999,woodTensorNetworksGraphical2015,liuApproximateSymmetriesQuantum2023} with the identity channel,
\begin{equation}
    \bar{F}\left(\mathcal{C}, \mathbbm{1}\right) = F \left((\mathcal{C} \otimes \mathbbm{1})(\Psi),\left(\mathbbm{1}
    \otimes \mathbbm{1}\right)(\Psi)\right),
\end{equation}
where $\ket{\Psi} =(1/\sqrt{2^{n}}) \sum_{\{z_i\}} \ket{\{z_i\}} \otimes \ket{\{z_i\}}$ is the maximally entangled state between the physical system and a reference system, $\mathcal{C}$ is an arbitrary completely positive trace-preserving (CPTP) channel, $\mathbbm{1}$ is the identity channel and $F$ is the fidelity of quantum states. It is shown to be equivalent to the averaged channel fidelity \cite{horodeckiGeneralTeleportationChannel1999,woodTensorNetworksGraphical2015},
\begin{equation}
    \bar{F}\left(\mathcal{C}, \mathbbm{1}\right) = (1+2^{-n})F_{\text{av}} (\mathcal{C}) - 2^{-n},
\end{equation}
where
\begin{equation}
    F_{\text{av}}(\mathcal{C}) = \int d\psi \bra{\psi}\mathcal{C}(\ket{\psi}\bra{\psi}) \ket{\psi}.
\end{equation}
The above $\psi$ belongs to the $2^n$ dimensional Hilbert space and the integration is over the Fubini-Study measure. 

To simultaneously characterize both entanglement and SWSSB order, we construct a novel partition function from the channel fidelity:
\begin{equation}
\begin{aligned}
        \mathcal{Z}^{(1)} &\equiv \bar{F}\left(\exp(t_f\mathcal{L}^{(1)}), \mathbbm{1}\right)\\
        & = \bra{\Psi} (\exp(t_f\mathcal{L}^{(1)}) \otimes \mathbbm{1})(\Psi) \ket{\Psi} = P(\Psi)\\
        & = \tr\left(e^{t_f L^{(1)}}\right).
\end{aligned}
\label{eq:periodic}
\end{equation}
The first two lines view $\exp(t_f\mathcal{L}^{(1)})$ as a quantum channel, while the trace operation in the last line takes $L^{(1)}$ as a super operator.
The second line illustrates its operational meaning. Starting from a Bell state between the system and the reference, the system is subject to the evolution $\mathcal{L}^{(1)}$ while the reference is left untouched. Slightly later than the final time $t \rightarrow t_f^+$ , we apply Bell measurement and $P(\Psi)$ is the probability of post-selecting the $\Psi$ state. 
The third expression resembles a canonical partition function, with a periodic boundary condition imposed on the temporal direction. 
{To accurately detect SWSSB, we insert a symmetry-probing operator acting simultaneously on both the forward and backward branches of the Liouville space, building upon the definitions established above.}
Then, the corresponding order parameter for SWSSB is expressed as 
\begin{equation}
\begin{aligned}
    Q_i^{+-}(t) &= \frac{\bar{F}\left(\exp((t_f-t)\mathcal{L}^{(1)}) \circ \mathcal{Z}_i \circ\exp(t\mathcal{L}^{(1)}), \mathbbm{1}\right)}{\bar{F}\left(\exp(t_f\mathcal{L}^{(1)}), \mathbbm{1}\right)}\\
    &= \tr\left(e^{(t_f-t) L^{(1)}} Z^+_i Z^-_i e^{t L^{(1)}}\right)/\tr\left(e^{t_f L^{(1)}}\right),
\end{aligned}
\label{eq:Qpm}
\end{equation}
where $\mathcal{Z}_i(\rho) = Z_i \rho Z_i$ is a channel of operators.
{The resulting expectation value defines an order parameter which acquires a nonzero value in the presence of SWSSB and vanishes otherwise.}

This stands in sharp contrast to the standard Keldysh partition function~\cite{SCHWINGER1960169,schwingerBrownian,Keldysh:1964ud,siebererKeldyshFieldTheory2016,altlandCondensedMatterField2023a,kamenevFieldTheoryNonequilibrium2023,thompson_field_2023,sieberer_universality_2023}
\begin{equation}
    \sbra{I} \exp\left(t_f L^{(1)}\right) \sket{\rho(0)}, 
    \label{eq:keldysh}
\end{equation}
{which is normalized to unity for any trace-preserving Liouvillian. This normalization forces every one-point function of the Keldysh ``quantum'' fields to vanish and does not acquire a non-zero expectation value (or \emph{condenses}); and therefore, the standard Keldysh framework~\cite{sieberer_universality_2023} cannot support a non-trivial SWSSB order parameter~\cite{lessaStrongtoWeakSpontaneousSymmetry2024,salaSpontaneousStrongSymmetry2024,huang_hydrodynamics_2024}.}
From a complementary perspective, it results from the different temporal boundary conditions when evaluating the two quantities. In Eq. \eqref{eq:keldysh} the $\pm$ branches are connected by $\sbra{I}$ at the final time, while in Eq. \eqref{eq:periodic} they are isolated. 
Because SWSSB is encoded in the correlated order parameter between the $+$ and $-$ branches, sewing them together at the final time smears out the relevant structure and forces a trivial outcome.
In fact, Eq. \eqref{eq:periodic} is in analogy to the way of defining strange correlator~\cite{You_2014} which is shown to detect SWSSB in Ref. ~\cite{salaSpontaneousStrongSymmetry2024}. Meanwhile, Eq. \eqref{eq:periodic} is independent of the choice of the initial state and provides an intrinsic characterization of the dynamics.

With the mean-field decoupling, we restrict Eq. \eqref{eq:HS} to the $R=1$ case without replica indices, and consequently, the effective Liouvillian takes the form 
\begin{equation}
    L_{\text{eff}}^{(1)}(t) = 2J\sum_{ij}  Q_i^{+-}(t) K_{ij} Z_j^+  Z_j^- + 2h \sum_i X^+_i X^-_i.
    \label{eq:Leff1}
\end{equation}
Substituting in Eq. \eqref{eq:periodic}, the partition function follows
\begin{equation}
    \mathcal{Z}^{(1)} = \int D Q^{+-}(t) \exp\left({-S_l[Q^{+-}(t)]}\right),
\end{equation}
where the effective action is given by 
\begin{align}
    \mathcal{S}_s[Q^{+-}(t)] = &\int_0^{t_f}dt J\sum_{ij} Q_i^{+-}(t) K_{ij} Q_j^{+-}(t) \nonumber\\
    &- \log \tr \left[\exp \left( \int_0^{t_f} dtL_{\text{eff}}^{(1)}(t)\right)\right],
    \label{eq:action-averaged}
\end{align}   
and $Q^{+-}$ captures the coupling between $\pm$ branches (in a single replica copy).

In the no-readout case described by Eq.~\eqref{eq:Leff1}, the $X$-decoherence channel (represented by the $X^+ X^-$ term) commutes with the $Z^+ Z^-$ term and therefore does not compete with the mechanism responsible for SWSSB. Its only effect is to scale the partition function by an overall purity-reducing factor, which shifts the free-energy baseline without affecting the minimization condition. Consequently, $X$-decoherence is dynamically irrelevant for the SWSSB transition and can be omitted without loss of accuracy.
Meanwhile, since the effective Liouvillian in this regime contains no time-dependent noise, the auxiliary field $Q^{\alpha\beta}_i(t)$ becomes rigid along the temporal direction. Accordingly, we introduce the time-averaged order parameter 
\begin{equation}
    Q^{s}_i = (1/t_f) \int_0^{t_f} dt Q^{+-}_i(t),
\end{equation}
which yields the Landau's free energy 
\begin{equation}
\begin{aligned}
    f_s &= \frac{J t_f}{n} \sum_{ij} Q^s_i K_{ij} Q^s_j - \frac{1}{n} \sum_j \log \cosh \big(2J t_f\sum_i  Q^s_i K_{ij}\big)\\
    &\sim 2d J t_f Q^{s2} - \log \cosh \left(4d J t_f Q^s\right),
\end{aligned}
\label{eq:free-energy-averaged}
\end{equation}
where we assume spatial uniformness for $Q^s_i$ in the second line, which is accurate if we approach the phase transition point from the ordered phase. The expression is nothing but the mean-field Ising free energy, and gives the mean-field equation 
\begin{equation}
    Q^s = \tanh(4dJt_f Q^s).
    \label{eq:meanfield-s}
\end{equation}

Following the standard routine (e.g. Ref. \cite{altlandCondensedMatterField2023a}), one can also apply the spatial-continuum approximation up to second order derivative (see also Appendix \ref{app:3}),
\begin{equation}
\begin{aligned}
        &K_{ii'} \sim (2d+\partial^2_{\vec x_i})\delta(\vec x_i - \vec x_{i'}),\\
        &\sum_{j} K_{ij} K_{i'j} \sim (4d^2+4d\partial^2_{\vec x_i})\delta(\vec x_i - \vec x_{i'}).
\end{aligned}
\label{eq:spatial-continuum}
\end{equation}
In the vicinity of the critical point, the order parameter is sufficiently small. So we expand the Landau functional to quadratic order of $Q$ and gradient terms up to $\partial^{2}$, and obtain the Gaussian action
\begin{align}
         &\mathcal{S}_s[Q^s(\vec x)]=       (2d J t_f - 8d^2 J^2 t_f^2 ) \int d^d \vec x Q^s (\vec x)^2 \nonumber\\
         &+ (8dJ^2 t_f^2 - J t_f) \int d^d \vec x \left(\partial_{\vec x} Q^s (\vec x)\right)^2 + \mathcal{O}(\partial^4,Q^4).
         \label{eq:h0}
\end{align}
The mean-field critical point is determined by the positivity of the quadratic Hessian (the coefficients of the $Q^2$ term), which is simply the sign of mass term here. If the mass is positive,
\begin{equation}
    (2d J t_f - 8d^2 J^2 t_f^2 ) > 0,
\end{equation}
the Gaussian integral is stable and the system stays in the disordered phase. If the mass is negative, 
\begin{equation}
    (2d J t_f - 8d^2 J^2 t_f^2 ) < 0,
\end{equation}
the Gaussian integral is unstable. The behavior of the effective action $S[Q^{s}(\vec x)]$ is dominated by higher order terms of $Q$, thus leading to an ordered phase. 
The critical time for the second-order phase transition is
\begin{equation}
    t_c = \frac{1}{4dJ},
\end{equation}
across which the system evolves from a disordered state into an ordered state. The phase diagram is shown in Fig. \ref{fig:phase} (b).

In the vicinity of $t_c$, the coefficient of the $(\partial Q)^2$ term remains positive. Assume $t_f < t_c$, the two-point correlation of the order parameter field $Q^{s}(\vec x)$ has a finite correlation length
\begin{equation}
    \xi^s = \sqrt{\frac{8dJ t_f - 1 }{2d -8d^2J t_f}} \sim \frac{1}{\sqrt{8d^2 J}} (t_c - t_f)^{-\nu},
    \label{eq:correlation-h0}
\end{equation}
with the Gaussian critical exponent $\nu = 1/2$. In contrast in the ordered phase $t_f > t_c$, the order parameter field is long-range correlated $\braket{Q^{s}(\vec x)}$ saturates to a non-zero constant determined by quartic and higher interactions. Since the $Q^4$ interaction is non-vanishing, the upper critical dimension of the mean-field analysis is $d_c = 4$, above which the interaction is irrelevant and the validity of the mean-field predictions is guaranteed. For $d<d_c$, the SWSSB phase transition is controlled by the Wilson-Fisher critical point \cite{altlandCondensedMatterField2023a}, and could be determined by substituting in the coefficient of the $Q^4$ term.

In summary, in the no readout case when both the $ZZ$ and $X$ measurement outcomes are discarded, the monitoring evolution experience a SWSSB transition at a critical time $t_c$, which is independent of the $X$ measurement strenght $h$ (Fig. \ref{fig:phase} (b)). We then turn to the partial readout case and complete readout case which are much more non-trivial.

\begin{figure}[t]
    \includegraphics[width=1\columnwidth]{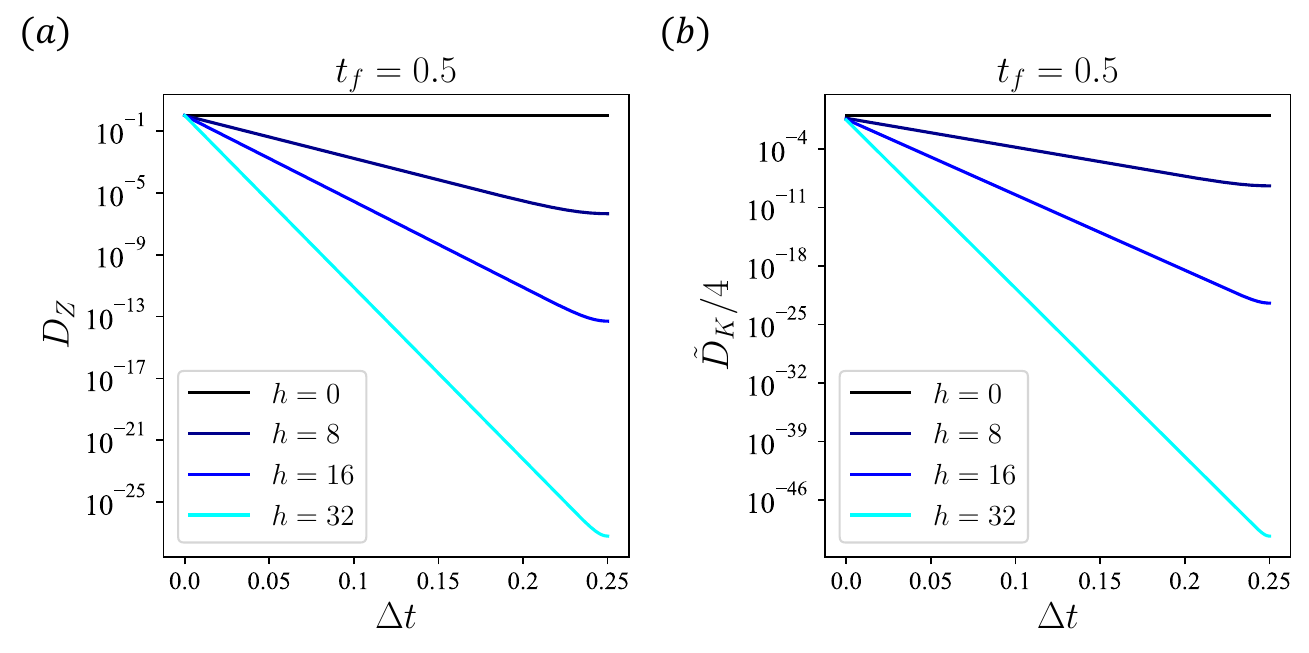}
    \caption{\textbf{Examples of the temporal coupling function setting different $X$ measurement rates $h$ (assuming $J=1$).} (a) $D^{Z}(\Delta t)$ in the partial readout case. (b) $\tilde{D}^{K}(\Delta t)$ in the complete readout case evaluated with saddle-point approximation of $\xi$. The final time is chosen as $t_f=0.5$.}
    \label{fig:correlation}
\end{figure}

\begin{widetext}
\subsection{Partial readout and mixed state phase transition} 
\label{sec:partial-readout}
We now examine the partial readout scenario, in which the outcomes of the $X$ measurements are averaged over while those of the $ZZ$ measurements are explicitly recorded, as described by Eq.~\eqref{eq:average-X}.

\textit{\textbf{Partition function and order parameters.}} In this context, we consider a suitable partition function that manifests both the SRE-to-LRE transition and the SWSSB transition.
Guided by the preceding discussion, we prepare the combined system–ancilla in the maximally entangled Bell state $\ket{\Psi}$, apply the measurement protocol of Eq.~\eqref{eq:average-X} exclusively to the system, and, at $t = t_f^{+}$, perform a Bell measurement and post-selects the same state $\ket{\Psi}$. Suppose that the overall evolution channel given a particular quantum trajectory $\xi$ is 
\begin{equation}
    \mathcal{C}_Z [\xi;t,t'] = \mathcal{T} \exp\left(\int_t^{t'}dt\mathcal{L}_Z[\xi]\right),
\end{equation}
the joint probability of both the trajectory $\xi$ and Bell measurement $\Psi$ is then expressed as
\begin{equation}
    P[\xi, \Psi] = P_G[\xi] \bra{\Psi} \left( \mathcal{C}_Z [\xi;0,t_f]\otimes \mathbbm{1}\right)(\Psi) \ket{\Psi},
\end{equation}
where $\mathcal{L}_Z[\xi]$ denotes the r.h.s. of Eq. \eqref{eq:average-X-master}.
We define the corresponding replica partition function as
\begin{equation}
    \mathcal{Z}^{(R)}_Z = \int D\xi P_G[\xi] \left[\bra{\Psi} \left(\mathcal{C}_Z [\xi;0,t_f] \otimes \mathbbm{1}\right)(\Psi) \ket{\Psi}\right]^R
    =\tr\left(e^{t_f \mathcal{L}^{(R)}_Z}\right),
    \label{eq:partition-def}
\end{equation}
This choice of partition function allows us to extract the LRE order and the SWSSB order simultaneously. Here for each individual trajectory $\xi$, the SWSSB order does not manifest in linear quantities like $\tr[O \mathcal{C}_Z [\xi;0,t_f] (\rho)]$ where $\pm$ branches are connected at the final time. This problem is resolved by considering the fidelity-type quantity (see Eq. \eqref{eq:order_s}) which captures the coupling between $\pm$ branches. Meanwhile, introducing replicas provides access to the Edwards-Anderson-type quantities for the LRE order.
Specifically, the partition function is associated with two distinct types of order parameters $Q_i^{\alpha\beta}$. The first type, $Q_i^{(+a)(-a)}$, measures correlations between forward $(+)$ and backward $(-)$ evolution branches within the same replica copy and serves as the order parameter for SWSSB: 
\begin{align}
    Q^{(+a)(-a)}_i(t) &=
    \frac{\displaystyle\int D\xi P_G[\xi]
     \bra{\Psi}^{\otimes R}\left(\mathcal{C}_Z [\xi;t,t_f] \right)^{\otimes R} \circ \mathcal{Z}_i^{a}\circ \left(\mathcal{C}_Z [\xi;0,t]\right)^{\otimes R}(\Psi^{\otimes R}) \ket{\Psi}^{\otimes R}}{\displaystyle\int D\xi P_G[\xi]
     \bra{\Psi}^{\otimes R} \left(\mathcal{C}_Z [\xi;0,t_f] \right)^{\otimes R}(\Psi^{\otimes R}) \ket{\Psi}^{\otimes R}} \nonumber\\
     &\xrightarrow{R\rightarrow 1}\frac{\bar{F}\left(\exp((t_f-t)\mathcal{L}^{(1)}) \circ \mathcal{Z}_i \circ\exp(t\mathcal{L}^{(1)}), \mathbbm{1}\right)}{\bar{F}\left(\exp(t_f\mathcal{L}^{(1)}), \mathbbm{1}\right)}. \label{eq:order_s}
\end{align}
where $\mathcal{Z}_i^a(\rho)=Z_i^a\rho Z_i^a$, and we omit the explicit reference part $\otimes\mathbbm{1}$ for simplicity. In the replica limit $R\rightarrow 1$, this quantity reduces precisely to the previously defined SWSSB order parameter in Eq.~\eqref{eq:Qpm}.
The second type of order parameter, $Q_i^{(\sigma_1 a)(\sigma_2 b)}$, characterizes correlations between different replica copies and serves as an order parameter for the LRE (or strong symmetry breaking) phase:
\begin{align}
    Q^{(\sigma_1 a)(\sigma_2 b)}_i(t) &=
    \frac{\displaystyle\int D\xi P_G[\xi]
     \bra{\Psi}^{\otimes R}\left(\mathcal{C}_Z [\xi;t,t_f] \right)^{\otimes R} \circ \mathcal{Z}_i^{(\sigma_1 a)}\circ \mathcal{Z}_i^{(\sigma_2 b)}\circ  \left(\mathcal{C}_Z [\xi;0,t]\right)^{\otimes R}(\Psi^{\otimes R}) \ket{\Psi}^{\otimes R}}{\displaystyle\int D\xi P_G[\xi]
     \bra{\Psi}^{\otimes R} \left(\mathcal{C}_Z [\xi;0,t_f] \right)^{\otimes R}(\Psi^{\otimes R}) \ket{\Psi}^{\otimes R}}\nonumber\\
     &\xrightarrow{R\rightarrow 1} \int D\xi P[\xi|\Psi] \prod_{m=1,2}
     \frac{\bar{F}\left(\mathcal{C}_Z [\xi;t,t_f] \circ \mathcal{Z}_i^{\sigma_m} \circ \mathcal{C}_Z [\xi;0,t], \mathbbm{1}\right)}{\bar{F}\left( \mathcal{C}_Z [\xi;0,t_f], \mathbbm{1}\right)}.
\end{align}
Here, $P[\xi|\Psi]=P[\xi,\Psi]/P(\Psi)$ is the conditional probability for the trajectories given post-selection on $\Psi$. The superscripts $\sigma_1,\sigma_2\in\{+,-\}$ specify the branch on which the operator $\mathcal{Z}_i$ acts, e.g.\ $\mathcal{Z}_i^{(+a)}(\rho)=Z_i^a\rho$ and $\mathcal{Z}_i^{(-a)}(\rho)=\rho Z_i^a$. Notice that we slightly abuse the notation $\bar{F}$ here, as its arguments are not strictly CPTP maps. In the replica limit, $Q_i^{(\sigma_1 a)(\sigma_2 b)}$ reduces to an Edwards-Anderson order parameter incorporating contributions from both $\pm$ branches of two replica copies $m=1,2$.

\textit{\textbf{Derivation of mean-field theory.}} To this end, we derive the mean-field effective action by decoupling the $ZZZZ$ term in Eq. \eqref{eq:average-X}, from which it follows that
\begin{equation}
    \mathcal{Z}^{(R)}_Z \propto \int DQ^{\alpha\beta}_i \tr \exp\left({t_f L_{eff,Z}^{(R)}}\right)
    \exp \int_0^{t_f} dt \left(  -J\sum_{\alpha<\beta,ij} Q_i^{\alpha\beta}(t) K_{ij} Q_j^{\alpha\beta}(t)\right),\label{eq:partition-Z}
\end{equation}
where
\begin{equation}
    e^{t_f L_{eff,Z}^{(R)}} =\mathcal{T}\exp\int_0^{t_f} dt\left(  2J\sum_{\alpha<\beta,ij}  Q_i^{\alpha\beta}(t) K_{ij} Z_j^\alpha  Z_j^\beta
    + 2h  \sum_i \sum_{a=1}^R X_i^{(+,a)}X_i^{(-,a)}  \right).
    \label{eq:eff-Z}
\end{equation}
First, we assume that in the vicinity of the second-order phase transition, the order parameter $Q^{\alpha\beta}_i(t)$ is sufficiently small so that we can expand Eq. \eqref{eq:eff} with respect to $Q$. Up to the second order, the effective action is expressed as (see Sec. SII of SM \cite{SupMat})
\begin{align}
    \mathcal{S}_Z^{(R)}[Q^{\alpha\beta}_i(t)]
    =& J\sum_{\alpha<\beta,ii'} \int_0^{t_f} dt Q_i^{\alpha\beta}(t) K_{ii'} Q_{i'}^{\alpha\beta}(t)
    -2J^2\sum_{a,ii'} \int_0^{t_f} dt\int_0^{t_f}dt' \Big(\sum_{j} K_{ij} K_{i'j}\Big)
    Q_i^{(+a)(-a)}(t) Q_{i'}^{(+a)(-a)}(t')  \notag\\
    & -2J^2\sum_{a<b,\sigma_1\sigma_2,ii'} \int_0^{t_f} dt\int_0^{t_f}dt'\Big(\sum_{j} K_{ij} K_{i'j}\Big) 
     D_{Z}(|t-t'|) Q_i^{(\sigma_1 a)(\sigma_2 b)}(t) Q_{i'}^{(\sigma_1 a)(\sigma_2 b)}(t')+ \mathcal{O}(Q^3),
    \label{eq:action-Z}
\end{align}
\end{widetext}
where $a, b =1, \cdots R$ are the replica indices, $\sigma_1 \sigma_2 = \pm$ denotes the $\pm$ branches and the propagator has the form
\begin{equation}
    D_{Z}(\Delta t) = \cosh^2(2h(t_f-2\Delta t)) \sech^2(2ht_f).
    \label{eq:DZ}
\end{equation}
Here, we denoted $\Delta t = |t-t'|$. This propagator function is visualized in Fig. \ref{fig:correlation} (a), which decays exponentially with respect to $\Delta t$ at finite $h$.
The temporal boundary condition Eq. \eqref{eq:periodic} ensures the invariance under time translation. The function $D_Z(\Delta t)$ is defined on $\Delta t\in [0, t_f]$ and satisfies
\begin{equation}
    \quad D_Z(t_f/2 + \Delta t) = D_Z(t_f/2 - \Delta t),
    \label{eq:reflection}
\end{equation}
as a consequence of temporal periodicity introduced by the trace operation.

Notably, two kinds of order parameters, i.e. $Q^{(\sigma_1 a)(\sigma_2 b)}$ and $Q^{(+a)(-a)}$, decouple at second order and display distinct temporal-coupling patterns.
\begin{itemize}
  \item \textbf{LRE order parameter.}  $Q^{(\sigma_1 a)(\sigma_2 b)}$ characterises the long-range-entangled (LRE) phase and is governed by the non-trivial kernel $D_Z(\Delta t)$, which encodes quantum fluctuations introduced by the non-commutativity between the $ZZ$ measurement and $X$-type decoherence.
  \item \textbf{SWSSB order parameter.}  $Q^{(+a)(-a)}$ characterizes spontaneous SWSSB and remains temporally rigid, as $X$ decoherence leaves it inert.  
\end{itemize}
Even for replica number $R>1$, their differing temporal responses imply that the two phases occupy separate regions of the same dynamical phase diagram: retaining $R>1$ preserves part of the measurement-trajectory information in the replica action, so both phases are realised in the dynamics of Eq.~\eqref{eq:average-X-master} once $ZZ$ measurement outcomes are collected.

\textit{\textbf{Replica limit.}}
Now we analyze the replica limit $R\rightarrow 1$. To extract the precise information of the mean-field phase transition, we cannot naively set $R=1$ in the action Eq. \eqref{eq:action-Z}, especially when $t_f$ is large. Accordingly, in the stationary limit $t_f\rightarrow +\infty$, the LRE phase survives for small but finite $h$, where $D_{Z}(\Delta t)$ is close to $1$ and thus $Q^{(\sigma_1 a)(\sigma_2 b)}$ is finite;
however, LRE phase vanishes for sufficiently large $h$ as $D_Z(\Delta t)\!\to\!0$ (where $0<\Delta t<t_f$), leading to the complete suppression of temporal correlations and thus $Q^{(\sigma_1 a)(\sigma_2 b)} = 0$. Please refer to Fig.~\ref{fig:correlation} for the numerical value of $D_{Z}(\Delta t)$.
Conversely, $Q^{(+a)(-a)}$ obeys the same effective action as in the fully averaged case, Eq.~\eqref{eq:h0}, and remains finite for arbitrary $h$ after the critical time, so SWSSB persists even at large~$h$.
The separation of the two transitions (see Fig. \ref{fig:phase} (c)) indicates that $\mathcal{S}_Z^{(R=1)}$ is reliable only for $t_f\!\approx\!t_c$ and $h\!\approx\!0$; deep inside the ordered regime ($t_f\!\gg\!t_c$) the mutual interaction of the two order parameters prevents a simultaneous perturbative expansion of the two order parameters. So we need to find a way of incorporating this interaction into the replica limit.


To clarify the interplay between the two order parameters, we analyze their behaviors in detail. For any finite $h$ and $t_f$, the SWSSB order parameter $Q^{(+a)(-a)}$ is governed by the effective action of Eq.~\eqref{eq:action-averaged}, whereas $Q^{(\sigma_1 a)(\sigma_2 b)}$ is influenced by the finite background value of $Q^{(+a)(-a)}$ in the replica limit. For convenience, we adopt the replica-symmetric ansatz, that assumes $Q^{ab}$ is the same for all pairs of distinct replicas $(a,b)$:
\begin{equation}
  Q^{(+a)(-a)}_i(t)=Q_i^{s}(t), \,
  Q^{(\sigma_1 a)(\sigma_2 b)}_i(t)=Q_i^{r,\sigma_1\sigma_2}(t),
  \label{eq:QRS_ansatz}
\end{equation}
where we label the two order parameters by $s$ and $r$, respectively. Our analysis proceeds by examining the structure of the total action, $\mathcal{S}_Z$, in the replica limit $R\rightarrow 1$.

First, we consider the SWSSB order parameter $Q_i^{s}(t)$. The analysis hinges on the behavior of the full action from Eq.~\eqref{eq:action-Z} in the replica limit, $R\rightarrow 1$. The crucial insight is derived from how terms in the action scale with the replica indices. Assuming replica symmetry, terms that are exclusively functions of $Q_i^{s}(t) = Q^{(+a)(-a)}_i(t)$ involve a summation over a single replica index, $\sum_a$, which yields a factor of $R$. In the $R\rightarrow 1$ limit, this contribution is of order $\mathcal{O}(1)$. In contrast, terms involving the SRE-to-LRE order parameter, $Q^{r,\sigma_1 \sigma_2}_i(t) = Q^{(\sigma_1 a)(\sigma_2 b)}_i(t)$, require a summation over a pair of distinct indices, $\sum_{a<b}$. This sum yields a factor of $R(R-1)/2$, which is of order $\mathcal{O}(R-1)$. Consequently, when considering the leading, $\mathcal{O}(1)$, part of the action, the dynamics of the SWSSB order parameter $Q^s$ are entirely decoupled from $Q^{r,\sigma_1 \sigma_2}$. This formal decoupling aligns with the physical picture in which the SWSSB transition occurs first (i.e., when $Q^{r,\sigma_1\sigma_2}=0$). Therefore, the SWSSB dynamics are effectively governed by the single-variable action $S_s[Q^s]$ of Eq.~\eqref{eq:action-averaged}.

Secondly, we consider the SRE-to-LRE order paramter $Q^{r,\sigma_1 \sigma_2}_i(t)$. From the above analysis, the vanishing of the action for $Q^{r,\sigma_1 \sigma_2}_i(t)$ in $O(1)$ of the replica limit reflects that the SRE-to-LRE transition only manifests in non-linear quantities of the trajectory-dependent density matrices.
As a consequence, when looking at the $\mathcal{O}(R-1)$ part of the action, the finite $Q^s$contributes by coupling to the SRE-to-LRE order parameter $Q^{r,\sigma_1 \sigma_2}$. 
Formally the action decouples into 
\begin{equation}
    \mathcal{S}_Z \rightarrow \mathcal{S}_s[Q^s] + \mathcal{O}(R-1) \mathcal{S}_r[Q^{r,\sigma_1 \sigma_2},Q^s]
    \label{eq:decoupling}
\end{equation}
when $R\rightarrow 1$, and the information of the SRE-to-LRE transition is captured by $\mathcal{S}_r$. Although we have only discussed the quadratic terms above, the formula of decoupling Eq. \eqref{eq:decoupling} is expected to hold for higher orders of $Q$ (by checking the summation of replica index order by order). It also holds for the complete readout case as we will see later in the Sec. \ref{sec:complete-readout}.

Here we remark that the $R(R-1)/2$ indicates the replica symmetry does not spontaneously break when $R\rightarrow 1$.
Roughly speaking, in conventional spin glasses the instability of replica symmetric solution results from the ``$-$' sign of the leading order $-R/2$ when taking $R\rightarrow 0$ \cite{nishimoriStatisticalPhysicsSpin2001}. However, in the measurement problem, the factor stays non-negative as $R$ approaches $1^+$, thus preserving the replica symmetry. This is consistent with the replica symmetric ansatz.

\begin{widetext}

{To investigate the critical behavior of the SRE-to-LRE transition, we examine the limit $R\!\to\!1^{+}$, where the SWSSB order parameter $Q^{s}$ is already finite while $Q^{r,\sigma_{1}\sigma_{2}}$ remains small, refer to the ansatz in Eq.~(\ref{eq:QRS_ansatz}) and discussion above.
In that sense, we keep $Q^s$ finite while expanding $Q^{r,\sigma_1 \sigma_2}$ to the second order, leading to the action in the replica limit (we have dropped the $\mathcal{O}(R-1)$ factor), and yields the quadratic action for the SRE-to-LRE transition}:
\begin{align}
    \mathcal{S}_r[Q^{r,\sigma_1 \sigma_2}_i(t)]
    &= \sum_{ii'} \left[ \sum_{\sigma_1\sigma_2}J\int_0^{t_f} dt Q^{r,\sigma_1 \sigma_2}_i(t) K_{ii'} Q^{r,\sigma_1 \sigma_2}_i(t)\right.\notag\\
    &-2J^2 \Big(\sum_{j} K_{ij} K_{i'j}\Big) \int_0^{t_f} dt\int_0^{t_f}dt'   \left.\sum_{\sigma_1\sigma_2,\sigma_1'\sigma_2'} \tilde{D}_{Z}^{\sigma_1\sigma_2,\sigma_1'\sigma_2'}(|t-t'|) Q^{r,\sigma_1 \sigma_2}_i(t) Q^{r,\sigma_1' \sigma_2'}_{i'}(t')\right]+ \mathcal{O}(Q^{r3}),
    \label{eq:action-r}
\end{align}
with a modified expression for 
the temporal coupling $\tilde{D}_{Z}^{\sigma_1\sigma_2,\sigma_1'\sigma_2'}(\Delta t)$,
\begin{equation}
    \tilde{D}_{Z}^{\sigma_1\sigma_2,\sigma_1'\sigma_2'}(\Delta t) = \left(Q^s \right)^{\frac{2-\sigma_1\sigma_1'-\sigma_2\sigma_2'}{2}} D_Z(\Delta t),
    \label{eq:DtildeZ}
\end{equation}
where $Q^s$ is the solution of Eq. \eqref{eq:meanfield-s}. Unlike Eq. \eqref{eq:action-Z}, now the action is not diagonal in the space of $\pm$ branches, so we apply the Keldysh rotation on the field $Q^r$ in the action Eq.~(\ref{eq:action-r}),
\begin{equation}
    {Q}^r \rightarrow H Q^r H^T, \quad \text{where} \quad
    H=\frac{1}{\sqrt{2}}\begin{pmatrix}
    1 & 1 \\
    1 & -1
    \end{pmatrix}, \quad
    Q^r = \begin{pmatrix}
    Q^{r,++} & Q^{r,+-} \\
    Q^{r,-+} & Q^{r,--}
    \end{pmatrix},
\label{eq:Keldysh-rotation}
\end{equation}
which diagonalizes the coupling $\tilde{D}_{Z}$ and hence the Hessian matrix (depending on $Q^s$),
\begin{equation}
        \tilde{D}_{Z} = D_Z \begin{pmatrix}
            1 & Q^s & Q^s & Q^{s2}\\
            Q^s & 1 & Q^{s2} & Q^s \\
            Q^s & Q^{s2} & 1 & Q^s \\
            Q^{s2} & Q^s & Q^s & 1
        \end{pmatrix} \rightarrow
        D_Z \begin{pmatrix}
            (1+Q^s)^2 & 0 & 0 & 0\\
            0 & 1-Q^{s2} & 0 & 0 \\
            0 & 0 & 1-Q^{s2} & 0 \\
            0 & 0 & 0 & (1-Q^s)^2
        \end{pmatrix}.
\end{equation}
The second order phase transition is determined by the sign change of the largest eigenvalue of $\tilde{D}_Z$, i.e. $(1+|Q^s|)^2$, which is associated with the eigenvector $Q^{r,K} = Q^{r,++} + Q^{r,--} + \sgn(Q^s) (Q^{r,+-} + Q^{r,-+})$. Consequently, the effective theory of the SRE-to-LRE transition is given by 
\begin{align}
    &\mathcal{S}_r[Q^{r,K}(\vec x,t)] =  \int d^d \vec x\int_0^{t_f} dt Q^{r,K}(\vec x, t)  (2dJ+ J\partial_{\vec x}^2) Q^{r,K}(\vec x t)\notag\\
    &\quad\quad\quad\quad -\int d^d \vec x\int_0^{t_f} dt\int_0^{t_f}dt' (1+|Q^s|)^2 D_{Z}(|t-t'|)   Q^{r,K}(\vec x, t) (8d^2J^2+ 8dJ^2\partial_{\vec x}^2) Q^{r,K}(\vec x, t') + \mathcal{O}(\partial^4,(Q^{r})^3),
    \label{eq:action-K}
\end{align}
\end{widetext}
where we assumed a spatial continuum. In the vicinity of $h=0$, $t \sim t_c$, we have $D_{Z}(|t-t'|)=1$ and $Q^s\sim 0$, thus the above formula coincides with Eq. \eqref{eq:h0} in the quadratic order, yielding the same phase transition point. However, one can check that Eq. \eqref{eq:action-K} has a nonzero $(Q)^3$ term even when $h=0$, leading to an upper critical dimension of $d_c = 6$. In addition, the negativity of the coefficient of $(Q)^3$ rules out the possibility of a first-order transition.

\textit{\textbf{Stationary phase transition.}} We then consider the stationary state.
Taking the $t_f \rightarrow + \infty$ limit while keeping $\Delta t$ finite, we have
\begin{equation}
    \lim_{t_f \rightarrow + \infty}D_Z(\Delta t) = e^{-8 h\Delta t},
    \label{eq:coupling-stationary}
\end{equation}
implying that the coupling has only a finite correlation range characterized by $ \tau= 1/8h$. This finite range allows us to perform a long-wave approximation, retaining terms up to second-order temporal derivatives (see Appendix \ref{app:3}): 
\begin{equation}
    D_Z(\Delta t) \sim 2(\tau + \tau^3 \partial_t^2)\delta(t-t').
    \label{eq:temporal-expansion}
\end{equation}
Notice that the additional factor $2$ complements the finite contribution near $t_f$.  In the stationary limit, the $Q^s$ order parameter reaches $Q^s = 1$, leading to the effective action:
\begin{align}
    &\mathcal{S}_r[Q^{r,K} (\vec x, t)] =  \int d^d \vec x\int dt \left\{(2dJ - 64 d^2 J^2 \tau) \left(Q^{r,K} \right)^2 \right. \notag\\
    &\quad\quad \left.+( 64 dJ^2\tau-J) \left(\partial_{\vec x}Q^{r,K} \right)^2 + 64 d^2 J^2 \tau^3 \left(\partial_{t}Q^{r,K} \right)^2\right\}\notag\\
    &\quad\quad +\mathcal{O}(\partial^4,(Q^{r})^3).
    \label{eq:stationary}
\end{align}
The vanishing of the mass term determines the critical non-commutative measurement strength, that is
\begin{equation}
    \tau_c = \frac{1}{32 dJ},\text{ or }\quad h_c = 4 d J.
    \label{eq:critical-h}
\end{equation}
For $h<h_c$ the system stays in the strong $\mathbb{Z}_2$ symmetry-breaking ordered phase. However, when $h>h_c$, the $X$ measurement destroys this order.
Approaching the critical point from the disordered phase, the spatial and temporal correlation length behave as:
\begin{equation}
    \begin{aligned}
        &\xi_{\vec x}^r = \sqrt\frac{64dJ\tau-1}{2d-64d^2 J\tau} \sim \frac{1}{8\sqrt{d^2 J}} (\tau_c - \tau)^{-\nu},\\
        &\xi_{t}^r = \sqrt\frac{32dJ\tau^3}{1-32d J\tau}\sim \frac{1}{(32 dJ)^{3/2}} (\tau_c - \tau)^{-\nu},
    \end{aligned}
    \label{eq:correlation-stationary}
\end{equation}
where the critical exponent remains $\nu = 1/2$. The upper critical dimension for the mean-field description of the stationary state, as given in Eq. \eqref{eq:stationary}, is $d_c + 1= 6$ or $d_c = 5$, since the additional time dimension is taken into consideration compared with Eq. \eqref{eq:h0}. This is a manifestation that quantum fluctuation has a milder effect than thermal fluctuation.

\textit{\textbf{Finite-time phase transition.}}
For the most general case where both $t_f$ and $h$ are finite, the dimension $d\geq 6$ is sufficient to guarantee the validity of the mean-field solution. We assume a temporally static and spatially uniform order parameter $Q^{r,K} (\vec x,t) = Q^{r,K}$ near the phase transition, which yields the Landau free energy
\begin{align}
        f = &\Big[2dJt_f - 8 d^2 J^2 t_f (1+|Q^s|)^2\int_0^{t_f} dtD_z (t)\Big] (Q^{r,K})^2\nonumber\\
        & + \mathcal{O}\Big[(Q^{r,K})^3\Big].
\end{align}
The quadratic coefficient determines the phase diagram as shown in Fig. \ref{fig:phase} (c). Such dynamics exhibit three different phases. 
\begin{enumerate}
    \item \textbf{Trivial phase.} When $Q^s = Q^{r,K} = 0$, the system is in the trivial phase with no symmetry breaking.
    \item  \textbf{SWSSB phase.} When $Q^s > 0$ and $Q^{r,K}=0$, the strong $\mathbb{Z}_2\times\mathbb{Z}_2$ symmetry spontaneously breaks down to weak $\mathbb{Z}_2$ symmetry. However, this is only a classical order with SRE.
    \item \textbf{LRE phase.} When both $Q^s>0$ and $Q^{r,K} >0$, the remaining weak $\mathbb{Z}_2$ symmetry is further broken, leaving no symmetry. This corresponds to the LRE phase. 
\end{enumerate} 
The above patterns are also summarized in Tab. \ref{tab}.
Notably, the LRE phase here is a mixed-state phase and the transition from the SWSSB phase to the LRE phase ($Q^{r,K}=0$ to $Q^{r,K} > 0$) is a mixed-state phase transition. A mixed-state LRE phase implies that despite the presence of decoherence or noise, the system retains nontrivial quantum entanglement over long distances \cite{fanDiagnosticsMixedstateTopological2023}. This is distinct from a pure-state LRE phase, where the entanglement is associated with a single quantum state (e.g., a GHZ state). The mixed-state phase transition here results from the competition between $X$ decoherence that tends to thermalize the system, and the $ZZ$ measurement, which tends to cool the system down to a $GHZ$ state.

As shown in Fig.~\ref{fig:phase} (c), when the strength of the non-commutativity $h$ takes different value, the finite time behavior differs:
\begin{enumerate}
    \item For $h=0$: The strong symmetry directly breaks down to no symmetry.
    \item For $0<h<h_c$: The symmetry-breaking pattern is successively $\mathbb{Z}_2 \times \mathbb{Z}_2 \rightarrow \mathbb{Z}_2 \rightarrow 0$, i.e. there exists an intermediate region of time $t_f$ where the dynamics exhibit weak symmetry before the system transitions to the fully broken symmetry state. 
    \item For $h=h_c$: At the quantum critical point, the Edwards-Anderson correlation length scales as $\xi^r_{\vec x} \sim \mathcal O(t_f^\nu)$, with $\nu =1/2$. 
    \item For $h>h_c$: The strong symmetry only breaks down to weak symmetry $\mathbb{Z}_2 \times \mathbb{Z}_2 \rightarrow \mathbb{Z}_2$ with no further symmetry breaking as time evolves.
\end{enumerate}
Notice that in the partial readout case, the direct strong symmetry breaking pattern $\mathbb{Z}_2 \times \mathbb{Z}_2 \rightarrow 0$ is unstable once the decoherence is turned on ($h>0$), and becomes $\mathbb{Z}_2 \times \mathbb{Z}_2 \rightarrow \mathbb{Z}_2  \rightarrow 0$ with an intermediate SWSSB region even for sufficiently small $h$. In comparison, the SRE-to-LRE transition in the complete readout case in Sec. \ref{sec:complete-readout} is a pure state transition averaged over disorder configurations (measurement trajectories), where the strong symmetry always directly breaks down to no symmetry $\mathbb{Z}_2 \times \mathbb{Z}_2 \rightarrow 0$ when $h<h_c$. 

From the perspective of quantum error correction (QEC), the $ZZ$ measurement serves as a preparation of repetition code, while the $X$ decoherence can be viewed as stochastic Pauli $X$ noise. The mixed-state phase transition to the LRE phase is conceptually the error threshold of repetition code under Pauli noise \cite{fanDiagnosticsMixedstateTopological2023,zhao2023extractingerrorthresholdsframework,niwa2024coherentinformationcsscodes}.
The difference from the common setup is that the noise occurs during the code preparation process, where a perfect code is not fully reached. Such imperfect codes belong to approximate QEC codes \cite{beny_general_2010,tyson_two-sided_2010,
Ng_transpose,Ng_unified,Beny_perturb,Beny_channel,BrandaoETH,
wangQuasiexactQuantumComputation2020,yiComplexityOrderApproximate2023,zhao2023extractingerrorthresholdsframework}. So the error threshold we are concerned with here is essentially an approximate QEC threshold \cite{zhao2023extractingerrorthresholdsframework}, which assesses whether the quantum information can be recovered from errors starting with an imperfect code subspace.

In summary, in the partial readout case when collecting $ZZ$ outcomes and discarding $X$ outcomes, the SWSSB transition still appears at the critical time $t_c$, while the LRE phase only exist for small enough $X$ measurement strength $h$ and long enough time $t_f$ after the SWSSB transition (Fig. \ref{fig:phase} (c)).

\begin{widetext}
\subsection{Complete readout and measurement phase transition}
\label{sec:complete-readout}

\textit{\textbf{Derivation of mean-field theory.}} 
We now analyze the complete readout case as described in Eq.~\eqref{eq:replica-evolution}. The derivation mainly follows the procedure introduced in Sec. \ref{sec:partial-readout}, so we briefly go through it and present the details in Sec. SII of SM \cite{SupMat}. In this scenario, all measurement outcomes are kept, enabling a pure-state description. To derive the mean-field theory, we compute the partition function $\tr e^{t_f\mathcal{L}_{\text{eff}}^{(R)}}$ following the formalism in Eq.~\eqref{eq:eff}. For $R > 1$, we expand the action up to the second order in $Q$, resulting in the following expression: 
\begin{align}
    &\mathcal{S}^{(R)}[Q^{\alpha\beta}_i(t)] = \sum_{\alpha<\beta,ii'}\left[  J \int_0^{t_f} dt Q_i^{\alpha\beta}(t) K_{ii'} Q_{i'}^{\alpha\beta}(t) - 2J^2       \Big(\sum_{j} K_{ij} K_{i'j}  \Big)  \int_0^{t_f} \int_0^{t_f}  dt dt' D^{(R)}(|t-t'|)  Q_i^{\alpha\beta}(t) Q_{i'}^{\alpha\beta}(t')\right].  \notag\\    
\label{eq:space-discrete}
\end{align}
The expression of $ D^{(R)}(\Delta t)$ can be found in Appendix~\ref{app:R-replica}.
This formula serves as a $R$-replica approximation for analyzing the measurement phase transition (see Appendix \ref{app:R-replica} for further details).
The time-dependence of the temporal coupling $D^{(R)}(\Delta t)$ again reflects the quantum fluctuation induced by the non-commutativity of $ZZ$ and $X$ measurements. Importantly, we still have the symmetry
\begin{equation}
    \quad D^{(R)}(t_f/2 + \Delta t) = D^{(R)}(t_f/2 - \Delta t).
    \label{eq:reflection2}
\end{equation}
Compared with the partial readout case (Eq. \eqref{eq:action-Z}), a notable distinction here is that $D^{(R)}(\Delta t)$ depends explicitly on the number of replica copies $R$. Furthermore, as long as $R>1$, the form of Eq. \eqref{eq:space-discrete} is symmetric in the index $\alpha\beta$. Consequently, the order parameters possess the symmetry: $Q^{(\sigma_1 a)(\sigma_2 b)}$ are identical for any replica pair $(a,b)$ and any branch pair $(\sigma_1,\sigma_2)$. Therefore, the two order parameters are all equal:  $Q^{(\sigma_1 a)(\sigma_2 b)}=Q^{(+a)(-a)}$, unlike the partial readout case. This coupling implies that the SWSSB order, characterized by $Q^{(\sigma_1 a)(\sigma_2 b)}=0$ and $Q^{(+a)(-a)}>0$, does not appear in the complete readout scenario. This absence of SWSSB order is intuitive because collecting all measurement results ensures that the final state is always pure. In such cases, there is no room for mixed-state order parameters.

\textbf{\textit{Replica limit.}} We now consider the replica limit $R \rightarrow 1$. A subtlety arises because directly setting $R = 1$ in the temporal coupling function yields $D^{(1)}(\Delta t) = 1$, reducing the second-order action to the fully averaged (no readout) case in Eq.~\eqref{eq:action-averaged}. This occurs because the $\mathcal{O}(1)$ contribution to the action arises solely from Lindblad evolution as shown in Eq.~\eqref{eq:lindblad}, which lacks measurement-trajectory information. To resolve this and capture the measurement phase transition, we must instead extract the subleading $\mathcal{O}(R - 1)$ contributions, as indicated by the decoupling in Eq.~\eqref{eq:decoupling}.
In this $R \rightarrow 1$ limit, the order parameter $Q^s \equiv Q^{(+a)(-a)}$ from the $\mathcal{O}(1)$ action loses sensitivity to the measurement trajectory. Its value converges to that of the no-readout case as shown in Eq.~\eqref{eq:meanfield-s}, reflecting only averaged dynamics. In contrast, the order parameter $Q^{r, \sigma_1 \sigma_2} \equiv Q^{(\sigma_1 a)(\sigma_2 b)}$ governs the measurement transition: it encodes \emph{nonlinear} dependencies on the density matrix and thus captures the distinct phases of post-measurement states across typical quantum trajectories. For clarity, we summarize the roles of these order parameters in Tab.~\ref{tab}.

To proceed, we apply a re-summation procedure similar to Eq.~\eqref{eq:action-r}, in which \(Q^s\) is treated as a finite constant while expanding in terms of \(Q^{r, \sigma_1 \sigma_2}\). This yields the effective action (see Appendix~\ref{app:complete} and Sec. SII of SM \cite{SupMat} for details):
\begin{align}
    &\mathcal{S}_r[Q^{r,\sigma_1 \sigma_2}_i(t)] =  \sum_{ii'} \Big[ \sum_{\sigma_1\sigma_2}J\int_0^{t_f} dt Q^{r,\sigma_1 \sigma_2}_i(t) K_{ii'} Q^{r,\sigma_1 \sigma_2}_i(t)\notag\\
    &\quad\quad\quad\quad -2J^2 \Big(\sum_{j} K_{ij} K_{i'j}\Big) \int_0^{t_f} dt\int_0^{t_f}dt'  \sum_{\sigma_1\sigma_2,\sigma_1'\sigma_2'}\tilde{D}^{(1),\sigma_1\sigma_2,\sigma_1'\sigma_2'}(t.t') Q^{r,\sigma_1 \sigma_2}_i(t) Q^{r,\sigma_1' \sigma_2'}_{i'}(t')\Big]+ \mathcal{O}(Q^{r3}),
    \label{eq:action-RL}
\end{align}
where we have dropped the $\sum_{a<b} = R(R-1)/2 = \mathcal{O}(R-1)$ factor when taking the replica limit.
The temporal coupling function is now expressed as, 
\begin{equation}  \tilde{D}^{(1),\sigma_1\sigma_2,\sigma_1'\sigma_2'}(t,t') =
    \frac{\int D\xi P_G[\xi] K^{\sigma_1 \sigma_1'}[\xi;t_>,t_<] K^{\sigma_2 \sigma_2'}[\xi;t_>,t_<] K_0[\xi]^{-1}}{\int D\xi P_G[\xi] K_0[\xi]^{R}},
\label{eq:Dtilde}
\end{equation}
which involves the solution of a disordered $(0+1)$ dimensional  effective model,
\begin{align}
    &L_0(t) = 4dJ Q^s Z^+ Z^- + \sqrt{2h} \xi(t) (X^+ + X^-), \notag\\
    &K_0[\xi] = \tr \mathcal{T} \exp\left( \int_0^{t_f} dt L_0(t)\right), \notag\\
    &K^{\sigma \sigma'}[\xi;t_>,t_<] =\tr\left[\mathcal{T} \exp\left( \int_{t_>}^{t_f} dt L_0(t)\right) Z^{\sigma}  \mathcal{T} \exp\left( \int_{t_<}^{t_>} dt L_0(t)\right)  Z^{\sigma'} \mathcal{T} \exp\left( \int_0^{t_<} dt L_0(t)\right) \right].
\end{align}
Here we have denoted $t_> = \max \{t,t'\}$, $t_< = \min \{t,t'\}$ and $\xi(t)$ follows the independent standard distribution $\mathbb{E}_G \{\xi(t) \xi(t')\} = \delta(t-t')$. The model $L_0$ describes simultaneous $X$ measurement and $Z$ decoherence on a single qubit. Notice that the exchange symmetry between $\pm$ branches enforces that
\begin{equation}
    K^{\sigma, \sigma'}[\xi;t_>,t_<] = K^{-\sigma, -\sigma'}[\xi;t_>,t_<].
\end{equation}
Then as we did in Eq. \eqref{eq:action-K}, we diagonalize the Hessian matrix similarly through the Keldysh rotation Eq. \eqref{eq:Keldysh-rotation} to determine the second-order phase transition. 
The largest eigenvalue is again associated with the Keldysh component of the $Q^r$ field, $Q^{r,K} = Q^{r,++} + Q^{r,--} + \sgn(Q^s) (Q^{r,+-} + Q^{r,-+})$, and is expressed as
\begin{equation}
\begin{aligned}
    \tilde{D}^{K}(t,t') &= \int D\xi \frac{ P_G[\xi] K_0[\xi]^{-1}}{\int D\xi P_G[\xi] K_0[\xi]} \left( K^{++}[\xi;t_>,t_<] +| K^{+-}[\xi;t_>,t_<]|\right)^2\\
    &= \int D\xi P[\xi] \left(\braket{Z^+(t_>)Z^+(t_<)}+\braket{Z^+(t_>)Z^-(t_<)}\right)^2
\end{aligned}
\label{eq:DK}
\end{equation}
 We then see that the above propagator takes the form of Edwards-Anderson-type correlation of  $\braket{Z^\sigma(t_>)Z^{\sigma'}(t_<)} = K^{\sigma \sigma'}[\xi;t_>,t_<]/K_0[\xi]$ averaged by the measurement probability $P[\xi]=P_G[\xi]K_0[\xi]/\int D\xi P_G[\xi] K_0[\xi]$. The effective action of the measurement phase transition is given by
\begin{align}
    &\mathcal{S}_r[Q^{r,K}(\vec x,t)] = \int d^d \vec x\int_0^{t_f} dt Q^{r,K}(\vec x, t)  (2dJ+ J\partial_{\vec x}^2) Q^{r,K}(\vec x t)\notag\\
    &-\int d^d \vec x\int_0^{t_f} dt\int_0^{t_f}dt' \tilde{D}^{K}(t,t')    Q^{r,K}(\vec x, t) (8d^2J^2+ 8dJ^2\partial_{\vec x}^2) Q^{r,K}(\vec x, t') + \mathcal{O}(\partial^4,Q^{r3}),
    \label{eq:action-RL-K}
\end{align}
where we again applied the continuum approximation. This yields the Landau free energy 
\begin{equation}
\begin{aligned}
        &f =  \left(2dJt_f - 8 d^2 J^2  \int_0^{t_f} dt \int_0^{t_f} dt' \tilde{D}^{K}(t,t')\right) (Q^{r,K})^2 + \mathcal{O}((Q^{r,K})^3).
\end{aligned}
\label{eq:f-RL-K}
\end{equation}
 
\end{widetext}

\begin{table*}[t]
\centering
\begin{tabular*}{\textwidth}{@{\extracolsep{\fill}}llccc}
\toprule
Phase & Condition & \multicolumn{3}{c}{Readout type} \\
\cmidrule(lr){3-5}
 & & Complete readout & No readout & Partial readout \\
\midrule
\multirow{4}{*}{Trivial} 
& $R \geq 2$ & $Q^s = Q^r = 0$ & \multirow{2}{*}{$Q^s = 0$} & \multirow{2}{*}{$Q^s = Q^r = 0$} \\
& $R \rightarrow 1$ & $Q^r = 0$ & & \\
& Symmetry breaking & No SSB & No SSB & No SSB \\
\cmidrule{1-5}
\multirow{4}{*}{SWSSB} 
& $R \geq 2$ & \multirow{4}{*}{None} & \multirow{2}{*}{$Q^s > 0$} & \multirow{2}{*}{$Q^s > 0$, $Q^r = 0$} \\
& $R \rightarrow 1$ & & & \\
& Symmetry breaking & & $\mathbb{Z}_2 \times \mathbb{Z}_2 \rightarrow \mathbb{Z}_2$ & $\mathbb{Z}_2 \times \mathbb{Z}_2 \rightarrow \mathbb{Z}_2$ \\
& Critical dimension & & $d_c=4$ & $d_c=4$\\
\cmidrule{1-5}
\multirow{6}{*}{LRE} 
& $R \geq 2$ & $Q^s = Q^r > 0$ & \multirow{6}{*}{None} & \multirow{2}{*}{$Q^s > 0$, $Q^r > 0$} \\
& $R \rightarrow 1$ & $Q^r > 0$ & & \\
& Symmetry breaking & $\mathbb{Z}_2 \times \mathbb{Z}_2 \rightarrow 0$ & & $\mathbb{Z}_2 \rightarrow 0$ \\
& Critical point ($t_f \rightarrow +\infty$) & $h_c \simeq 1.42dJ$ & & $h_c = 4dJ$ \\
& \multirow{2}{*}{Critical dimension} & {$d_c=6$ when $h\rightarrow 0$} & & {$d_c=6$ when $h\rightarrow 0$}\\
& & $d_c=5$ when $t_f\rightarrow +\infty$ & & $d_c=5$ when $t_f\rightarrow +\infty$\\
\bottomrule
\end{tabular*}
\caption{Comparison of order parameters, symmetry breaking, and critical behavior across different phases and readout conditions. Notice that in the complete readout case, $R\geq 2$ and $R\rightarrow 1$ yield different quantitative phase boundaries, but the qualitative behaviors are the same. $R=2$ for the no readout case trivially refers to replicating the partition function Eq. \eqref{eq:periodic} to $R$ copies and is included for completeness.}
\label{tab}
\end{table*}

\textit{\textbf{Calculating the propagator to estimate phase transition.}} 
In contrast to Eq. \eqref{eq:action-r}, the temporal coupling function $\tilde{D}^{K}(t,t')$ lacks an exact analytical expression. Consequently, we employ a mean-field approximation for the measurement trajectory $\xi(t)$. The main steps of this analysis are presented here, with a more detailed derivation provided in Sec. SIII of SM \cite{SupMat}. The expression for $\tilde{D}^{K}(t,t')$ in Eq. \eqref{eq:DK} is analogous to the expectation value of an observable with respect to the partition function $\mathcal{Z}_0 = \int D\xi P_G[\xi]K_0[\xi]$. This formulation allows for an approximation based on the saddle-point value of $\xi$. To formalize this, we define the Landau free energy as $f_0[\xi]=-(1/t_f) \log P_G[\xi]K_0[\xi]$. The resulting saddle-point equation is then given by:
\begin{equation}
    \frac{\delta}{\delta \xi(t)} f[\xi] = \xi(t) - \frac{1}{t_f} \frac{\delta}{\delta \xi(t)}\log K_0[\xi]=0.
    \label{mean-field variance}
\end{equation}
We consider the stationary solution $\xi(t)=\xi$ which yields an analytical form of the free energy,
\begin{equation}
    f_0(\xi) = \frac{\xi^2}{2} - \frac{1}{t_f} \log \left[ 2\cosh\left(2 t_f \Omega(\xi)\right) + 2\cosh\left(t_f \Gamma\right) \right].
\label{eq:meanfield-xi}
\end{equation}
Here, $\Omega(\xi)\equiv \sqrt{4 d^2 J^2 (Q^s)^2 + 2 h \xi^2}$ represents the $\xi$-dependent effective frequency, while $\Gamma=4 d J Q^s$ is a constant term. Consequently, the saddle-point equation, derived from $\partial f_0 / \partial \xi = 0$, takes the  form:
\begin{equation}
    \Omega(\xi) = \frac{4 h \sinh\left( 2t_f \Omega(\xi) \right)}{\cosh\left(t_f \Gamma\right) + \cosh\left( 2t_f \Omega(\xi)\right)}.
\label{eq:saddle-xi}
\end{equation}

At the mean-field level, we find that $f_0$ undergoes a phase transition, with $\xi = 0$ for small $h$ and $t_f$, and $\xi > 0$ when both parameters are sufficiently large. Substituting the solution of Eq.~\eqref{eq:saddle-xi} into Eq.~\eqref{eq:DK} yields the approximate $\tilde{D}^K$ values shown in Fig.~\ref{fig:correlation}(b), which continue to exhibit exponential decay with $\Delta t$, analogous to the partial readout case. Further evaluation of Eq.~\eqref{eq:f-RL-K} produces the phase diagram in Fig.~\ref{fig:phase}(a).
In contrast to the partial readout case, where three distinct phases are present, here only two phases are observed:
\begin{enumerate}
    \item \textbf{Trivial phase.} $Q^{r,K} = 0$ indicates the trivial phase.
    \item \textbf{LRE phase.} When $Q^{r,K} >0$, the strong symmetry $\mathbb{Z}_2 \times \mathbb{Z}_2$ breaks down to no symmetry, corresponding to the LRE phase. 
\end{enumerate} 
To be noted, in the compelte readout case the measurement phase transition from the SRE phase to the LRE phase exhibits a direct strong symmetry-breaking $\mathbb{Z}_2 \times \mathbb{Z}_2 \rightarrow 0$ since the post-readout state is always pure. Here only the order parameter $Q^{r,K}$ plays the role of distinguishing different phases and the nonzero mixed-state order parameter $Q^s$ does not indicates SWSSB phase.
Consider fixing the $X$ measurement rate $h$ and increasing the overall time $t_f$.
\begin{enumerate}
    \item $0\leq h<h_c$: The system always evolves into the LRE phase after a critical time, where the symmetry breaking pattern is $\mathbb{Z}_2 \times \mathbb{Z}_2 \rightarrow 0$.
    \item $h=h_c$: The quantum critical point where the Edwards-Anderson corelation length $\xi^r_{\vec x} \sim \mathcal O(t_f^\nu)$, $\nu =1/2$. 
    \item $h > h_c$: The system eventually stays in the trivial phase of SRE product states where no symmetry spontaneously breaks.
\end{enumerate}

In the stationary limit $t_f \rightarrow +\infty$, the saddle point value of $\xi$ takes
\begin{equation}
    \xi = \left\{ \begin{aligned}
        &\pm\sqrt{8h -  2 d^2 J^2/h}, &h>h_0,\\
        &0, &h\leq h_0,
    \end{aligned}\right.
\end{equation}
where the critical point of $f_0$ is $h_0 = dJ/2$. As a result, the temporal coupling function takes the form 
\begin{equation}
    \tilde{D}^{K}(t,t') = \left\{ \begin{aligned}
        &\left(1+\frac{dJ}{2h}\right)^2 e^{- 8 (2h - dJ) \Delta t}, &h>h_0,\\
        &4, &h\leq h_0,
        \end{aligned}\right.
\end{equation}
when $h> h_0$, it is exponentially decaying with characteristic time scale $\tau = 1/(16h-8dJ)$. Similar to Eq. \eqref{eq:temporal-expansion}, we expand it according to derivatives 
\begin{equation}
    \tilde{D}^{K}(t,t')  \sim  2 \left(1+\frac{dJ}{2h}\right)^2 \left(\tau + \tau^3 \partial_t^2\right)\delta(t-t'),
\end{equation}
which leads to the effective action of stationary limit the complete readout case
\begin{align}
    \mathcal{S}[Q^{r,K} (\vec x, t)] = &\frac{a}{2}  \int d^d \vec x \int dt
    \Big[\frac{1}{c^2}\left(\partial_{t}Q^{r,K} \right)^2+\left(\partial_{\vec x}Q^{r,K} \right)^2\notag\\
    &+m^2c^2\left(Q^{r,K} \right)^2 + \mathcal{O}(\partial^4,Q^{r3})\Big],
    \label{eq:stationary-cp}
\end{align}
where the coefficients are
\begin{equation}
    \begin{aligned}
        &a = \frac{dJ^2(2h+dJ)}{h^2(2h-dJ)}-2J,\\
        &c =8(2h-dJ) \sqrt{\frac{1}{d}-\frac{ 2h^2(2h-dJ)}{d^2J(2h+dJ)}},\\
        & m = \frac{d \sqrt{(2 h + d J) (4 dJ h^2 (2 h - d J) - 
    d^2 J^2 (2 h + d J))}}{8 (2 h - d J) (-2 h^2 (2 h - d J)+  d J (2 h + d J))}.
    \end{aligned}
\end{equation}
Consequently, the stationary phase transition point is given by the vanishing point of the mass term, that is
\begin{equation}
    h_c \simeq 1.42dJ.
\end{equation} 
Notice that $0<h_0<h_c$ when $d>0, J>0$, which makes the analysis self-consistent. Meanwhile, it can be checked that the spatial and temporal correlation length also scales like
\begin{equation}
    \xi_{\vec x} \sim \xi_{t} \sim \mathcal{O}\left((h-h_c)^{-v}\right),
\end{equation}
with $\nu = 1/2$.

Throughout the analysis above, the measurement phase transition induced by non-commutative measurements must be mediated by nonzero $Q^s$ in the replica limit when taking the channel fidelity as the partition function. If one considers a specific initial state $\rho_0$ of the system (without ancilla), the replica Keldysh partition function can still capture the measurement transition. In that case, the nonzero-ness of $Q^s$ is automatically imposed by the temporal boundary condition $Q^s \sim \tr( Z \rho_0 Z) = 1$ rather than a spontaneous symmetry breaking. The calculation of the mean-field phase transition should be similar without providing more insights, hence it is skipped in the work.

In summary, in the complete readout case when collecting both $ZZ$ and $X$ outcomes, the evolution enters the LRE phase for sufficiently long time $t_f$ at small $h$, while the $SWSSB$ is absent here (Fig. \ref{fig:phase} (a)).

\section{Discussion}
\label{sec:discussion}
In this work, we studied the effect of simultaneously monitoring two types of non-commutative operators, that is Ising coupling $Z_i Z_j$ and single-site $X_i$, concerning whether their outcomes are collected. We treated the randomness of measurement as a Nishimori-type disorder and applied the replica method. We derived the effective theory in the replica limit at the mean-field level, which enabled us to calculate the finite-time phase diagram. We identified the lRE-to-SRE transition in the complete readout case, the SWSSB phase transition in the no readout case, and both transitions in the partial readout case.

Although we have focused on mean-field results, our method naturally enables further perturbative renormalization calculations concerning higher orders of $Q$ to get a more accurate estimation of the phase transition below the upper critical dimension. It is also straightforward to generalize to other temporal boundary conditions instead of the initial-state-independent one discussed above. Importantly, we incorporate the non-commutativity between the two distinct classes of operators, noting that operators within the same class continue to commute. Another scenario of interest involves a single class of operators in which each operator does not commute with its spatially adjacent counterparts; a representative example is the monitoring of composite operators such as $c_1 Z_i Z_j + c_2 X_i$. 
This situation may result from more general coherent noise mechanisms in the system–detector coupling. Nevertheless, the proposed methods are, in principle, still applicable in such cases.

We anticipate that our work will provide insights into more sophisticated and more realistic measurement transitions in state preparations. Nonetheless, there are further difficulties to be overcome. For example, in the noisy preparation of topological code one encounters the spontaneous breaking of higher-form symmetries \cite{gaiottoGeneralizedGlobalSymmetries2015} or subsystem symmetries \cite{nandkishoreFractons2019}, both of which lack local order parameters and therefore fall outside the scope of a simple mean-field treatment, necessitating further investigation. Furthermore, 
recent studies have reported that the honeycomb Floquet code with discrete weak measurements exhibits exhibits a Majorana-liquid-like behavior~\cite{zhu_qubit_2023}. Developing an analytical framework capable of capturing such liquid-like behavior remains an open and challenging problem.

\begin{acknowledgements}
The author thanks Sebastian Diehl, Xiao Chen and Chong Wang for helpful discussions. This work is supported by the National Natural Science Foundation of China (Grant No.~92365111), Shanghai Municipal Science and Technology (Grant No.~25LZ2600200), Beijing Natural Science Foundation (Grants No.~Z220002), and the Innovation Program for Quantum Science and Technology (Grant No.~2021ZD0302400).
\end{acknowledgements}

\begin{appendix}
\section{Review of Weak Measurement}
\label{app:1}
The formalism of projective measurement does not suffice to describe real-world measurement processes, since they unavoidably suffer from noise. Experimental realization of quantum measurement always requires coupling the system to an external detector, which could be a rather sophisticated multi-layer system. A noisy coupling eventually leads to imperfect weak measurement. For example, measurements in atomic or superconducting qubit systems utilize resonators, and realistic noises in these processes are of Gaussian type \cite{clerkIntroductionQuantumNoise2010,krantz_quantum_2019}. In general, one needs to enhance the overall time of monitoring to achieve a more precise measurement, and such a process is described by continuous weak measurement \cite{wisemanQuantumMeasurementControl2009,jacobsQuantumMeasurementTheory2014},
\begin{equation}
\begin{aligned}
    &\partial_t \hat{\rho} = - k [O,[O,\hat{\rho}]] + 4 k \xi(t) \{O,\hat{\rho}\},\\
    &\mathbb{E}_G\{\xi(t)\} = 0, \quad \mathbb{E}_G\{\xi(t)\xi(t')\} = \frac{1}{8k} \delta(t-t'),
\end{aligned}
\label{eq:linear-stochastic}
\end{equation}
where $O$ is the measured operator and $\hat{\rho}$ is the unnormalized state. Such a linear differential equation is solved by 
\begin{equation}
\begin{aligned}
       & \hat{\rho}[\xi,t_f] =  K[\xi] \rho (0) K[\xi]^\dagger,\\
       & K[\xi] = \exp{\left[ \int_0^{t_f} dt \left(4k\xi(t)O-2k O^2\right)\right ]},   
\end{aligned}
\label{eq:linear-final}
\end{equation}
and each unnormalized state appears in the ensemble with probability 
\begin{equation}
    P_G[\xi] = \frac{1}{\mathcal{N}} \exp \left[- \int_0^{t_f} dt 4 k \xi (t)^2\right].
\end{equation}
Here $\mathcal{N} = \int D\xi \exp \left[- \int_0^{t_f} dt 4 k \xi (t)^2\right]$ is the normalization of the functional integral. One can check that the measurement quantum channel
\begin{equation}
    \mathcal{M}(\cdot ) = \int D\xi P_G[\xi]  \left(K[\xi] \cdot K[\xi]^\dagger\right) \otimes \ket{\xi}_c \bra{\xi}_c 
\end{equation}
is trace-preserving, thus probability is conserved. Here $\ket{\xi}_c$ formally represents the classical register that stores the information of the measurement outcome. This quantum channel incorporates the complete information of the measurement process. Given a specific measurement trajectory $\xi$, it is clear from the above expression that the physical normalized state of the system part is
\begin{align}
    &\rho [\xi, t_f] =  \frac{\hat{\rho} [\xi, t_f]}{\tr\left(\hat{\rho} [\xi, t_f]\right)} = \frac{K[\xi] \rho(0) K[\xi]^\dagger}{\tr(K[\xi] \rho(0) K[\xi]^\dagger)}.
\end{align}
The corresponding physical measurement probability is given by
\begin{equation}
    P[\xi] = P_G [\xi] \tr(\hat{\rho}[\xi,t_f]) = P_G [\xi] \tr(K[\xi] \rho(0) K[\xi]^\dagger).
\end{equation}
Ignoring measurement outcomes means tracing over the classical register, 
\begin{equation}
    \tr_c \mathcal{M}(\cdot ) = \int D\xi P_G[\xi] K[\xi] \cdot K[\xi]^\dagger.
\end{equation}
The Kraus operator of the measurement channels is written as 
\begin{equation}
\begin{aligned}
    M[\xi] &= \sqrt{P_G[\xi]} K[\xi] \\&=  \frac{1}{\mathcal{N}^{1/2}}  \exp{\left[ -\int_0^{t_f} dt 2k \left(\xi(t)-O\right)^2\right ]}.
    \label{eq:Kraus-Gaussian}
\end{aligned}
\end{equation}
One could check that such a Kraus operator defines a set of positive operator-valued measurements (POVM) \cite{nielsenQuantumComputationQuantum2010}, which normalizes as 
\begin{equation}
    \int D\xi E[\xi] = I, E[\xi] =  M[\xi]^\dagger M[\xi].
\end{equation}
$E[\xi]$ is called POVM operator.
For the case where only a single operator (or multiple commuting operators) is measured, the final state Eq. \eqref{eq:linear-final} only depends on the temporal-averaged measurement result
\begin{equation}
    s = \frac{1}{t_f} \int^{t_f}_0 dt \xi(t)
\end{equation}
instead of the details of the trajectory $\xi(t)$. 
Specifically, the central limit theorem states that for a Gaussian distribution $P_G[\xi] = (1/\mathcal{N})\exp\left(-(1/2)\int_{t_1}^{t_2} dt \xi(t)^2  \right)$, the statistical average $\xi= ({1}/({t_2-t_1}) )\int_{t_1}^{t_2} dt \xi(t)$ also follows from a Gaussian distribution, which has mean value $0$ and variance $1/(t_2-t_1)$. Thus for an arbitrary function $f$ that depend only on the average of $\xi(t)$, we have
\begin{equation}
\begin{aligned}
    &\frac{1}{\mathcal{N}}\int D\xi(t)\exp\left(-\frac{1}{2}\int_{t_1}^{t_2} dt \xi(t)^2  \right) f\left(\frac{1}{t_2-t_1} \int_{t_1}^{t_2} dt \xi(t)\right)\\
    &= \sqrt{\frac{t_2-t_1}{2\pi}}\int_{-\infty}^{+\infty} d\xi \exp\left(-\frac{1}{2}(t_2-t_1)\xi^2\right) f(\xi).    
\end{aligned}
\label{eq:central-limit}
\end{equation}
The above relation could be applied to any physical quantity evaluated with $\hat{\rho}[\xi, t_f]$ in Eq. \eqref{eq:linear-final},
so we have an equivalent description of the final state
\begin{equation}
    \begin{aligned}
    &\hat{\rho} (s, t_f) = K(s) \rho (0) K(s)^\dagger,\\
    &K(s) =  \exp{\left[ kt_f \left(4 sO -2O^2\right)\right ]},\\
    &P_G (s) = \sqrt{\frac{4k t_f}{\pi}} \exp \left(-4k t_f s^2\right).
    \end{aligned}
    \label{eq:average-final}
\end{equation}
The corresponding POVM operator is written as 
\begin{equation}
\begin{aligned}
    &M(s) = \left(\frac{4k t_f}{\pi} \right)^{1/4} \exp{\left[ 2kt_f \left( s -O\right)^2\right ]}\\
    &E(s) = \sqrt{\frac{4k t_f}{\pi}} \exp{\left[ 4kt_f \left( s -O\right)^2\right ]}.
\end{aligned}
\label{eq:POVM-single}
\end{equation}
Intuitively, the measurement result $s$ concentrates around the eigenvalues of $O$ as Gaussian distributions with variance $1/8kt_f$. In the large time limit $t_f \rightarrow$, the variance vanishes, and the distribution becomes a delta function, 
\begin{equation}
    \lim_{t_f \rightarrow +\infty} E(s) = \delta(s - O),
\end{equation}
therefore the projective measurement is recovered.

 As for simultaneously measuring multiple non-commuting operators $\{ O_i\}_i$, the dynamics still obeys the differential equation 
\begin{equation}
\begin{aligned}
    &\partial_t \hat{\rho} = -\sum_i k_i [O_i,[O_i,\hat{\rho}]] + 4\sum_i k_i \xi_i(t) \{O_i,\hat{\rho}\},\\
    &\mathbb{E}_G\{\xi_i(t)\} = 0, \quad \mathbb{E}_G\{\xi_i(t)\xi_j(t')\} = \frac{1}{8k_i}\delta_{ij} \delta(t-t'),
\end{aligned}
\end{equation}
where $k_i$ and $\xi_i$ is the measurement rate and trajectory for each $O_i$. This is because the contribution of the commutator in each infinitesimal time step $dt$ is of order $o(dt)$ and hence does not change the form of Eq. \eqref{eq:linear-stochastic} \cite{jacobsQuantumMeasurementTheory2014}.
However Eq. \eqref{eq:average-final} does not apply to this case, as the final state spends on the complete information of the trajectories $\{\xi_i(t)\}_i$. This can be seen from the solution of the evolution 
\begin{equation}
    K[\xi] = \mathcal{T}  \exp{\left[ \int_0^{t_f} dt\sum_i \left(4k_i\xi_i(t)O_i-2k_i O_i^2\right)\right ]},
\end{equation}
where we must keep track of the time ordering $\mathcal{T}$. As a result, we cannot naively evaluate the time integral as $s_i = (1/t_f)\int_0^{t_f} dt \xi_i(t)$. 

Aside from the microscopic origin, the problem of weak measurement can also emerge at the circuit level. One commonly needs to measure multi-qubit operators to prepare many-body quantum states such as stabilizer codes. Because of the limitation that we can only perform single-qubit measurements at the experimental level, we have to entangle the system qubits to ancilla qubits and then measure ancilla qubits to achieve multi-qubit measurement of the system. In this setup, weak measurement of the system qubits emerges due to coherent errors on the entanglement gate even if the ancilla measurement is perfect (projective). For example, suppose that we are measuring a multi-qubit operator $O$ satisfying $O^2=I$ using a single ancilla qubit, then the Kraus operator of POVM is given by \cite{ref:guoyi}
\begin{equation}
    M(s) = \frac{\exp(\beta s O/2)}{\sqrt{2\cosh \beta}}, \quad \sum_{s=\pm}M(s)^\dagger M(s) = I,
\end{equation}
where $s=\pm 1$ is the measurement outcome provided by the ancilla qubit, and $\beta$ is the discrete weak measurement's strength (or rate). The noise here is binary type instead of Gaussian type in Eq. \eqref{eq:Kraus-Gaussian}. Higher precision can also be achieved through multiple rounds of the same measurement circuit. In that case, the binary noises accumulate and approach Gaussian-type according to the central limit theorem.

Concretely, we label the rounds with discrete time $t = 1, \cdots, T$, 
the overall weak measurement operator is given by 
\begin{equation}
   M(\{s_t\}) = \prod_t M(s_t) = \frac{\exp(\beta \sum_t s_t O/2)}{\sqrt{2\cosh \beta}^t},
\end{equation}
where $s_t$ is the outcome at time step $t$. Notice that it only depends on the temporal-averaged result 
\begin{equation}
    s = \frac{1}{T} \sum_t s_t.
\end{equation}
Similar to Eq. \eqref{eq:average-final}, we can apply the central limit theorem for discrete variables. Consider the POVM operator for each $s_t$
\begin{equation}
    E(s_t) = \frac{\exp(\beta s_t O)}{2\cosh\beta}.
\end{equation}
Since we are now dealing with a single measurement operator $O$ (also for commutative measurements), the overall POVM measurement operator $E(\{s_t\}) = M(\{s_t\})^\dagger M(\{s_t\})$ factorizes into products of $E(s_t)$. Thus each $s_t$ can be viewed as an independent and identically distributed random variable with  operator-valued mean $\tanh\beta O$ and variance $\sigma^2 = 1$. According to the central limit theorem, the average outcome $s$ 
in the $T \rightarrow \infty$ limit approaches a Gaussian random variable  with mean value $\tanh \beta O$ and variance $1/T$. 
As a result, for sufficiently large $T$ the weak measurement is equivalently described by
\begin{equation}
    E(s) = \sqrt{ \frac{T}{2\pi}} \exp\left[ - \frac{T}{2} (s - \tanh \beta O)^2\right].
\end{equation}
Redefine the variable as $s \rightarrow s \tanh \beta $, we arrive at the form of Eq. \eqref{eq:POVM-single} with the identification of parameters 
\begin{equation}
    T\tanh^2 \beta \sim 8 k t_f.
\end{equation}
So for a sufficiently long time, binary weak measurement approximates Gaussian weak measurement. 
The precision of this approximation is controlled by Chebyshev's inequality, which leads to the requirement $T \gg \sigma^2 = 1$.
For an infinitely long measurement process, if we look at a coarse-grained time scale $\Delta T$ which is sufficiently long for the Gaussian approximation but sufficiently small compared with the intrinsic time scale $1/\beta^2 $ for continuous approximation, i.e. $1 \ll \Delta T \ll 1/\beta^2$, then such a process can be described by Eq. \eqref{eq:linear-stochastic}.

As for the non-commutative measurements case, suppose that the different kinds of measurements are applied periodically one layer after another, and each layer is of width $\Delta T$. Then for sufficiently small measurement strength $\beta^2 \Delta T \sim k dt$ the contribution of the commutator is of order $o(dt )$ and hence ignored in the differential equation. 
In conclusion, multiple rounds of discrete weak measurements at a coarse-grained level can be approximated by continuous weak measurements for a sufficiently large time and a sufficiently small measurement strength.

\section{Spatial continuum and temporal long-wave approximation}
\label{app:3}
Here we explain the standard procedure for taking Spatial continuum and temporal long-wave approximation, see for example Ref. \cite{altlandCondensedMatterField2023a}. For the continuum approximation, we Fourier transform the order parameter together with the spatial connectivity matrix $K_{ij}$,
\begin{equation}
    \begin{aligned}
    &Q_i = \frac{1}{\sqrt{n}} \sum_{\vec k} e^{-i \vec k \cdot \vec r_i} Q_{\vec k}, \quad Q_{\vec k} = \frac{1}{\sqrt{n}} \sum_i e^{-i\vec k \cdot \vec r_i} Q_i,\\
    &K_{ij} = \frac{1}{n} \sum_n e^{i \vec k \cdot (\vec r_i - \vec r_j)} K(\vec k), \quad K(\vec k) =  \sum_{i} e^{i \vec k \cdot \vec r_i} K_{0,i}.
    \end{aligned}
\end{equation}
Consequently 
\begin{equation}
    K(\vec k) = \sum_{e} 2\cos k_e \sim 2d - k^2 + \mathcal{O}(k^4),
\end{equation}
where $k_e$ denotes the elements of $\vec k$ and $k = |\vec k|$.
Similarly, denote $(KK)_{ii'} = \sum_{j} K_{ij} K_{i'j}$, its Fourier transformation yields
\begin{equation}
\begin{aligned}
    &(KK)(\vec k) = \sum_{i} e^{i \vec k \cdot \vec r_i} (KK)_{0,i} \\&= K(\vec k)^2 \sim 4d^2 - 4d k^2 + \mathcal{O}(k^4).
\end{aligned}
\end{equation}
Reverse Fourier transformation then yields Eq. \eqref{eq:spatial-continuum}.

In the stationary limit $t_f \rightarrow +\infty$, we apply the long-wave approximation and expand $D_Z(\Delta t)$ according to frequency.
Recall that $D_Z(\Delta t)$ is defined on $\Delta t \in [0,t_f]$. Thanks to the reflection symmetry with respect to $\Delta t = t_f /2$ Eq. \eqref{eq:reflection}, we can extend $D_Z(\Delta t)$ to a periodic function defined on $\Delta t \in \mathbb{R}$, 
\begin{equation}
    D_Z( t + m t_f) = D_Z(t), \quad m\in \mathbb{Z}, \quad t \in [0,t_f],
\end{equation}
It follows that 
\begin{equation}
    D_Z(t - t') = D_Z(|t-t'|\mod t_f), \quad t,t'\in \mathbb{R},
\end{equation}
and $D_Z$ is periodic with respect to both $t$ and $t'$. In the $t_f \rightarrow +\infty$ limit, we have
\begin{equation}
    \lim_{t_f \rightarrow + \infty} \tilde{D}_Z(t) = e^{-8h |t| }.
    \label{eq:stationary-coupling-app}
\end{equation}
Notice that the $t<0$ contribution has to be included here to reproduce the $\Delta t \sim t_f$ contribution to the function $D_Z(\Delta t)$, which eventually leads to the factor $2$ in the Eq. \eqref{eq:temporal-expansion}. Concerning the periodic boundary condition for the order parameter field $Q(t+t_f)=Q(t)$, we apply the Fourier transformation as
\begin{equation}
    \begin{aligned}
    &Q(t) = \frac{1}{\sqrt{t_f}} \sum_n e^{i\omega_n t} Q_n, \quad Q_n = \frac{1}{\sqrt{t_f}} \int_0^{t_f} dt e^{-i\omega_n t} Q(t),\\
    &{D}_Z(t) = \frac{1}{t_f} \sum_n e^{i\omega_n t} {D}_Z(\omega_n),\\ & {D}_Z(\omega_n) =  \int_0^{t_f} dt e^{-i\omega_n t} {D}_Z(t),
    \end{aligned}
\end{equation}
with $\omega_n = 2\pi n /t_f$, $n \in \mathbb{Z}$ the Matsubara frequency. It satisfies $D_Z(\omega_n) = D_Z(-\omega_n)$. As a consequence, we re-express the quadratic action
\begin{equation}
    \begin{aligned}
        &\int_0^{t_f} dt \int_0^{t_f} dt' Q(t) D_Z(|t-t'|) Q(t') \\&= \int_0^{t_f} dt\int_0^{t_f}   dt' Q(t) {D}_Z(t-t') Q(t')\\
        &= \sum_n  D_Z(\omega_n)Q_{-n} Q_n.
    \end{aligned}
\end{equation}
Notice that we have suppressed the replica and spatial indices here for convenience. The frequency representation of the temporal coupling function is expressed as 
\begin{equation}
    D_Z(\omega_n)= \frac{16 h \tanh 2 ht_f}{64 h^2 + \omega_n^2}.
    \label{eq:frequency}
\end{equation}

Consider the stationary limit $t_f \rightarrow +\infty$, where we replace the Matsubara frequency as real frequency $\omega_n \rightarrow \omega$, and the Fourier transformation becomes continuous $ (1/t_f) \sum_n \rightarrow (1/2\pi) \int d\omega$, $\int_0^{t_f}dt \rightarrow \int_{-\infty}^{+\infty} dt$. 
The frequency representation of Eq. \eqref{eq:stationary-coupling-app} is
\begin{equation}
    \tilde{D}_Z(\omega) = \frac{16 h}{64 h^2+ \omega^2} \sim 2(\tau + \tau^3 \omega^2),
\end{equation}
and $\tau = 1/8h$, which is exactly Eq. \eqref{eq:temporal-expansion}. Also, taking the $t_f \rightarrow + \infty$ limit in Eq. \eqref{eq:frequency} yields the same expression.

\section{Replica limit of the complete readout case}
\begin{widetext}
\label{app:complete}
Here we present the overall procedure of taking the replica limit of the complete readout case as we did in the partial readout case. At the same time, detailed derivations can be found in Sec SII of SM \cite{SupMat}. To begin with, we should keep $Q^{(+a)(-a)}_i(t) = Q^s$ a finite constant and variate with respect to $Q^{(\sigma_1 a)(\sigma_2 b)}_i(t) = Q^{r, \sigma_1 \sigma_2}_i(t)$ up to the second order, which yields the form for the effective action of SRE-to-LRE transition
\begin{align}
    &\mathcal{S}_r^{(R)}[Q^{(\sigma_1 a)(\sigma_2 b)}_i(t)] = \sum_{a<b,ii'} \Big[ \sum_{\sigma_1\sigma_2}J\int_0^{t_f} dt Q^{(\sigma_1 a)(\sigma_2 b)}_i(t) K_{ii'} Q^{(\sigma_1 a)(\sigma_2 b)}_i(t) \notag\\
    &\quad\quad\quad\quad-2J^2 \Big(\sum_{j} K_{ij} K_{i'j}\Big) \int_0^{t_f} dt\int_0^{t_f}dt'  \sum_{\sigma_1\sigma_2,\sigma_1'\sigma_2'}\tilde{D}^{(R),\sigma_1\sigma_2,\sigma_1'\sigma_2'}(t,t') Q^{(\sigma_1 a)(\sigma_2 b)}_i(t) Q^{(\sigma_1' a)(\sigma_2' b)}_{i'}(t')\Big]+ \mathcal{O}(Q^{3})
    \label{eq:action-R-app}
\end{align}
To determine the coefficient $\tilde{D}^{(R)}$, we set $Q^s$ finite and $Q^{(\sigma_1 a)(\sigma_2 b)}_i(t) = 0$ in Eq. \eqref{eq:eff}, the effective model on each replica copy becomes 
\begin{equation}
    L_0 (t) = 4dJ Q^s Z^+ Z^- + \sqrt{2h}\xi(t) (X^+ +X^-),
\end{equation}
 then the second-order variation of $Q^{(\sigma_1 a)(\sigma_2 b)}_i(t)$ is given by the correlation function of $L_0$ (here $a<b$ and $a'<b'$ are implied)
\begin{equation}
\begin{aligned}
    &\braket{\mathcal{T}(Z^{(\sigma_1 a)}Z^{(\sigma_2 b)})(t)(Z^{(\sigma_1' a')}Z^{(\sigma_2' b')})(t')}_R\\& = \delta_{aa'} \delta_{bb'} \tilde{D}^{(R),\sigma_1\sigma_2,\sigma_1'\sigma_2'}(t,t').
\end{aligned}  
\end{equation}
Notice that it enforces $a=a'$, $b=b'$ in Eq. \eqref{eq:action-R-app} while leaving $\sigma_1$, $\sigma_1'$, $\sigma_2$, $\sigma_2'$ arbitrary. Evaluating it immediately leads to Eq. \eqref{eq:Dtilde}. 
This leads to 
\begin{equation}  \tilde{D}^{(R),\sigma_1\sigma_2,\sigma_1'\sigma_2'}(t,t') =
    \frac{\int D\xi P_G[\xi] K^{\sigma_1 \sigma_1'}[\xi;t_>,t_<] K^{\sigma_2 \sigma_2'}[\xi;t_>,t_<] K_0[\xi]^{R-2}}{\int D\xi P_G[\xi] K_0[\xi]^{R}}.
\label{eq:Dtilde-R-app}
\end{equation}
Notice that unlike the partial readout case, the action $\mathcal{S}_r^{(R)}$ for the complete readout SRE-to-LRE transition explicitly depend on the replica number $R$, so in the replica limit we have
\begin{align}
    &\mathcal{S}_r^{(R)}[Q^{(\sigma_1 a)(\sigma_2 b)}_i(t)]  \xrightarrow{R\rightarrow 1} \sum_{ii'} \Big[ \sum_{\sigma_1\sigma_2}J\int_0^{t_f} dt Q^{r,\sigma_1 \sigma_2}_i(t) K_{ii'} Q^{r,\sigma_1 \sigma_2}_i(t)\notag\\
    &\quad\quad\quad\quad -2J^2 \Big(\sum_{j} K_{ij} K_{i'j}\Big) \int_0^{t_f} dt\int_0^{t_f}dt'  \sum_{\sigma_1\sigma_2,\sigma_1'\sigma_2'}\tilde{D}^{(1),\sigma_1\sigma_2,\sigma_1'\sigma_2'}(t.t') Q^{r,\sigma_1 \sigma_2}_i(t) Q^{r,\sigma_1' \sigma_2'}_{i'}(t')\Big]+ \mathcal{O}(Q^{r3}),
    \label{eq:action-RL-app}
\end{align}
and
\begin{equation}  \tilde{D}^{(1),\sigma_1\sigma_2,\sigma_1'\sigma_2'}(t,t') =
    \frac{\int D\xi P_G[\xi] K^{\sigma_1 \sigma_1'}[\xi;t_>,t_<] K^{\sigma_2 \sigma_2'}[\xi;t_>,t_<] K_0[\xi]^{-1}}{\int D\xi P_G[\xi] K_0[\xi]}.
\label{eq:Dtilde-app}
\end{equation}
We then diagonalize the Hessian matrix similarly through the Keldysh rotation Eq. \eqref{eq:Keldysh-rotation} to determine the second-order phase transition, 
\begin{equation}
    \begin{aligned}
        &\tilde{D}^{(1)} = \int D\xi \frac{ P_G[\xi] K_0[\xi]^{-1}}{\int D\xi P_G[\xi] K_0[\xi]} \begin{pmatrix}
            K^{++}K^{++} & K^{++} K^{+-} & K^{+-} K^{++} & K^{+-} K^{+-}\\
            K^{++} K^{-+}& K^{++} K^{--}& K^{+-} K^{-+}& K^{+-} K^{--}\\
            K^{-+} K^{++}& K^{-+}K^{+-}& K^{--} K^{++}& K^{--} K^{+-}\\
            K^{-+}K^{-+} & K^{-+} K^{--}& K^{--}K^{-+} & K^{--} K^{--}
        \end{pmatrix} \\&\rightarrow
        \int D\xi \frac{ P_G[\xi] K_0[\xi]^{-1}}{\int D\xi P_G[\xi] K_0[\xi]} 
        \begin{pmatrix}
            (K^{++} + K^{+-})^2 & 0 & 0 & 0\\
            0 & K^{++ 2} - K^{+- 2} & 0 & 0 \\
            0 & 0 & K^{++ 2} - K^{+- 2} & 0 \\
            0 & 0 & 0 & (K^{++} - K^{+-})^2
        \end{pmatrix}.
    \end{aligned}
\end{equation}
where we omitted the arguments of $K^{\sigma, \sigma'}[\xi;t_>,t_<]$ for notational simplicity. By noticing that the sign change $Q^s \rightarrow -Q^s$ leads to $K^{++} \rightarrow K^{++}$ and $K^{+-} \rightarrow -K^{+-}$, the largest eigenvalue is the one given by Eq. \eqref{eq:DK}.

\end{widetext}

\section{$R$-replica approximation of the measurement transition}
\label{app:R-replica}
Here we sort out the details of Eq. \eqref{eq:space-discrete}, in which we can obtain an analytical solution to approximate the measurement transition in $R$-replica level.
First, we analyze the properties of the temporal coupling function
\begin{widetext}
\begin{equation}
    \begin{aligned}
        &D^{(R)}(\Delta t) =    \sqrt{\frac{(t_f-\Delta t)\Delta t}{2\pi t_f}}  \frac{1}{\displaystyle\int d\xi \exp\left(-\frac{t_f}{2} \xi^2\right)\cosh^{2R}  \left(\sqrt{2h}t_f\xi  \right)} \times\notag\\
    & \int d\xi_1 d\xi_2\left\{ \exp\left(-\frac{t_f-\Delta t}{2}\xi_1^2-\frac{\Delta t}{2}\xi_2^2\right) \cosh^2  \left[\sqrt{2h} ((t_f-\Delta t)\xi_1 -\Delta t \xi_2)\right]  \cosh^{2R-2} \left[\sqrt{2h} ((t_f-\Delta t)\xi_1 +\Delta t \xi_2)\right] \right\},
    \end{aligned}
\end{equation}
which depend only on the relative time $\Delta t = |t - t'| $.
In this section, the integer $R>1$ is assumed. We separately consider the numerator and denominator. We denote the time-dependent numerator as $D^{(R)}_{-}(\Delta t)$ and the time-independent denominator $D^{(R)}_{+}$.
The denominator is expressed as
\begin{equation}
\begin{aligned}
     D^{(R)}_{+} = \frac{1}{2^{2R} }\sum_{a=0}^{2R} \binom{2R}{a} \exp\left(ht_f(2a-2R)^2\right),
\end{aligned}
\end{equation}
where we expanded the $\cosh$ function and used the binomial theorem before integrate out $\xi$. 
The numerator is evaluated as 

\begin{equation}
    \begin{aligned}
    & D^{(R)}_{-}(\Delta t)= \frac{1}{2^{2R} }\left\{2\binom{2R-2}{R-1}\left(1+e^{4ht_f}\right) + 4 \sum_{a=0}^{R-2} \binom{2R-2}{a}\exp\left[4ht_f(a+1-R)^2\right]\right.\\  &\left. + 4 \sum_{a=0}^{R-2} \binom{2R-2}{a}\exp\left[4ht_f(a+1-R)^2+4ht_f\right] \cosh[8(ht_f-2h\Delta t)(a+1-R)]\right\}.
\end{aligned}
\end{equation}
\end{widetext}
In the fourth line, we explicitly performed the $b$ summation and separated the $\Delta t$-independent (the fourth line) and $\Delta t$-dependent (the fifth line) terms. For sufficiently large $t_f$ and near the critical point $h_c$, the $\Delta t$-independent contribution of the above equation is sufficiently small, while the $\Delta t$-dependent part can be approximated by a second-order derivative. To this end, we first consider fixing $h$ and $\Delta t$ and take the stationary limit $t_f \rightarrow +\infty$. The $\Delta t$-dependent part approaches $(2/2^{2R}) \exp(4ht_f R^2 - 16 h \Delta t (R-1))$, while the $\Delta t$-independent is exponentially small in comparison. Dividing the $D^{(R)}_{+} \sim (2/2^{2R}) \exp(4ht_f R^2)$ leads to Eq. \eqref{eq:coupling-stationary}. 

Then we apply Fourier transformation to $D^{(R)}(\Delta t)$,
\begin{widetext}
\begin{equation}
    \begin{aligned}
    & \tilde{D}^{(R)}(\omega_n)= \frac{1}{2^{2R} D^{(R)}_+}\left\{ \left[2\binom{2R-2}{R-1}\left(1+e^{4ht_f}\right) + 4 \sum_{a=0}^{R-2} \binom{2R-2}{a}\exp\left[4ht_f(R-a-1)^2\right] \right] t_f\delta_{\omega_n, 0}\right.\\  &\left. + 4 \sum_{a=0}^{R-2} \binom{2R-2}{a}\exp\left[4ht_f(R-a-1)^2+4ht_f\right] \frac{32 h (R-a-1)\sinh[8h t_f(R-a-1)]}{256 h^2 (R-a-1)^2 + \omega_n^2}\right\}.
\end{aligned}
\end{equation}
\end{widetext}
The first term is a temporal rigid contribution. However, as long as $h>0$, it exponentially decays with the final time $t_f$. The second term is 'smooth' (in the stationary limit) in the frequency domain, in which we can perform frequency expansion to get the most relevant contributions.
In the stationary limit $t_f \rightarrow +\infty$,
the frequency representation becomes
\begin{equation}
    \tilde{D}^{(R)}(\omega) = \frac{32h(R-1)}{256h^2(R-1)^2+ \omega^2} \sim 2(\tau + \tau^3 \omega^2),
\end{equation}
with $\tau =1/ 16 h (R-1)$.

When $h>0$, the function $D^{(R)}(|t-t'|)$ is peaked at $t=t'$.
The spatial continuum theory is written as 
\begin{align}
    &S[Q^{\alpha\beta}(\vec x, t)] = \notag\\
    &\sum_{\alpha<\beta} \int d^d \vec x\int_0^{t_f} dt Q^{\alpha\beta}(\vec x,t) (2dJ+ J\partial_{\vec x}^2) Q^{\alpha\beta}(\vec x,t) \notag\\
    &-\sum_{\alpha<\beta} \int d^d \vec x \int_0^{t_f} dt \int_0^{t_f}dt'   D^{(R)}(|t-t'|)  \notag\\
    &\times Q^{\alpha\beta}(\vec x,t) (8d^2J^2+ 8dJ^2\partial_{\vec x}^2) Q^{\alpha\beta}(\vec x,t') + \mathcal{O}(\partial^4,Q^3).\label{eq:continuum}
\end{align}
In the extreme case $h\rightarrow +\infty$, keeping $R\geq 2$, $D^{(R)}(\Delta t)$ is proportional to delta function $\delta(\Delta t) + \delta(\Delta t - t_f)$, indicating that there is no temporal coupling between the order parameter field at different times. This reproduces the intuitive observation that the symmetry-breaking order or the LRE is unable to persist in the infinitely large $X$ measurement case.

We now derive the effective model for the stationary state.
Taking the $t_f \rightarrow + \infty$ limit while keeping $\Delta t$ finite, we have
\begin{equation}
    \lim_{t_f \rightarrow + \infty}D^{(R)}(\Delta t) = e^{-16h(R-1)\Delta t},
    \label{eq:coupling-stationary-R}
\end{equation}
which possesses only a finite range of  coupling $ \tau= 1/(16h(R-1))$. Therefore we could perform the long-wave approximation the same as Eq. \eqref{eq:temporal-expansion},
and keep up to the second-order spatial and temporal derivatives in the effective action, leading to
\begin{align}
    &S[Q(\vec x, t)] = R(2R-1)\notag\\
    &\times \int d^d \vec x\int dt \left\{(2dJ - 16 d^2 J^2 \tau) \left(Q(\vec x,t)\right)^2 \right. \notag\\
    &\left.+(16dJ^2\tau-J) \left(\partial_{\vec x}Q(\vec x,t)\right)^2 + 16d^2 J^2 \tau^3 \left(\partial_{t}Q(\vec x,t)\right)^2\right\}\notag\\
    &+\mathcal{O}(\partial^4,Q^3).
    \label{eq:stationary-R}
\end{align}
Here a replica symmetric solution $Q^{\alpha\beta}(\vec x,t) = Q(\vec x, t)$ is expected, and the vanishing of the mass term gives the critical non-commutative measurement strength for the $t_f\rightarrow +\infty$ stationary state, that is
\begin{equation}
    \tau_c = \frac{1}{8dJ},\text{ or }\quad h_c = \frac{dJ}{2(R-1)}.
    \label{eq:critical-h-R}
\end{equation}
The best estimation now for the SRE-to-LRE phase transition is the $2$-replica result $h_c^{(2)} = \frac{dJ}{2}$, since smaller $R$ is believed to yield better results \cite{fanDiagnosticsMixedstateTopological2023}. 
Approaching the critical point from the disordered phase, the spatial and temporal correlation length is given by
\begin{equation}
    \begin{aligned}
        &\xi_{\vec x} = \sqrt\frac{16dJ\tau-1}{2d-16d^2 J\tau} \sim \frac{1}{4\sqrt{d^2 J}} (\tau_c - \tau)^{-\nu},\\
        &\xi_{t} = \sqrt\frac{8dJ\tau^3}{1-8d J\tau}\sim \frac{1}{(8dJ)^{3/2}} (\tau_c - \tau)^{-\nu},
    \end{aligned}
    \label{eq:correlation-stationary-R}
\end{equation}
where still $\nu = 1/2$. The upper critical dimension for the mean-field description of the stationary state is $d_c = 5$. 

For the finite $t_f$ and $h$ case, we assume a temporally static, spatially uniform, and replica symmetric order parameter $Q^{\alpha \beta} (\vec x,t) = Q$ near the phase transition, which yields the Landau free energy
\begin{equation}
\begin{aligned}
        &f = \\&R(2R-1) \left(2dJt_f - 8 d^2 J^2 t_f \int_0^{t_f} dtD^{(R)}(t)\right) Q^2\\& + \mathcal{O}(Q^3).
\end{aligned}
\end{equation}

\end{appendix}

\bibliographystyle{apsrev4-1-titles} 
\bibliography{ref.bib}

\onecolumngrid
\clearpage


\section*{Supplementary material}

# Supplemental Material for "Non-Commutative weak measurements: Entanglement, Symmetry Breaking, and the Role of Readout"

In this supplementary Material, we present the lengthy derivations as a complement of the main text. Specifically, we provide details on

- Sec. SI: Statistical mechanical mapping for the 1D system without $X$ measurement.

- Sec. SII: Construction for the mean-field effective theory for higher-dimensional systems.

- Sec. SIII: Saddle point solution of the measurement trajectory distribution for the complete readout case.

## SI. DERIVATION OF SM MAPPING

Here we present the SM mapping derivation for the purely $ZZ$ measured 1D system. For example, consider the physical measurement outcome probability

$$P(s) = P_G(s) \operatorname{tr}(\hat{\rho}(s, t_f)) = P_G(s) \langle +|^{\otimes L} K(s)^\dagger K(s) |+\rangle^{\otimes L} \tag{S1}$$

To compute the expectation value, we use $|+\rangle^{\otimes L} = \sum_{\{\sigma_i\}} |\{\sigma_i\}\rangle / \sqrt{2^L}$ where $|\{\sigma_i\}\rangle$ denotes the Pauli $Z$ basis. Thus we have

$$\begin{aligned}
P(s) &= P_G(s) \frac{e^{-4Jt_f L}}{2^L} \sum_{\{\sigma_i\}} \exp\left[8Jt_f \sum_i s_{i,i+1} \sigma_i \sigma_{i+1}\right] \\
&= \left(\sqrt{\frac{4Jt_f}{\pi}}\right)^L \exp\left[-4Jt_f \sum_i s_{i,i+1}^2 - 4Jt_f L\right] \left(\prod_i \cosh(8Jt_f s_{i,i+1}) + \prod_i \sinh(8Jt_f s_{i,i+1})\right).
\end{aligned} \tag{S2}$$

In the first line the expression acquires the form of the partition function of 1D Ising model with bond disorder. Similarly, the $Z_i Z_j$ correlation given a particular set of measurement outcomes $\{s_{i,i+1}\}$ (suppose that $i < j$) can be evaluated as the spin correlation function of the Ising model, which yields,

$$\langle Z_i Z_j \rangle_s = \frac{\operatorname{tr}(\hat{\rho}(s, t_f) Z_i Z_j)}{\operatorname{tr}(\hat{\rho}(s, t_f))} = \frac{\prod_{i \leq l < j} \tanh(8Jt_f s_{l,l+1}) + \prod_{l \geq j \text{ or } l < i} \tanh(8Jt_f s_{l,l+1})}{1 + \prod_i \tanh(8Jt_f s_{l,l+1})}, \tag{S3}$$

where $r = |i - j|$ is the distance between the two spins. It is easy to check that the averaged correlation function $\mathbb{E} \langle Z_i Z_j \rangle_s$ always yields 0,

$$\begin{aligned}
\mathbb{E} \langle Z_i Z_j \rangle_s &= \int ds P(s) \langle Z_i Z_j \rangle_s \\
&= \int \prod_i ds_{i,i+1} \left(\sqrt{\frac{4Jt_f}{\pi}}\right)^L \exp\left[-4Jt_f \sum_i s_{i,i+1}^2\right] e^{-4Jt_f L} \left(\prod_i \cosh(8Jt_f s_{i,i+1})\right) \\
&\quad \times \left(\prod_{i \leq l < j} \tanh(8Jt_f s_{l,l+1}) + \prod_{l \geq j \text{ or } l < i} \tanh(8Jt_f s_{l,l+1})\right) = 0.
\end{aligned} \tag{S4}$$

We can similarly evaluated the nonlinear Edwards-Anderson correlation function

$$\mathbb{E}\left\{\langle Z_i Z_j\rangle_s^2\right\} = \int ds P(s) \langle Z_i Z_j\rangle_s^2$$

$$= \int \prod_i ds_{i,i+1} \left(\sqrt{\frac{4Jt_f}{\pi}}\right)^L \exp\left[-4Jt_f \sum_i s_{i,i+1}^2\right]$$

$$\times e^{-4Jt_f L} \left(\prod_i \cosh(8Jt_f s_{i,i+1})\right) \frac{\left(\prod_{i\leq l<j} \tanh(8Jt_f s_{l,l+1}) + \prod_{l\geq j \text{ or } l<i} \tanh(8Jt_f s_{l,l+1})\right)^2}{1 + \prod_i \tanh(8Jt_f s_{l,l+1})} \quad (S5)$$

$$\xrightarrow{L\to\infty} \int \prod_i ds_{i,i+1} \left(\sqrt{\frac{4Jt_f}{\pi}}\right)^L \exp\left[-4Jt_f \sum_i s_{i,i+1}^2\right] e^{-4Jt_f L} \left(\prod_i \cosh(8Jt_f s_{i,i+1})\right) \prod_{i\leq l<j} \tanh(8Jt_f s_{l,l+1})^2$$

$$= \left[\int ds \sqrt{\frac{4Jt_f}{\pi}} e^{-4Jt_f s^2 - 4Jt_f} \cosh(8Jt_f s) \tanh^2(8Jt_f s)\right]^r,$$

where $r = |i-j|$. Reorganize the above result in a more explicit form,

$$\mathbb{E}\left\{\langle Z_i Z_j\rangle_s^2\right\} \xrightarrow{L\to\infty} e^{-r/\xi}$$

$$\xi = -1/\log\left[\int ds \sqrt{\frac{4Jt_f}{\pi}} \exp\left(-4Jt_f s^2 - 4Jt_f\right) \cosh(8Jt_f s) \tanh^2(8Jt_f s)\right], \quad (S6)$$

where we kept $r$ finite and took the thermodynamic limit $L \to \infty$. The Rényi-2 correlation function for the LRE order has a more explicit form

$$R^{(2)}(i,j) = \frac{\int ds P(s)^2 \langle Z_i Z_j\rangle_s^2}{\int ds P(s)^2} = \frac{(\tanh(4Jt_f))^{|i-j|} + (\tanh(4Jt_f))^{L-|i-j|}}{1 + \tanh^L(4Jt_f)} \to (\tanh 4Jt_f)^{|i-j|}. \quad (S7)$$

Turning to the Lindblad evolution

$$\partial_t \rho_l = 2J \sum_i (Z_i Z_{i+1} \rho_l Z_i Z_{i+1} - \rho_l), \quad (S8)$$

the differential equiation is solved by

$$\rho_l = \int ds P_G(s) K(s) \rho(0) K(s)^\dagger = \prod_i \left[\frac{1+e^{-4Jt_f}}{2}\rho(0) + \frac{1-e^{-4Jt_f}}{2} Z_i Z_{i+1} \rho(0) Z_i Z_{i+1}\right], \quad (S9)$$

where the initial state $\rho(0)$ is send through a stochastic $ZZ$ noise channel. We evaluate the Rényi-2 correlator for the SWSSB order

$$R_l^{(2)}(i,j) = \frac{\text{tr}(\rho_l Z_i Z_j \rho_l Z_i Z_j)}{\text{tr}\,\rho_l^2} = \frac{\langle\langle\rho_l|\, Z_i^+ Z_j^+ Z_i^- Z_j^-\, |\rho_l\rangle\rangle}{\langle\langle\rho_l|\rho_l\rangle\rangle}, \quad (S10)$$

by expanding $|\rho(0)\rangle\rangle = \sum_{\{\sigma_i^+\},\{\sigma_i^-\}} |\{\sigma_i^+\}\rangle |\{\sigma_i^-\}\rangle/2^L$. It is straightforward to verify that we can map $|\rho_l\rangle\rangle$ to an exactly solvable 1D classical Ashkin-Teller (AT) model [1], such that

$$\langle\langle\rho_l|\, Z_i^+ Z_j^+ Z_i^- Z_j^-\, |\rho_l\rangle\rangle \propto \sum_{\{\sigma_i^\pm\}} \sigma_i^+ \sigma_j^+ \sigma_i^- \sigma_j^- \exp\left[4Jt_f \sum_i \sigma_i^+ \sigma_{i+1}^+ \sigma_i^- \sigma_{i+1}^-\right]. \quad (S11)$$

The two layers of spins in the AT model come from the forward and backward evolution branches in the Lindblad

equation, which are coupled together. As a result,

$$R_l^{(2)}(i,j) = \frac{(\tanh(4Jt_f))^{|i-j|} + (\tanh(4Jt_f))^{L-|i-j|}}{1 + \tanh^L(4Jt_f)} \xrightarrow{L\to\infty} (\tanh(4Jt_f))^{|i-j|}. \tag{S12}$$

Notice that in this purely $ZZ$ measurement case the order parameter for LRE and SWSSB equals, $R_l^{(2)}(i,j) = R^{(2)}(i,j)$, so the two phases appears simultaneously, depending on whether the outcomes are read out.

## SII. DERIVATION OF MEAN-FIELD MODEL

Here we derive the mean-field effective theory for the measurement phase transition. We first consider the $R > 1$ case for the complete readout case. In the vicinity of the phase transition point, the order parameter $Q_i^{\alpha\beta}(t)$ is sufficiently small so that we expand $\mathcal{T}\exp\left(\int_0^{t_f} dt L_{\text{eff}}^{(R)}(t)\right)$ with respect to $Q$, where

$$L_{\text{eff}}^{(R)}(t) = 2J \sum_{\alpha<\beta,ij} Q_i^{\alpha\beta}(t) K_{ij} Z_j^\alpha Z_j^\beta + \sqrt{2h} \sum_{\alpha,j} \xi_j(t) X_j^\alpha, \tag{S13}$$

we have

$$\mathcal{T}\exp\left(\int_0^{t_f} dt L_{\text{eff}}^{(R)}(t)\right) = \hat{C}_0 + \int_0^{t_f} dt \sum_{i,\alpha<\beta} \hat{C}_1^{\alpha\beta}(i,t) Q_i^{\alpha\beta}(t)$$
$$+ \frac{1}{2} \int_0^{t_f} dt \int_0^{t_f} dt' \sum_{ii',\alpha<\beta,\alpha'<\beta'} \hat{C}_2^{\alpha\beta\alpha'\beta'}(i,t,i',t') Q_i^{\alpha\beta}(t) Q_{i'}^{\alpha'\beta'}(t') + \mathcal{O}(Q^3), \tag{S14}$$

where

$$\hat{C}_0 = e^{t_f L_{\text{eff}}}\big|_{Q=0} = \exp\left(\int_0^{t_f} dt \sqrt{2h} \sum_{\alpha,j} \xi_j(t) X_j^\alpha\right), \tag{S15}$$

$$\hat{C}_1^{\alpha\beta}(i,t) = \frac{\delta}{\delta Q_i^{\alpha\beta}(t)} e^{t_f L_{\text{eff}}}\bigg|_{Q=0} = \exp\left(\int_t^{t_f} dt \sqrt{2h} \sum_{\alpha,j} \xi_j(t) X_j^\alpha\right) \left(2J \sum_j K_{ij} Z_j^\alpha Z_j^\beta\right) \exp\left(\int_0^t dt \sqrt{2h} \sum_{\alpha,j} \xi_j(t) X_j^\alpha\right)$$
$$= 2J \sum_j K_{ij} Z_j^\alpha Z_j^\beta \prod_{\gamma\in\{\alpha,\beta\}} \exp\left(\sqrt{2h}\left(-\int_t^{t_f} dt \xi_j(t) + \int_0^t dt \xi_j(t)\right) X_l^\gamma\right) \prod_{\gamma\notin\{\alpha,\beta\}, l\neq j} \exp\left(\sqrt{2h} \int_0^{t_f} dt \xi_l(t) X_l^\gamma\right), \tag{S16}$$

$$\hat{C}_2^{\alpha\beta\alpha'\beta'}(i,t,i',t') = \frac{\delta}{\delta Q_j^{\alpha\beta}(t) \delta Q_{i'}^{\alpha'\beta'}(t')} e^{t_f L_{\text{eff}}}\bigg|_{Q=0} = \exp\left(\int_{t_>}^{t_f} dt \sqrt{2h} \sum_{\alpha,j} \xi_j(t) X_j^\alpha\right) \left(2J \sum_{j_>} K_{i_>,j_>} Z_{j_>}^{\alpha_>} Z_{j_>}^{\beta_>}\right)$$
$$\times \exp\left(\int_{t_<}^{t_>} dt \sqrt{2h} \sum_{\alpha,j} \xi_j(t) X_j^\alpha\right) \left(2J \sum_{j_<} K_{i_<,j_<} Z_{j_<}^{\alpha_<} Z_{j_<}^{\beta_<}\right) \exp\left(\int_0^{t_<} dt \sqrt{2h} \sum_{\alpha,j} \xi_j(t) X_j^\alpha\right). \tag{S17}$$

Notice that due to the presence of time ordering $\mathcal{T}$, the variation $\delta/\delta Q_i^{\alpha\beta}(t)$ brings down the Pauli $Z$ operators at time $t$ in Eq. (S16). In the last equation Eq. (S17), we have defined $t_> = \max\{t,t'\}$ and $t_< = \min\{t,t'\}$, and the labels $i_{>,<}, \alpha_{>,<}, \beta_{>,<}$ are associated with the corresponding time variable, for example $i_> = i$ and $i_< = i'$ if $t > t'$.

Tracing out the spin degrees of freedom, we obtain

$$\operatorname{tr}\hat{C}_0 = 2^{2Rn} \prod_j \cosh^{2R}\left(\sqrt{2h}\int_0^{t_f} dt\xi_j(t)\right) \tag{S18}$$

$$\operatorname{tr}\hat{C}_1^{\alpha\beta}(i,t) = 0 \tag{S19}$$

$$\operatorname{tr}\hat{C}_2^{\alpha\beta\alpha'\beta'}(i,t,i',t') = 2^{2Rn+2}J^2\delta_{\alpha\alpha'}\delta_{\beta\beta'}\sum_j \left[K_{i,j}K_{i',j}\cosh^2\left(\sqrt{2h}\left(\int_{t_>}^{t_f} dt\xi_j(t) - \int_{t_<}^{t_>} dt\xi_j(t) + \int_0^{t_<} dt\xi_j(t)\right)\right)\right.$$

$$\left.\times \cosh^{2R-2}\left(\sqrt{2h}\int_0^{t_f} dt\xi_j(t)\right) \prod_{l\neq j} \cosh^{2R}\left(\sqrt{2h}\int_0^{t_f} dt\xi_l(t)\right)\right]. \tag{S20}$$

Collect all the above results and substitute them in the definition of partition function,

$$\mathcal{Z}^{(R)} = \operatorname{tr} e^{t_f L^{(R)}} \propto \int DQ_i^{\alpha\beta} D\xi_i \operatorname{tr}\left[\mathcal{T}\exp\left(\int_0^{t_f} dt L_{\text{eff}}^{(R)}(t)\right)\right] \exp\int_0^{t_f} dt \left(-J\sum_{\alpha<\beta,ij} Q_i^{\alpha\beta}(t)K_{ij}Q_j^{\alpha\beta}(t) - \frac{1}{2}\sum_i \xi_i(t)^2\right), \tag{S21}$$

we obtain

$$\mathcal{Z}^{(R)} \propto \int DQ_i^{\alpha\beta} \exp\left(-J\sum_{\alpha<\beta,ij}\int_0^{t_f} dt Q_i^{\alpha\beta}(t)K_{ij}Q_j^{\alpha\beta}(t)\right.$$

$$\left.+2J^2\sum_{\alpha<\beta,ii'}\int_0^{t_f} dtdt' \left(\sum_j K_{i,j}K_{i',j}\right) D^{(R)}(t,t')Q_i^{\alpha\beta}(t)Q_{i'}^{\alpha\beta}(t') + \mathcal{O}(Q^3)\right) \tag{S22}$$

$$D^{(R)}(t,t') = \frac{\int D\xi_i e^{-\frac{1}{2}\sum_i \int dt\xi_i^2}\operatorname{tr}\hat{C}_2^{\alpha\beta\alpha\beta}(i,t,i',t')}{\int D\xi_i e^{-\frac{1}{2}\sum_i \int dt\xi_i^2}\operatorname{tr}\hat{C}_0}$$

$$= \frac{\int D\xi e^{-\frac{1}{2}\int dt\xi^2}\cosh^2\left(\sqrt{2h}\left(\int_{t_>}^{t_f} dt\xi(t) - \int_{t_<}^{t_>} dt\xi(t) + \int_0^{t_<} dt\xi(t)\right)\right)\cosh^{2R-2}\left(\sqrt{2h}\int_0^{t_f} dt\xi(t)\right)}{\int D\xi e^{-\frac{1}{2}\int dt\xi^2}\cosh^{2R}\left(\sqrt{2h}\int_0^{t_f} dt\xi(t)\right)}.$$

Here we performed the $\xi_i$ integration to each $\hat{C}$ and raised the quadratic term to the exponent. In the above expression, the function integral of $\xi(t)$ can be simplified by the central limit theorem ($\mathcal{N}$),

$$\frac{\int D\xi(t)\exp\left(-\frac{1}{2}\int_{t_1}^{t_2} dt\xi(t)^2\right)f\left(\frac{1}{t_2-t_1}\int_{t_1}^{t_2} dt\xi(t)\right)}{\int D\xi(t)\exp\left(-\frac{1}{2}\int_{t_1}^{t_2} dt\xi(t)^2\right)} = \sqrt{\frac{t_2-t_1}{2\pi}}\int_{-\infty}^{+\infty} d\xi \exp\left(-\frac{1}{2}(t_2-t_1)\xi^2\right)f(\xi). \tag{S23}$$

which directly leads to

$$D^{(R)}(\Delta t) = \sqrt{\frac{(t_f - \Delta t)\Delta t}{2\pi t_f}}\frac{1}{\int d\xi \exp\left(-\frac{t_f}{2}\xi^2\right)\cosh^{2R}\left(\sqrt{2h}t_f\xi\right)}\times$$

$$\int d\xi_1 d\xi_2 \left\{\exp\left(-\frac{t_f - \Delta t}{2}\xi_1^2 - \frac{\Delta t}{2}\xi_2^2\right)\cosh^2\left[\sqrt{2h}((t_f - \Delta t)\xi_1 - \Delta t\xi_2)\right]\cosh^{2R-2}\left[\sqrt{2h}((t_f - \Delta t)\xi_1 + \Delta t\xi_2)\right]\right\}.$$

It is clear in this form that $D^{(R)}$ only depend on $\Delta t = |t - t'| = t_> - t_<$, so we rewrite $D^{(R)}(t,t') = D^{(R)}(|t - t'|)$. The time correlation of $Q$ originates from the presence of $X$ measurement, and is absorbed into $D^{(R)}(|t - t'|)$. The temporal periodic boundary condition ensures the invariance under time translation.

Now recall the evaluation from Eq. (S15) and Eq. (S17) to Eq. (S22), one sees that the temporal coupling function $D^{(R)}(|t - t'|)$ is nothing but the normalized time-ordered correlation of a $0+1$d effective model

$$L_0(t) = \sqrt{2h}\xi(t)(X^+ + X^-), \tag{S24}$$

which is made in $R$ copies. Formally the correlation is expressed as

$$\langle \mathcal{T}(Z^\alpha Z^\beta)(t)(Z^{\alpha'} Z^{\beta'})(t') \rangle =$$
$$\frac{\int D\xi \operatorname{tr}\left[\mathcal{T}\exp\left(\int_{t_>}^{t_f} dt L_0(t)\right)^{\otimes R} Z^{\alpha_>} Z^{\beta_>} \mathcal{T}\exp\left(\int_{t_<}^{t_>} dt L_0(t)\right)^{\otimes R} Z^{\alpha_<} Z^{\beta_<} \mathcal{T}\exp\left(\int_0^{t_<} dt L_0(t)\right)^{\otimes R}\right]}{\int D\xi \operatorname{tr}\left[\mathcal{T}\exp\left(\int_0^{t_f} dt L_0(t)\right)^{\otimes R}\right]} \quad (S25)$$
$$= \delta_{\alpha\alpha'} \delta_{\beta\beta'} D^{(R)}(|t - t'|)$$

As for the partial readout case, i.e. averaging over $X$ measurement results and collecting only $ZZ$ measurement results, described by

$$\mathcal{Z}_Z^{(R)} \propto \int DQ_i^{\alpha\beta} \operatorname{tr} \exp\left(t_f L_{eff,Z}^{(R)}\right) \exp \int_0^{t_f} dt \left(-J \sum_{\alpha<\beta,ij} Q_i^{\alpha\beta}(t) K_{ij} Q_j^{\alpha\beta}(t)\right), \quad (S26)$$

where

$$e^{t_f L_{eff,Z}^{(R)}} = \mathcal{T} \exp \int_0^{t_f} dt \left(2J \sum_{\alpha<\beta,ij} Q_i^{\alpha\beta}(t) K_{ij} Z_j^\alpha Z_j^\beta + 2h \sum_i \sum_{a=1}^R X_i^{(+,a)} X_i^{(-,a)}\right), \quad (S27)$$

we similarly evaluate $\operatorname{tr}\exp(t_f \mathcal{L}_{eff,Z})$ by expanding the up to $Q^2$ as in Eq. (S14) while keeping track of the replica indices of the trajectories $\xi_j X_j^{(\sigma,a)} \to \xi_j^a X_j^{(\sigma,a)}$. Integrate $\xi_i^a(t)$ for each replica copy leads to the effective model

$$L_0(t) = 2h X^+ X^-. \quad (S28)$$

Unlike Eq. (S24), the $\pm$ branches are now distinguished from the replica index $a$, as the interaction $X^+X^-$ occurs only inside each copy. In contrast Integrate $\xi_i(t)$ in Eq. (S24) introduces interactions between replica copies. The correlation function in turn splits into two types,

$$\langle \mathcal{T}(Z^{(\sigma_1 a)} Z^{(\sigma_2 b)})(t)(Z^{(\sigma_1' a')} Z^{(\sigma_2' b')})(t') \rangle_R =$$
$$\delta_{a=a'=b=b'} \delta_{\sigma_1=\sigma_2=+} \delta_{\sigma_1'=\sigma_2'=-} D_Z^s(|t-t'|) + \delta_{aa'} \delta_{bb'} \delta_{a \neq b} \delta_{\sigma_1 \sigma_1'} \delta_{\sigma_2 \sigma_2'} D_Z^r(|t-t'|), \quad (S29)$$

which then distinguishes the two types of order parameter in the effective action

$$S_Z[Q_i^{\alpha\beta}(t)] = J \sum_{\alpha<\beta,ii'} \int_0^{t_f} dt Q_i^{\alpha\beta}(t) K_{ii'} Q_{i'}^{\alpha\beta}(t)$$
$$- 2J^2 \sum_{a,ii'} \int_0^{t_f} dt \int_0^{t_f} dt' Q_i^{(+a)(-a)}(t) Q_{i'}^{(+a)(-a)}(t') \left(\sum_j K_{ij} K_{i'j}\right) D_Z^s(|t-t'|) \quad (S30)$$
$$- 2J^2 \sum_{a<b,\sigma_1\sigma_2,ii'} \int_0^{t_f} dt \int_0^{t_f} dt' Q_i^{(\sigma_1 a)(\sigma_2 b)}(t) Q_{i'}^{(\sigma_1 a)(\sigma_2 b)}(t') \left(\sum_j K_{ij} K_{i'j}\right) D_Z^r(|t-t'|) + \mathcal{O}(Q^3).$$

We further evaluate the correlation function, which leads to

$$D_Z^s(|t - t'|) = 1, \quad (S31)$$
$$D_Z^r(|t - t'|) = \cosh^2(2h(t_f - 2\Delta t)) \operatorname{sech}^2(2h t_f). \quad (S32)$$

$D_Z^s(|t - t'|) = 1$ implies that the order parameter $Q_i^{(+a)(-a)}(t)$ is temporally rigid. In contrast $Q_i^{(\sigma_1 a)(\sigma_2 b)}(t)$ acquires

non-trivial temporal coupling through the time-dependent $D_Z^r(|t-t'|)$. This yields

$$\mathcal{S}_Z^{(R)}[Q_i^{\alpha\beta}(t)] = J \sum_{\alpha<\beta,ii'} \int_0^{t_f} dt Q_i^{\alpha\beta}(t) K_{ii'} Q_{i'}^{\alpha\beta}(t) - 2J^2 \sum_{a,ii'} \int_0^{t_f} dt \int_0^{t_f} dt' \Big(\sum_j K_{ij} K_{i'j}\Big) Q_i^{(+a)(-a)}(t) Q_{i'}^{(+a)(-a)}(t')$$
$$- 2J^2 \sum_{a<b,\sigma_1\sigma_2,ii'} \int_0^{t_f} dt \int_0^{t_f} dt' \Big(\sum_j K_{ij} K_{i'j}\Big) D_Z(|t-t'|) Q_i^{(\sigma_1 a)(\sigma_2 b)}(t) Q_{i'}^{(\sigma_1 a)(\sigma_2 b)}(t') + \mathcal{O}(Q^3).$$
(S33)

Now we consider the replica limit, in which we should keep $Q_i^{(+a)(-a)}(t) = Q^s$ a finite constant and variate with respect to $Q_i^{(\sigma_1 a)(\sigma_2 b)}(t) = Q_i^{r,\sigma_1\sigma_2}(t)$. Setting $Q^s$ finite and $Q_i^{(\sigma_1 a)(\sigma_2 b)}(t) = 0$ in Eq. (S13), the effective model on each replica copy becomes

$$L_0(t) = 4dJ Q^s Z^+ Z^- + \sqrt{2h}\xi(t)(X^+ + X^-).$$
(S34)

The second-order variation of $Q_i^{(\sigma_1 a)(\sigma_2 b)}(t)$ is again given by the correlation function (here $a<b$ is implied)

$$\langle \mathcal{T}(Z^{(\sigma_1 a)} Z^{(\sigma_2 b)})(t)(Z^{(\sigma_1' a')} Z^{(\sigma_2' b')})(t')\rangle_R = \delta_{aa'}\delta_{bb'} \tilde{D}^{(R),\sigma_1\sigma_2,\sigma_1'\sigma_2'}(t,t').$$
(S35)

It enforces $a=a'$, $b=b'$ while leaving $\sigma_1$, $\sigma_1'$, $\sigma_2$, $\sigma_2'$ arbitrary. Evaluating it immediately leads to Eq. (S37). Under this kind of resummation, we separate the $\mathcal{S}_r$ part from the action (S22), which leads to

$$\mathcal{S}_r[Q_i^{(\sigma_1 a)(\sigma_2 b)}(t)] = \sum_{a<b,ii'} \Big[ \sum_{\sigma_1\sigma_2} J \int_0^{t_f} dt Q_i^{(\sigma_1 a)(\sigma_2 b)}(t) K_{ii'} Q_i^{(\sigma_1 a)(\sigma_2 b)}(t)$$
$$- 2J^2 \Big(\sum_j K_{ij} K_{i'j}\Big) \int_0^{t_f} dt \int_0^{t_f} dt' \sum_{\sigma_1\sigma_2,\sigma_1'\sigma_2'} \tilde{D}^{(R),\sigma_1\sigma_2,\sigma_1'\sigma_2'}(t,t') Q_i^{(\sigma_1 a)(\sigma_2 b)}(t) Q_{i'}^{(\sigma_1' a)(\sigma_2' b)}(t')\Big] + \mathcal{O}(Q^3)$$
$$\xrightarrow{R\to 1} \sum_{ii'} \Big[ \sum_{\sigma_1\sigma_2} J \int_0^{t_f} dt Q_i^{r,\sigma_1\sigma_2}(t) K_{ii'} Q_i^{r,\sigma_1\sigma_2}(t)$$
$$- 2J^2 \Big(\sum_j K_{ij} K_{i'j}\Big) \int_0^{t_f} dt \int_0^{t_f} dt' \sum_{\sigma_1\sigma_2,\sigma_1'\sigma_2'} \tilde{D}^{(1),\sigma_1\sigma_2,\sigma_1'\sigma_2'}(t,t') Q_i^{r,\sigma_1\sigma_2}(t) Q_{i'}^{r,\sigma_1'\sigma_2'}(t')\Big] + \mathcal{O}(Q^{r3}),$$
(S36)

where

$$\tilde{D}^{(R),\sigma_1\sigma_2,\sigma_1'\sigma_2'}(t,t') = \frac{\int D\xi P_G[\xi] K^{\sigma_1\sigma_1'}[\xi;t_>,t_<] K^{\sigma_2\sigma_2'}[\xi;t_>,t_<] K_0[\xi]^{R-2}}{\int D\xi P_G[\xi] K_0[\xi]^R},$$
(S37)

We then diagonalize the Hessian matrix in the $R\to 1$ limit similarly through the Keldysh rotation

$$Q^r \to H Q^r H^T, \quad \text{where} \quad H = \frac{1}{\sqrt{2}}\begin{pmatrix} 1 & 1 \\ 1 & -1 \end{pmatrix}, \quad Q^r = \begin{pmatrix} Q^{r,++} & Q^{r,+-} \\ Q^{r,-+} & Q^{r,--} \end{pmatrix},$$
(S38)

to determine the second-order phase transition,

$$\tilde{D}^{(1)} = \int D\xi \frac{P_G[\xi] K_0[\xi]^{-1}}{\int D\xi P_G[\xi] K_0[\xi]} \begin{pmatrix} K^{++}K^{++} & K^{++}K^{+-} & K^{+-}K^{++} & K^{+-}K^{+-} \\ K^{++}K^{-+} & K^{++}K^{--} & K^{+-}K^{-+} & K^{+-}K^{--} \\ K^{-+}K^{++} & K^{-+}K^{+-} & K^{--}K^{++} & K^{--}K^{+-} \\ K^{-+}K^{-+} & K^{-+}K^{--} & K^{--}K^{-+} & K^{--}K^{--} \end{pmatrix}$$
$$\to \int D\xi \frac{P_G[\xi] K_0[\xi]^{-1}}{\int D\xi P_G[\xi] K_0[\xi]} \begin{pmatrix} (K^{++}+K^{+-})^2 & 0 & 0 & 0 \\ 0 & K^{++2}-K^{+-2} & 0 & 0 \\ 0 & 0 & K^{++2}-K^{+-2} & 0 \\ 0 & 0 & 0 & (K^{++}-K^{+-})^2 \end{pmatrix}.$$
(S39)

where we omitted the arguments of $K^{\sigma,\sigma'}[\xi;t_>,t_<]$ for notational simplicity. By noticing that the sign change $Q^s \to$

$-Q^s$ leads to $K^{++} \to K^{++}$ and $K^{+-} \to -K^{+-}$, the largest eigenvalue is the one given by

$$\tilde{D}^K(t,t') = \int D\xi \frac{P_G[\xi]K_0[\xi]^{-1}}{\int D\xi P_G[\xi]K_0[\xi]} \left(K^{++}[\xi;t_>,t_<] + |K^{+-}[\xi;t_>,t_<]|\right)^2$$
$$= \int D\xi P[\xi] \left(\langle Z^+(t_>)Z^+(t_<)\rangle + \langle Z^+(t_>)Z^-(t_<)\rangle\right)^2. \quad (S40)$$

For the partial readout case, the above derivation of replica limit is universal and the only difference is the form of $L_0$. The replacement $\xi_j X_j^{(\sigma,a)} \to \xi_j^a X_j^{(\sigma,a)}$ leads to

$$L_0(t) = 4dJQ^s Z^+ Z^- + 2hX^+ X^-, \quad (S41)$$

where we already integrated $\xi_j^a$ inside each replica copy. The model is analytically solvable,

$$\langle \mathcal{T}(Z^{(\sigma_1 a)} Z^{(\sigma_2 b)})(t)(Z^{(\sigma_1' a')} Z^{(\sigma_2' b')})(t')\rangle_R = \delta_{aa'}\delta_{bb'} \tilde{D}^{(R),\sigma_1\sigma_2,\sigma_1'\sigma_2'}(t,t')$$
$$= \delta_{aa'}\delta_{bb'} (\tanh 4dJQ^s)^{\frac{2-\sigma_1\sigma_1'-\sigma_2\sigma_2'}{2}} \frac{\cosh^2(2h(t_f - 2\Delta t))}{\cosh^2(2ht_f)}, \quad (S42)$$

which is now independent of $R$. We then substitute in the mean-field equation of $Q^s$ Eq. $Q^s = \tanh(4dJt_f Q^s)$, leading to Eq.

$$\tilde{D}_Z^{\sigma_1\sigma_2,\sigma_1'\sigma_2'}(\Delta t) = (Q^s)^{\frac{2-\sigma_1\sigma_1'-\sigma_2\sigma_2'}{2}} D_Z(\Delta t). \quad (S43)$$

## SIII.  SADDLE-POINT SOLUTION OF $f_0$

Here we study the measurement trajectory distribution of the reduced $0+1$d model $L_0$. As explained in the main text, the distribution follows from the effective Landau free energy,

$$f_0[\xi] = \int_0^{t_f} dt \frac{\xi(t)^2}{2} - \frac{1}{t_f} \log \text{tr}\, \mathcal{T} \exp\left[\int_0^{t_f} dt \left(4dJQ^s Z^+ Z^- + \sqrt{2h}\xi(t)(X^+ + X^-)\right)\right], \quad (S44)$$

and $P[\xi] = e^{-f_0[\xi]}/\int D\xi e^{-f_0[\xi]}$. It describes the measurement trajectory probability of simultaneous $X$ measurement and $Z$, and the globally minimal trajectory that satisfies $\delta f/\delta \xi(t) = 0$ is believed to contribute most at the mean-field level. Such a saddle-point equation

$$\frac{\delta}{\delta \xi(t)} f[\xi] = \xi(t) - \frac{1}{t_f} \frac{\delta}{\delta \xi(t)} \log K_0[\xi] = 0 \quad (S45)$$

has a stationary solution $\xi(t) = \xi$, which is more reliable when the outcomes at subsequent time steps are unlikely to change, that is for large $h$. In that case, the effective model $L_0$ becomes time translational invariant, and could be exactly solved to yield,

$$f_0(\xi) = \frac{\xi^2}{2} - \frac{1}{t_f} \log \left[2\left(\cosh\left(2t_f\sqrt{4d^2J^2Q^{s2} + 2h\xi^2}\right) + \cosh\left(4dJt_f Q^s\right)\right)\right]. \quad (S46)$$

The corresponding saddle-point equation becomes $\partial f_0/\partial \xi = 0$, resulting in

$$\sqrt{4d^2J^2Q^{s2} + 2h\xi^2} = \frac{4h\sinh\left(2t_f\sqrt{4d^2J^2Q^{s2} + 2h\xi^2}\right)}{\cosh(4dJt_f Q^s) + \cosh\left(2t_f\sqrt{4d^2J^2Q^{s2} + 2h\xi^2}\right)}. \quad (S47)$$

We numerically solve the equation and find that there is a phase transition separating an ordered phase $\xi > 0$ from a disordered case $\xi = 0$ (Fig. S1). For sufficiently small $t_f$ and $h$ the system stays in the disordered phase. Increasing $t_f$ and $h$, the system goes through a phase transition and enters the ordered phase. We numerically identify a critical point $(h_0, t_0)$. For $t_f < t_0$ the phase transition is second order, while for $t_0 < t_f < +\infty$ the phase transition is first

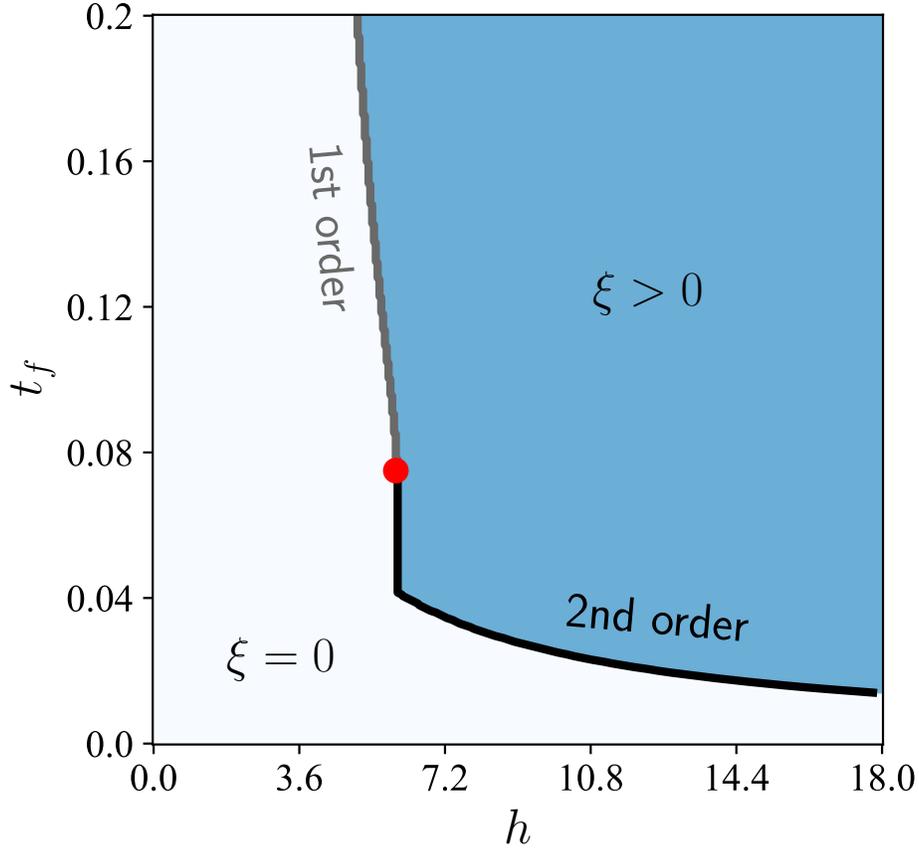

FIG. S1. Mean-field phase diagram of $f_0$ (assuming $d = 6$, $J = 1$). For sufficiently small $t_f$ and $h$ the system is in a $\xi = 0$ phase. For large $t_f$ and $h$ the system is in a $\xi > 0$ phase. The critical point ($h_0 \simeq 6, t_0 \simeq 0.075$) is labeled in Red, below which the phase transition is of second order (black line), and above which the phase transition is of first order (gray line). Notice that when $t_f = +\infty$ the phase transition again becomes a second-order one.

order. Consider the $t_f \to +\infty$ limit, the free energy becomes

$$f_0 = \frac{\xi^2}{2} - 2\sqrt{4d^2 J^2 + 2h\xi^2}, \tag{S48}$$

with the following global minimum,

$$\xi = \begin{cases} \pm \sqrt{8h - 2d^2 J^2/h}, & h > h_0, \\ 0, & h \leq h_0, \end{cases} \tag{S49}$$

Such a behavior describes a second-order phase transition at $h_0 = dJ/2$.

We then turning to calculating the propagator $\tilde{D}^K(t, t')$ by substituting the saddle-point value of $\xi$ in $(\langle Z^+(t_>)Z^+(t_<)\rangle + \langle Z^+(t_>$
Since $L_0$ is now time-independent, we are able to evaluate the analytical form of its correlation function. In particular,

the propagator again becomes time-translational invariant $\tilde{D}^K(t,t') \simeq \tilde{D}^K(|t-t'|)$. As a result

$$\tilde{D}^K(|t-t'|) = \left\{ \sqrt{2}dJQ^s \left( e^{4dJQ^s(t_f-\Delta t)} \sinh\left(2\Delta t\sqrt{4d^2J^2Q^{s2}+2h\xi^2}\right) - e^{4d\Delta t JQ^s} \sinh\left(2(\Delta t-t_f)\sqrt{4d^2J^2Q^{s2}+2h\xi^2}\right) \right) \right.$$
$$+ \sqrt{2d^2J^2Q^{s2}+h\xi^2} \left( \cosh(4dJQ^s(\Delta t-t_f)) \cosh\left(2\Delta t\sqrt{4d^2J^2Q^{s2}+2h\xi^2}\right) + e^{4d\Delta tJQ^s} \cosh\left(2(\Delta t-t_f)\sqrt{4d^2J^2Q^{s2}+2h\xi^2}\right) \right.$$
$$\left.\left. - \sinh(4dJQ^s(\Delta t-t_f)) \cosh\left(2\Delta t\sqrt{4d^2J^2Q^{s2}+2h\xi^2}\right) \right) \right\}^2$$
$$\Big/ (2d^2J^2Q^{s2}+h\xi^2) \left( \cosh\left(2t_f\sqrt{4d^2J^2Q^{s2}+2h\xi^2}\right) + \cosh(4dJt_fQ^s) \right)^2.$$
(S50)

With the above expression of the propagator, the phase boundary of the complete readout case can be determined by the equation $1 - 4dJ \int_0^{t_f} dt \tilde{D}^K(t) = 0$, which has the analytical form,

$$0 = 8\sqrt{2}dhJQ^s\xi^2 \left(2d^2J^2Q^{s2}+h\xi^2\right)^{3/2} \left(\cosh(4dJQ^s t_f) + \cosh\left(2\sqrt{4d^2J^2Q^{s2}+2h\xi^2}t_f\right)\right)^2$$
$$- 4de^{-4\sqrt{4d^2J^2Q^{s2}+2h\xi^2}t_f} J \left( dJ \left( -dhJ\xi^2 \left( 16e^{2\left(2dJQ^s+\sqrt{4d^2J^2Q^{s2}+2h\xi^2}\right)t_f} ht_f\xi^2 - 16e^{4dJt_fQ^s+6\sqrt{4d^2J^2Q^{s2}+2h\xi^2}t_f} ht_f\xi^2 \right.\right.\right.$$
$$- 3e^{8\sqrt{4d^2J^2Q^{s2}+2h\xi^2}t_f}\sqrt{4d^2J^2Q^{s2}+2h\xi^2} + 6e^{4\left(2dJQ^s+\sqrt{4d^2J^2Q^{s2}+2h\xi^2}\right)t_f}\sqrt{4d^2J^2Q^{s2}+2h\xi^2} - 3\sqrt{4d^2J^2Q^{s2}+2h\xi^2}\right) Q^s$$
$$+ 2d^2\left(1+e^{4\sqrt{4d^2J^2Q^{s2}+2h\xi^2}t_f}\right) hJ^2\xi^2 \left( 8e^{2\left(2dJQ^s+\sqrt{4d^2J^2Q^{s2}+2h\xi^2}\right)t_f}\sqrt{4d^2J^2Q^{s2}+2h\xi^2}t_f + 5e^{4\sqrt{4d^2J^2Q^{s2}+2h\xi^2}t_f} - 5 \right) Q^{s2}$$
$$- 8d^3J^3 \left( 4e^{2\left(2dJQ^s+\sqrt{4d^2J^2Q^{s2}+2h\xi^2}\right)t_f} ht_f\xi^2 - 4e^{4dJt_fQ^s+6\sqrt{4d^2J^2Q^{s2}+2h\xi^2}t_f} ht_f\xi^2 - e^{8\sqrt{4d^2J^2Q^{s2}+2h\xi^2}t_f}\sqrt{4d^2J^2Q^{s2}+2h\xi^2}\right.$$
$$\left.+ 2e^{4\left(2dJQ^s+\sqrt{4d^2J^2Q^{s2}+2h\xi^2}\right)t_f}\sqrt{4d^2J^2Q^{s2}+2h\xi^2} - \sqrt{4d^2J^2Q^{s2}+2h\xi^2}\right) Q^{3s} + 16d^4\left(-1+e^{8\sqrt{4d^2J^2Q^{s2}+2h\xi^2}t_f}\right) J^4Q^{s4}$$
$$+ h^2\xi^4 \left( 2e^{2\left(2dJQ^s+\sqrt{4d^2J^2Q^{s2}+2h\xi^2}\right)t_f} \left(2\sqrt{4d^2J^2Q^{s2}+2h\xi^2}t_f - 1\right) + e^{8\sqrt{4d^2J^2Q^{s2}+2h\xi^2}t_f}\right.$$
$$\left.\left.+ 2e^{4dJt_fQ^s+6\sqrt{4d^2J^2Q^{s2}+2h\xi^2}t_f} \left(2\sqrt{4d^2J^2Q^{s2}+2h\xi^2}t_f + 1\right) - 1 \right) Q^s$$
$$\left.\left.+ 2e^{4\left(dJQ^s+\sqrt{4d^2J^2Q^{s2}+2h\xi^2}\right)t_f} h^2\xi^4\sqrt{4d^2J^2Q^{s2}+2h\xi^2} \sinh(4dJQ^s t_f) \right).$$
(S51)

Again $Q^s$ and $\xi$ take their saddle-point values. In the stationary limit $t_f \to +\infty$, the propagator approaches

$$\tilde{D}^K(|t-t'|) \to \frac{\left(\sqrt{2d^2J^2+h\xi^2}+\sqrt{2}dJ\right)^2}{2d^2J^2+h\xi^2} e^{8d\Delta tJ - 4\Delta t\sqrt{4d^2J^2+2h\xi^2}}, \tag{S52}$$

where $\xi$ takes the value of Eq. (S49).

---

[1] J. Ashkin and E. Teller, Statistics of two-dimensional lattices with four components, Phys. Rev. **64**, 178 (1943).


\clearpage

\end{document}